\documentclass[aip, preprint, amsmath,amssymb
]{revtex4-1}

\usepackage{graphicx}
\usepackage{dcolumn}
\usepackage{bm}

\usepackage[utf8]{inputenc}
\usepackage[T1]{fontenc}
\usepackage{mathptmx}
\usepackage{etoolbox}
\makeatletter
\def\@email#1#2{%
 \endgroup
 \patchcmd{\titleblock@produce}
  {\frontmatter@RRAPformat}
  {\frontmatter@RRAPformat{\produce@RRAP{*#1\href{mailto:#2}{#2}}}\frontmatter@RRAPformat}
  {}{}
}%
\makeatother
\begin{document}

\preprint{AIP/123-QED}

\title[]{Background-Pressure Effects on Charge-Exchange Measurements in Plasma Flows at Elevated Pressures}
\author{I. Romadanov}
\email{iromada2@pppl.gov}
\author{S. Musikhin}
\affiliation{ Princeton Plasma Physics Laboratory, Princeton, NJ 08540, USA }%
\author{Je‑Hoi Mun}
\author{Sang Ki Nam}
\affiliation{ Samsung Electronics Co., Ltd. 1‑1 Samsungjeonja‑ro, Hwaseong‑si, Gyeonggi‑do, 18448, Republic of Korea }%
\author{Y. Raitses}
\affiliation{ Princeton Plasma Physics Laboratory, Princeton, NJ 08540, USA }%

\date{\today}
\begin{abstract}
Charge-exchange (CEX) collisions can affect measurements of plasma plumes and neutralized ion flows in vacuum facilities, particularly when the background gas pressure increases and the CEX mean free path becomes comparable to the characteristic plume or facility dimension. Here, we investigate that regime in the plasma plume of a gridded ion source operating with a 400 eV argon ion beam. The fast-ion flux and low-energy ion flux were measured using a retarding potential analyzer (RPA) and planar probes, while the fast-neutral flux was inferred from deposited-power measurements with a thermal flux probe using a power-balance analysis. The low-energy ion flux increases with both background gas pressure and axial distance and its detection also depends on probe geometry. After the fast-ion component is isolated, its attenuation is described more accurately by an analytical reduced semi-empirical quasi-2D model that includes charge exchange and the experimentally observed plume divergence than by a one-dimensional attenuation law. The inferred fast-neutral flux also increases with pressure; however, the model underpredicts it at small axial distance and overpredicts it at elevated pressure and larger axial distance. This discrepancy suggests additional angular and collisional effects, as well as possible fast-neutral production near or inside the ion source, that are not captured by the present model. These results show that background gas pressure affects both the plasma plume and the diagnostic response, and that complementary electrostatic, thermal, and energy-selective diagnostics are required to distinguish source behavior from facility-induced effects.
\end{abstract}

\maketitle

\section{\label{sec:intro}Introduction}

Charge-exchange (CEX) collisions are an important source of facility effects in laboratory studies of ion and plasma flows\cite{Brown_Facility}, including ion and neutral beam sources\cite{Hopf_2021} for fusion\cite{Stotler_NSTX} and materials processing\cite{Economou_2008}, as well as electric propulsion devices.\cite{Boyd_Plume_Model,Walker_Backpressure,Tajmar_SMART} In such systems, the flow propagates through a finite neutral background, so interactions with residual gas can alter both the flow itself and the diagnostics used to characterize it. This issue is especially relevant when measurements are performed in vacuum facilities where characteristic length scales are comparable to the collisional mean free path.

In such vacuum facilities, the chamber has finite size and finite pumping speed, while gas is continuously introduced through the plasma source.\cite{huang2016facility,passaro2004plasma,jin2022vacuum} As a result, a residual background gas remains in the chamber, and its pressure can become large enough to influence the measured flow.\cite{snyder2020effects,Nakles_Background_IVDF} These facilities enable performance evaluation,\cite{raitses2007enhanced} plume characterization,\cite{fisch2008plasma,azziz2007experimental} and long-duration operation or testing,\cite{frieman2018long} but they do not reproduce collisionless or free-space conditions exactly. Under these conditions, measured quantities such as the angular ion flux distribution, the ion energy distribution function, and plasma composition may reflect not only the characteristics of the plasma source, but also facility-induced effects associated with interactions of the plasma flow with background gas and nearby surfaces.\cite{Foster_ground_test_impact}

These effects complicate interpretation of the experiment: source properties must be distinguished from chamber-induced effects, and plume evolution must be separated from measurement bias introduced by the chamber environment. Standard electrostatic diagnostics like ion probes or energy analyzers\cite{dorf2004electrostatic, staack2004shielded} remain essential for characterizing charged-particle transport, but interpretation becomes more difficult as background gas pressure increases and multiple particle populations contribute to the measured signal.\cite{farnell2017recommended} Complementary diagnostics that respond to the energy fluxes \cite{rosenfeldt2021use,thornton1978substrate} rather than collected charge are therefore of interest for identifying facility-induced changes that are not captured by electrostatic measurements alone.\cite{Stahl_calorimetric} Although such diagnostic techniques are well established in plasma processing\cite{wiese2015energy,bornholdt2013transient,schlichting2023retarding} and plasma--surface interaction studies,\cite{schlichting2022energy,el2022correlation} their application to plume measurements in the context of the facility-induced CEX remains limited.

The present work investigates this problem for a gridded ion source operated with argon. In the present configuration, the extracted ion beam is neutralized downstream by electrons from the neutralizer, and the resulting quasineutral flow is referred to here as the plasma plume. The chamber background gas pressure is varied by controlled argon injection while the source operating conditions are otherwise held fixed. Measurements are performed for a commercial Kaufman-type source\cite{van1986characterization} at a beam energy near 400~eV over a pressure range of $10^{-5}$--$10^{-3}$~Torr, spanning conditions relevant to laboratory ion beam studies, plasma-source applications for microelectronics and ground-test facilities. The objective is to determine how increasing background gas pressure modifies the measured plume and to assess how consistently those changes can be interpreted using complementary diagnostics and a reduced quasi-2D model.

This study makes two main contributions. First, it measures the pressure dependence of fast-ion attenuation as a function of axial distance using retarding potential and planar probe diagnostics, over a pressure range spanning low pressures relevant to thruster test facilities and higher pressures relevant to ion beam processing applications.\cite{Vawter2000IonBE,Zhurin_Industrial_Ion_Sources} The measurements are compared with a reduced semi-empirical quasi-2D model that includes plume divergence and charge exchange. Second, the fast-neutral energy flux is inferred from thermal measurements using a power-balance approach, enabling direct comparison of the charged and neutral contributions to the plume. The combined results show that the corrected fast-ion attenuation is broadly consistent with the reduced model, whereas the inferred fast-neutral flux is underpredicted near the reference position and overpredicted at elevated pressure and larger axial distance, indicating that additional effects not captured by the simplified model influence the measured power balance.

The paper is organized as follows. Sections~\ref{sec:exper} and \ref{sec:diagn} describe the experimental facility, diagnostics, and data processing. Section~\ref{sec:model} presents the reduced quasi-2D model. Section~\ref{sec:results} presents the experimental results and comparison with the model. Section~\ref{sec:conclusion} summarizes the main conclusions.

\newpage

\section{\label{sec:CEX} General considerations of CEX in the plasma flows}
CEX collisions\cite{lichten1963resonant} modify a directed plasma plume by converting flux of fast ions into fast neutrals while creating a secondary population of low-energy ions. In the present context, ``fast ions'' denotes the directed high-energy flux component, whereas ``slow ions'' denotes the lower-energy ions produced by CEX and facility related background processes. Because the CEX rate scales with neutral density, the background gas pressure directly controls the importance of this process.

A first estimate of CEX-induced flux attenuation is described by the Beer--Lambert law\cite{swinehart1962beer} which is valid for a one-dimensional, monoenergetic model for an ion population propagating through a uniform neutral background,
\begin{equation}
\Gamma_i(L)=\Gamma_{i,0}\exp\left(-L/\lambda_{\mathrm{CEX}}\right),
\label{eq:BLL}
\end{equation}
where $\Gamma_{i,0}$ is the initial fast-ion flux, $\Gamma_i(L)$ is the fast-ion flux after propagation over distance $L$, and $\lambda_{\mathrm{CEX}}$ is the charge-exchange mean free path. For a uniform background gas,
\begin{equation}
\lambda_{\mathrm{CEX}}=1/n_g \sigma_{\mathrm{CEX}},
\label{eq:CEX_MFP}
\end{equation}
where $n_g$ is the neutral density and $\sigma_{\mathrm{CEX}}$ is the charge-exchange cross section (cross-section data were taken from Ref. \onlinecite{bromley2019symmetric}). In this approximation, the attenuation is governed by the parameter $n_g \sigma_{\mathrm{CEX}}L$. This model is useful as a baseline, but its interpretation in real plumes requires careful consideration. A reduction in the measured fast-ion flux can result not only from CEX losses, but also from plume divergence, spatial variations in ion density and energy, and collisional redistribution. In this paper, ``attenuation'' therefore refers specifically to the reduction of the fast-ion flux by CEX, while the measured decrease in the flux must be interpreted in the presence of multidimensional plume evolution.

The products of CEX have different transport and diagnostic signatures. A fast neutral retains most of the directed energy of the incident ion, but it no longer contributes to the collected ion current. The newly created low-energy ion remains detectable by electrostatic probes and can contribute to a low-energy plasma in the chamber volume. As a result, CEX affects both the physical evolution of the plume and the interpretation of probe measurements. Electrostatic probes\cite{godyak2015comparative} are well suited for measuring charged-particle flux and ion energy distributions, but they do not detect fast neutrals directly. In addition, at elevated background pressure, the collected current can include contributions from low-energy ions produced by the CEX process and from background plasma in the chamber. Ion-current measurements alone therefore do not uniquely determine the total directed energy flux carried by the plume. For this reason, electrostatic measurements are complemented by thermal diagnostics, which respond to the net energy deposited at a collector surface by both ions and neutrals.\cite{benedikt2021foundations} When electron heating is minimized, the measured thermal response is dominated by energetic particles reaching the surface. The combined use of electrostatic and thermal diagnostics therefore provides a basis for distinguishing fast-ion attenuation from fast-neutral production and for separating collisional CEX effects from purely geometric changes in the plume.

\newpage

\section{\label{sec:exper}Experimental facility}
\begin{figure}[ht]
\centering
\begin{minipage}[t]{0.49\linewidth}
\centering
a\includegraphics[height = 4cm]{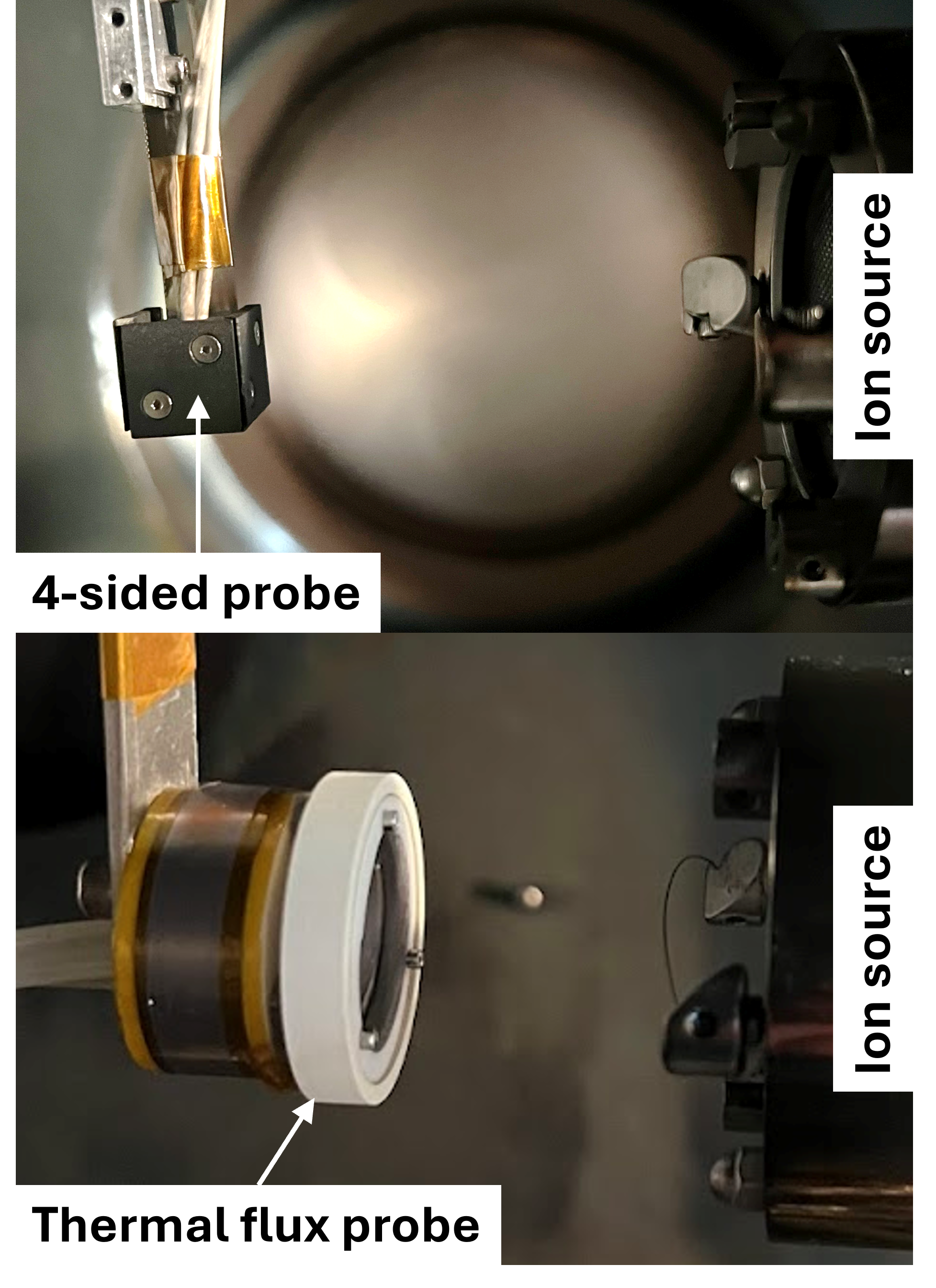}
\end{minipage}\hfill
\begin{minipage}[t]{0.49\linewidth}
\centering
b\includegraphics[height = 4cm]{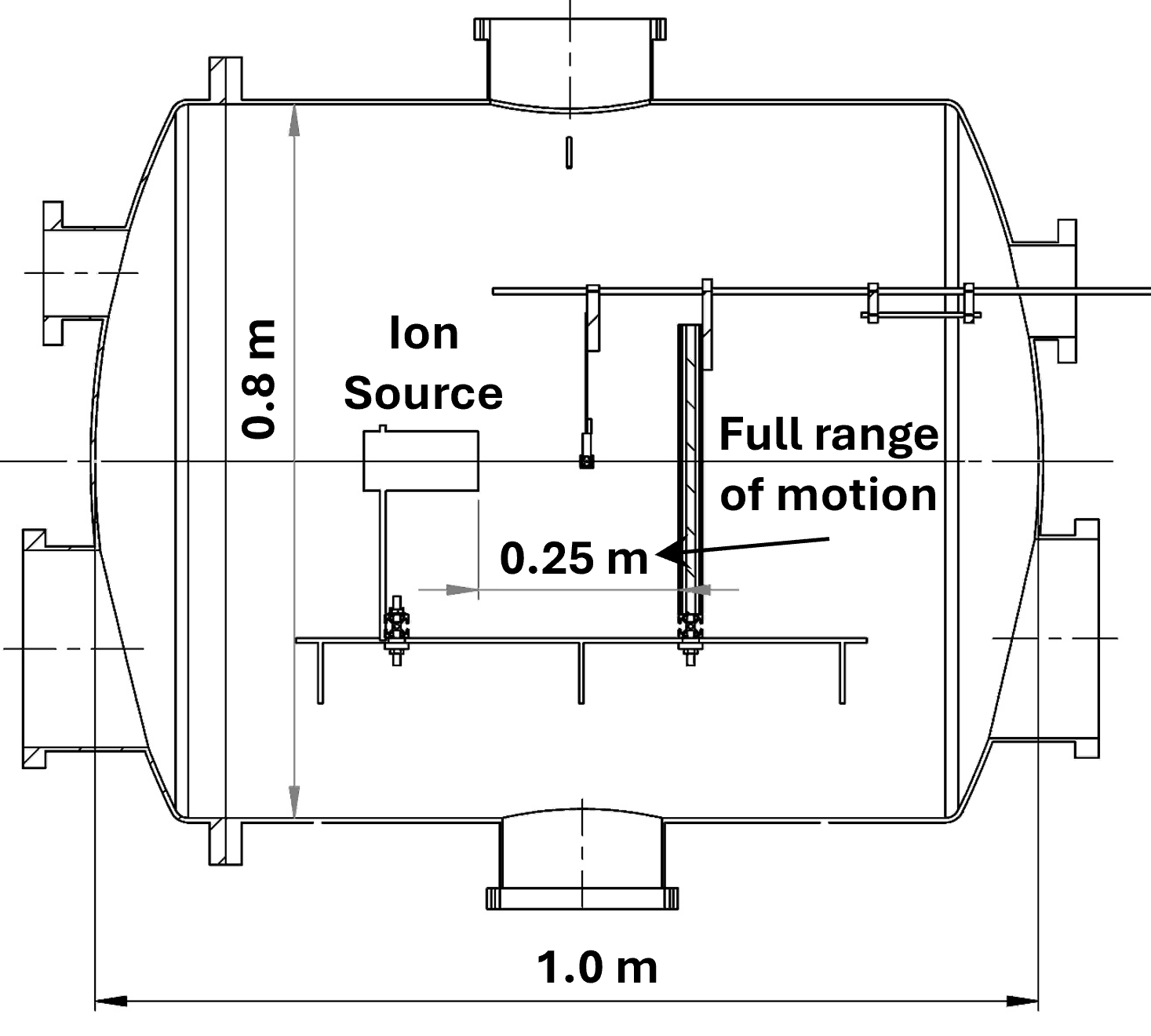}
\end{minipage}
\caption{\label{fig:SHTF} (a) Photographs of the four-sided planar probe (top) and thermal flux probe (bottom) positioned opposite the ion source; (b) characteristic dimensions and schematic of the movable stage.}
\end{figure}

Experiments were conducted in a Small Hall Thruster Facility (SHTF)\cite{Smirnov_CHT_Plasma} at the Princeton Plasma Physics Laboratory. The facility consists of a stainless-steel vacuum chamber approximately 1.0~m in length and 0.8~m in diameter, as shown in Fig. \ref{fig:SHTF}a. The chamber was electrically grounded and was used as the reference potential for all electrical measurements. The pumping system consisted of a blower and a mechanical forepump backing a turbomolecular pump (Osaka TG3203M). The base pressure prior to operation was about $3\times 10^{-6}$~Torr. Chamber pressure was monitored with an ion gauge mounted near the top of the vessel, whose reading was corrected for argon. The manufacturer-stated uncertainty of the pressure reading is $\pm 15\%$. The chamber pressure is denoted as $P$ and is reported in mTorr. Here, we assumed the pressure to be uniform along the measurement region, which is validated in Appendix~\ref{app:pressure}.

Argon was supplied using two mass flow controllers. The ion-source argon flow, denoted $Q_{\mathrm{IS}}$, was delivered to the ion source. An additional argon flow, denoted $Q_{\mathrm{add}}$, was injected through an inlet on the top chamber wall located approximately 10 cm downstream of the ion-source grid plane (ion-source exit plane) to vary the background gas pressure. The ion-source flow rate was typically $Q_{\mathrm{IS}}=3.7$--$4.6$~sccm. The additional argon flow was set to $Q_{\mathrm{add}}=0$, 18, 64, and 128~sccm. For these operating conditions, the pressure range was $P=0.03$--$0.82$~mTorr, corresponding to an effective pumping speed of $2000\pm 150$~L/s for argon.

A 3~cm diameter commercial Kaufman-type gridded ion source was used to produce a directed argon ion beam. The source utilizes two tungsten filaments. The cathode filament sustains the discharge inside the anode chamber, while an external neutralizer filament supplies electrons downstream in order to neutralize the extracted beam. Such neutralized ion beam is referred to here as the plasma plume. Plasma is generated by biasing the anode chamber positive with respect to the cathode filament. Ions are extracted and accelerated by biasing the anode chamber positive with respect to the acceleration grid. Typical operating settings of the ion source in this work were a discharge voltage of 40 V, a discharge current of 1.0--1.3 A, and a beam energy of 400 eV. The measured energy peak was about 400--410~eV. The cathode and neutralizer filament heating currents were typically 8.0~A and 7.5~A, respectively. A warm-up period of about 45~min was allowed for the ion beam current to reach a steady value before measurements.

Fig.~\ref{fig:SHTF}b shows the chamber dimensions and the axial measurement range. The axial coordinate $z$ is defined along the chamber centerline, with $z=0$ at the ion-source exit plane (ion-source acceleration grid location). A reference plane was defined 3~cm downstream of the ion-source exit plane, and all reported axial locations were converted to downstream distances relative to that plane. Diagnostics were mounted on a translation system aligned with the ion-source centerline, which enabled axial scans without breaking vacuum. The axial locations used in this work were $z=6$, 8, 13, 18, and 21~cm; the closest measurement location was $z=6$~cm.

\begin{figure}[ht]
\centering
\begin{minipage}[t]{0.49\linewidth}
\centering
\includegraphics[height=4cm]{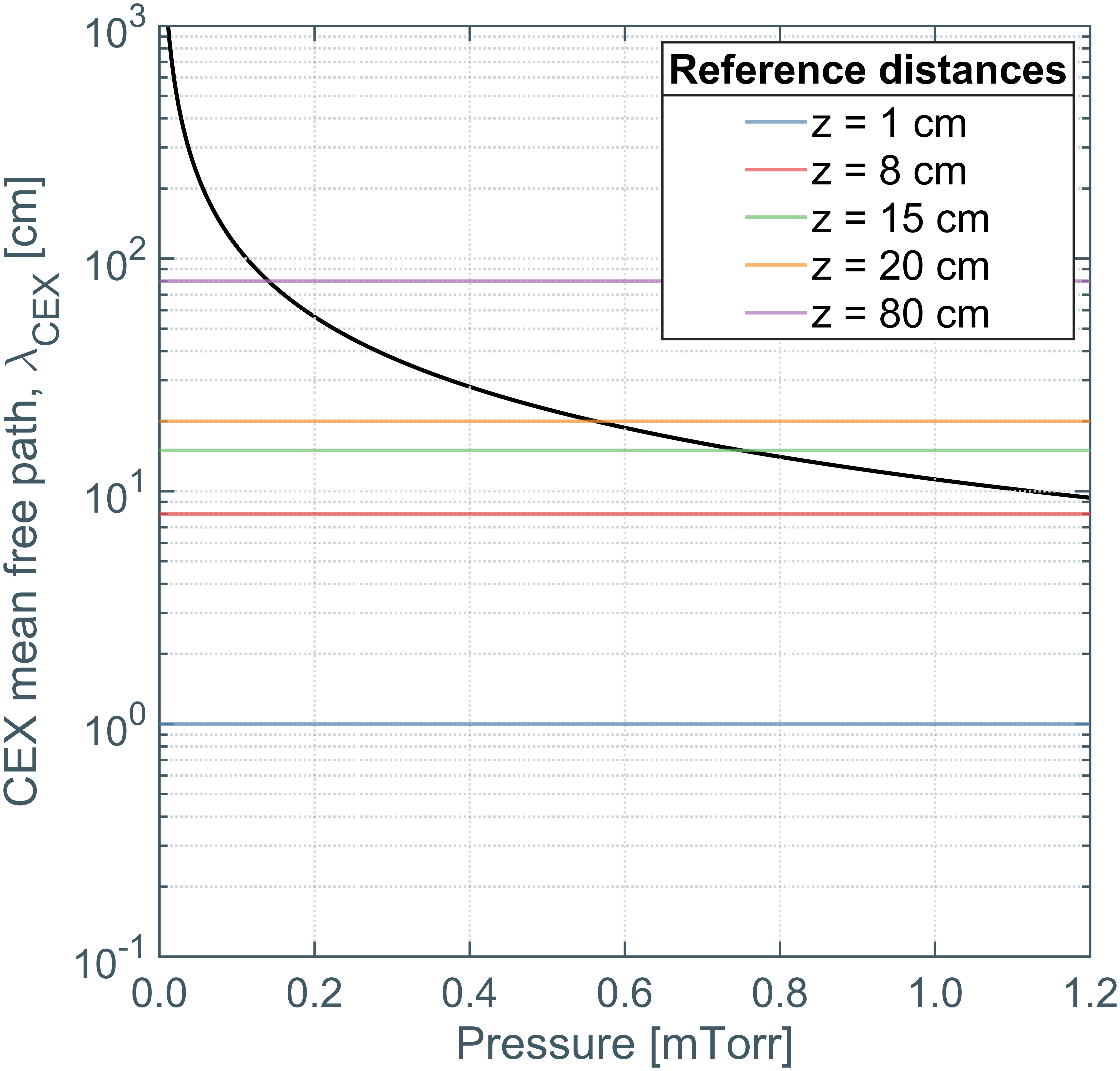}
\end{minipage}\hfill
\begin{minipage}[t]{0.49\linewidth}
\centering
\includegraphics[height=4cm]{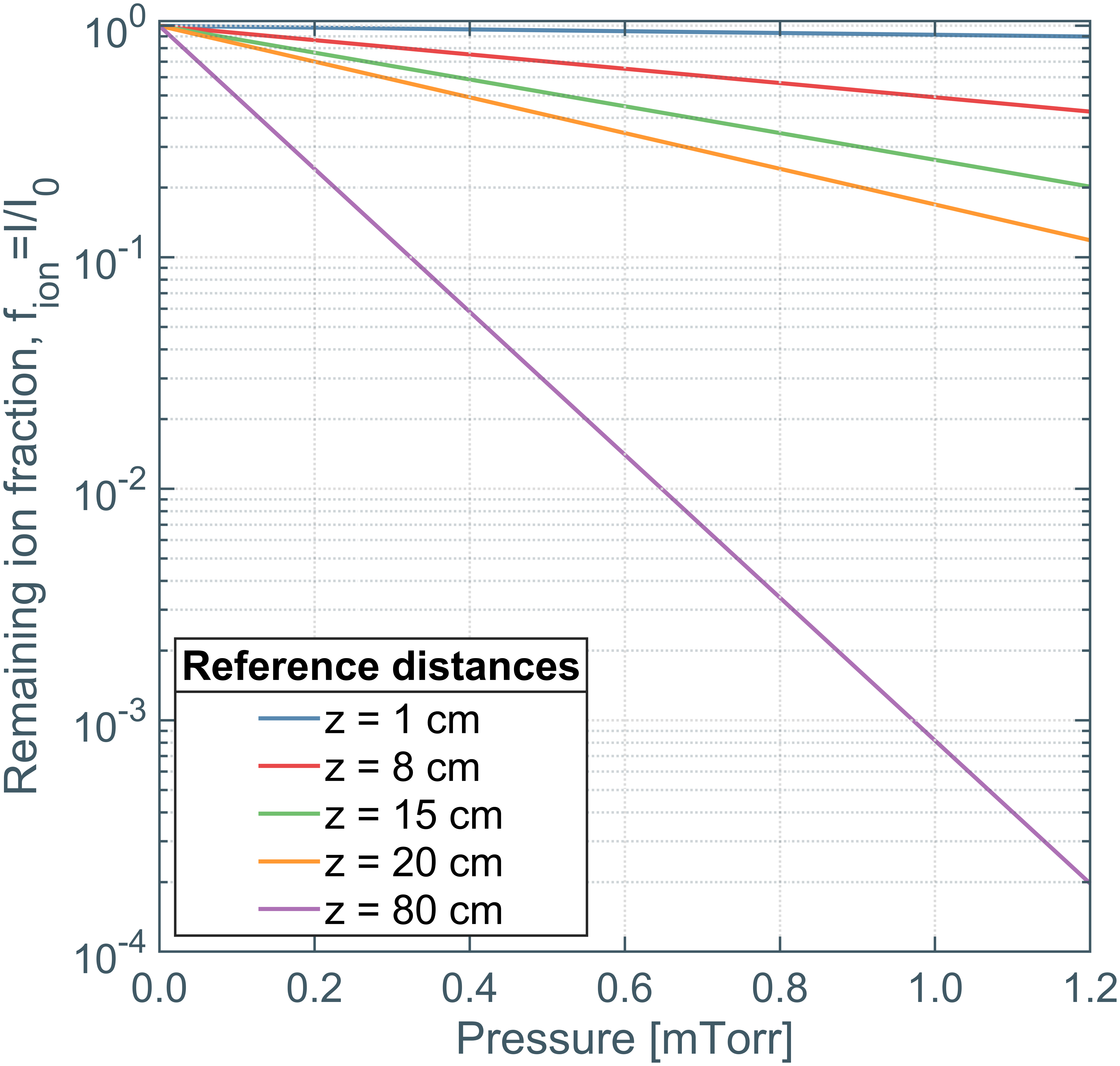}
\end{minipage}
\caption{\label{fig:cex_expect}Estimated charge-exchange mean free path and one-dimensional fast-ion survival fraction versus background gas pressure for $\mathrm{Ar}^+$ at $\approx 400$~eV in argon. Horizontal reference lines indicate characteristic length scales relevant to this work, including axial measurement distances and chamber dimensions.}
\end{figure}

Using Eqs (~\ref{eq:BLL}) and (\ref{eq:CEX_MFP}), the characteristic experimental length scales were compared with $\lambda_{\mathrm{CEX}}$ for $\mathrm{Ar}^+$ at $\approx 400$~eV in argon. Figure~\ref{fig:cex_expect} compares the estimated CEX mean free path, $\lambda_{\mathrm{CEX}}$, and the corresponding one-dimensional fast-ion survival fraction with characteristic distances relevant to the facility and diagnostics. At the lowest pressures, $\lambda_{\mathrm{CEX}}$ exceeds the characteristic chamber size, so only limited collision effects are expected during transport through the facility. As the background gas pressure increases, however, $\lambda_{\mathrm{CEX}}$ decreases to values comparable to the distances from the ion source exit plane to the probe locations. In this regime, CEX can substantially modify the downstream flow even when the source operating conditions are unchanged. The right panel of Figure~\ref{fig:cex_expect} shows the corresponding attenuation of the fast-ion fraction, $f_{\mathrm{ion}} = I/I_0$, predicted by the one-dimensional estimate (Eq.~\ref{eq:BLL}) for several reference distances. The trend indicates that, at the larger downstream locations used in this study, facility-induced CEX alone can produce a substantial reduction in the measured fast-ion flux as pressure increases.

\newpage

\section{\label{sec:diagn}Diagnostics and data processing}
Several probe diagnostics were used in order to characterize particle and energy fluxes in the ion-source plume and to quantify facility-background contributions. The diagnostics considered in this work are: planar probes, an RPA, and a thermal flux probe. The planar probes directly measure ion currents, and the RPA provides information about the energies of the arriving ions, while the thermal flux probe measures the deposited power and simultaneously collects ion current. The thermal flux probe analysis allows one to deduce the neutral energy flux, but it requires separation of the directed fast-ion flux from the low-energy ion flux associated with CEX and background plasma.

\paragraph{Planar ion probe measurements.}
Ion flux was measured with a movable planar probe made from graphite, $\approx$2~cm diameter, operated in a double-sided configuration. Two graphite collectors were mounted back-to-back and electrically insulated from each other. The front collector faced the ion source and measured the on-axis ion current dominated by the fast-ion flux. The rear collector faced away from the source and measured an ion current associated with the background plasma ions, including low-energy ions produced by CEX and ions redirected by collisions. Each collector was biased independently using separate supplies. The collector current was obtained from the voltage drop across a shunt resistor (1~k$\Omega$ for the front and 10~k$\Omega$ for the back in order to improve sensitivity). The probe was aligned with the ion-source axis and positioned at several axial distances. At each location, the bias was set negative with respect to the floating potential in order to collect ions.

\begin{figure}[ht]
\centering
\includegraphics[height=4cm]{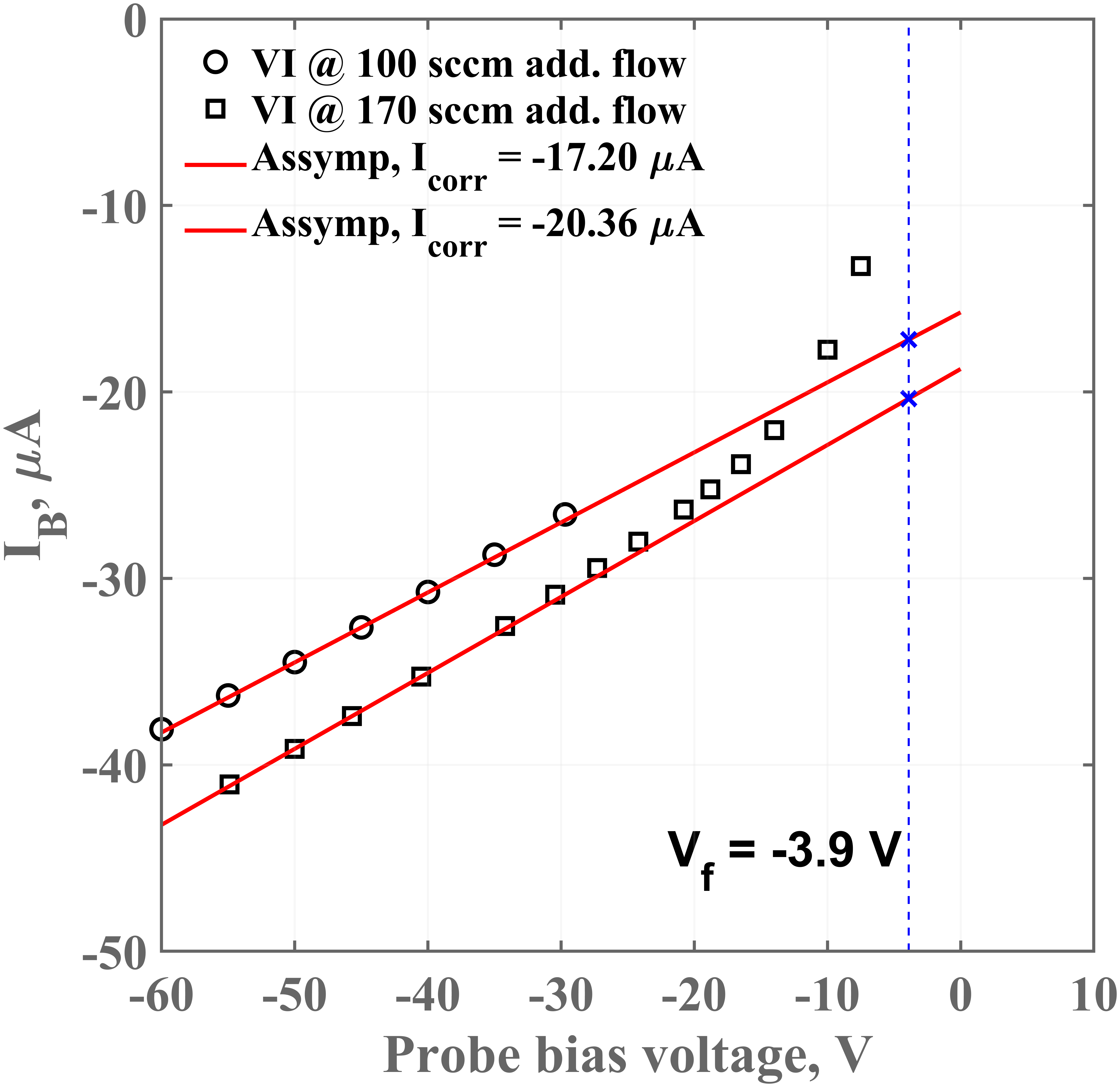}
\caption{\label{fig:vi_planar}Representative current--voltage characteristics of the back collector of the double-sided planar probe measured at two added chamber argon flow rates (100 and 170~sccm), corresponding to background gas pressures of $0.7\times 10^{-3}$~Torr and $1.0\times 10^{-3}$~Torr, respectively. The ion-source argon flow was 4.6~sccm.}
\end{figure}

Current--voltage (V-I) characteristics were measured in selected operating conditions in order to verify ion-collection conditions and to select fixed operating biases for scans. In most measurements, the biases were held constant at $-30$~V (front) and $-45$~V (back) during pressure scans. For a small planar collector, increasing the negative bias expands the ion sheath and increases the effective ion-collection area\cite{Langmuir_Sheridan,Langmuir_Hershkowitz}, so the ion current can continue to rise instead of reaching a clear saturation plateau\cite{Langmuir_Johnson}. This sheath-expansion behavior is well known for planar Langmuir probes\cite{raitses2007aedc, Smirnov_CHT_Plasma}. When a clear ion-saturation plateau was not reached for the background-facing collectors, the ion-collection branch of $I(V)$ was fit and extrapolated to a floating potential in order to define a V-I-corrected background current\cite{chen2003langmuir,lieberman1994principles} (an example is shown in Fig.~\ref{fig:vi_planar}). Current ratios such as $C_{\mathrm{back}}=I_{\mathrm{back}}/I_{\mathrm{front}}$ were evaluated using these V-I-corrected currents when needed.

\paragraph{Four-sided planar probe.}
In order to assess the influence of probe size and to sample directional asymmetries of the background plasma at a fixed axial location, a four-sided planar probe was implemented. The probe consisted of four graphite collectors mounted on a ceramic cube with a characteristic side length of about 1~cm. Adjacent collectors were electrically isolated by a narrow gap. The probe was installed on the axial translation system and aligned with the ion-source axis. Each collector was biased negative with respect to the grounded chamber in order to collect ions. The current from each collector was measured independently using a shunt resistor (1~k$\Omega$ for the front, side, and bottom collectors and 10~k$\Omega$ for the back collector in order to improve sensitivity). The collector facing the ion source measured the directed fast-ion current, while the lateral, bottom, and rear collectors provided simultaneous measurements of off-axis and background plasma ion currents at the same axial location.

\begin{figure}[ht]
\centering
\begin{minipage}[t]{0.49\linewidth}
\centering
\includegraphics[height=4cm]{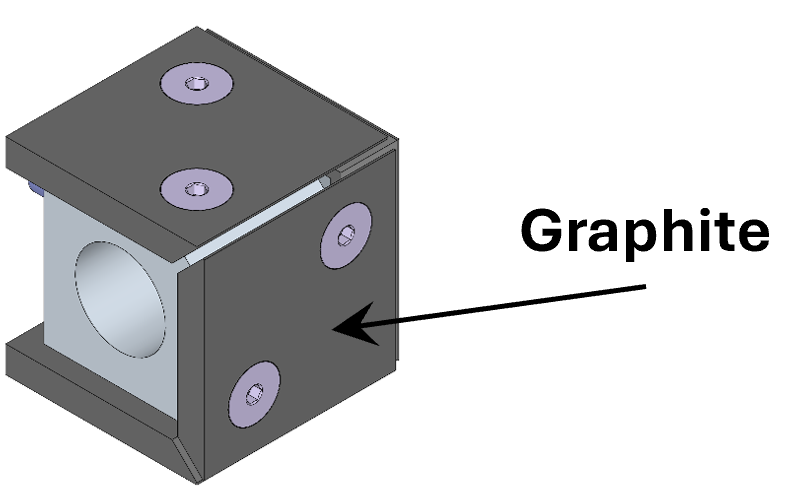}
\end{minipage}\hfill
\begin{minipage}[t]{0.49\linewidth}
\centering
\includegraphics[height=4cm]{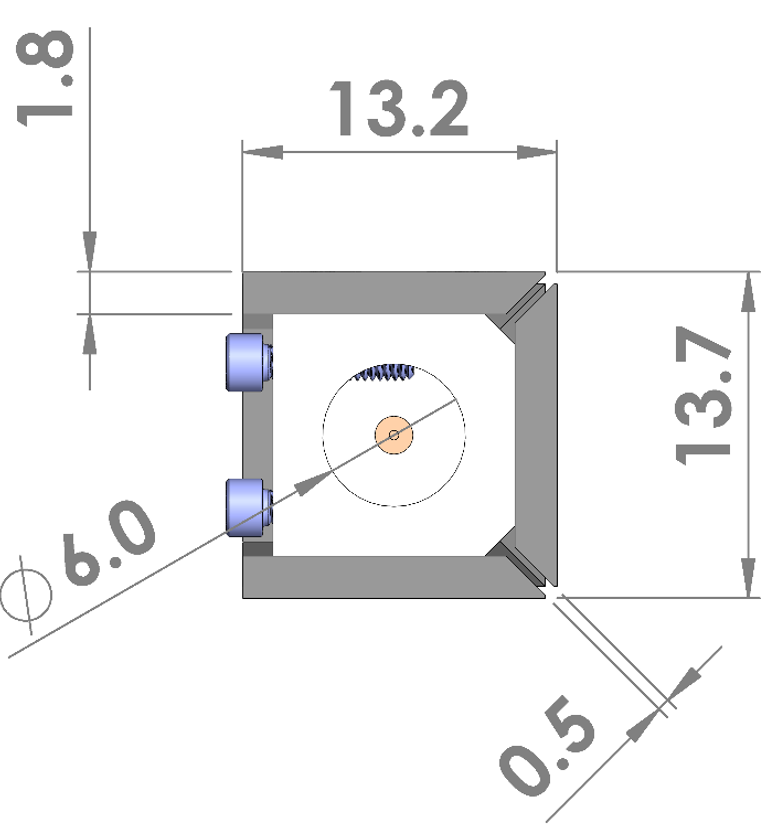}
\end{minipage}
\caption{\label{fig:4side_design}Four-sided probe: (a) CAD model; (b) characteristic dimensions in mm.}
\end{figure}

\begin{figure}[ht]
\centering
\includegraphics[height=4cm]{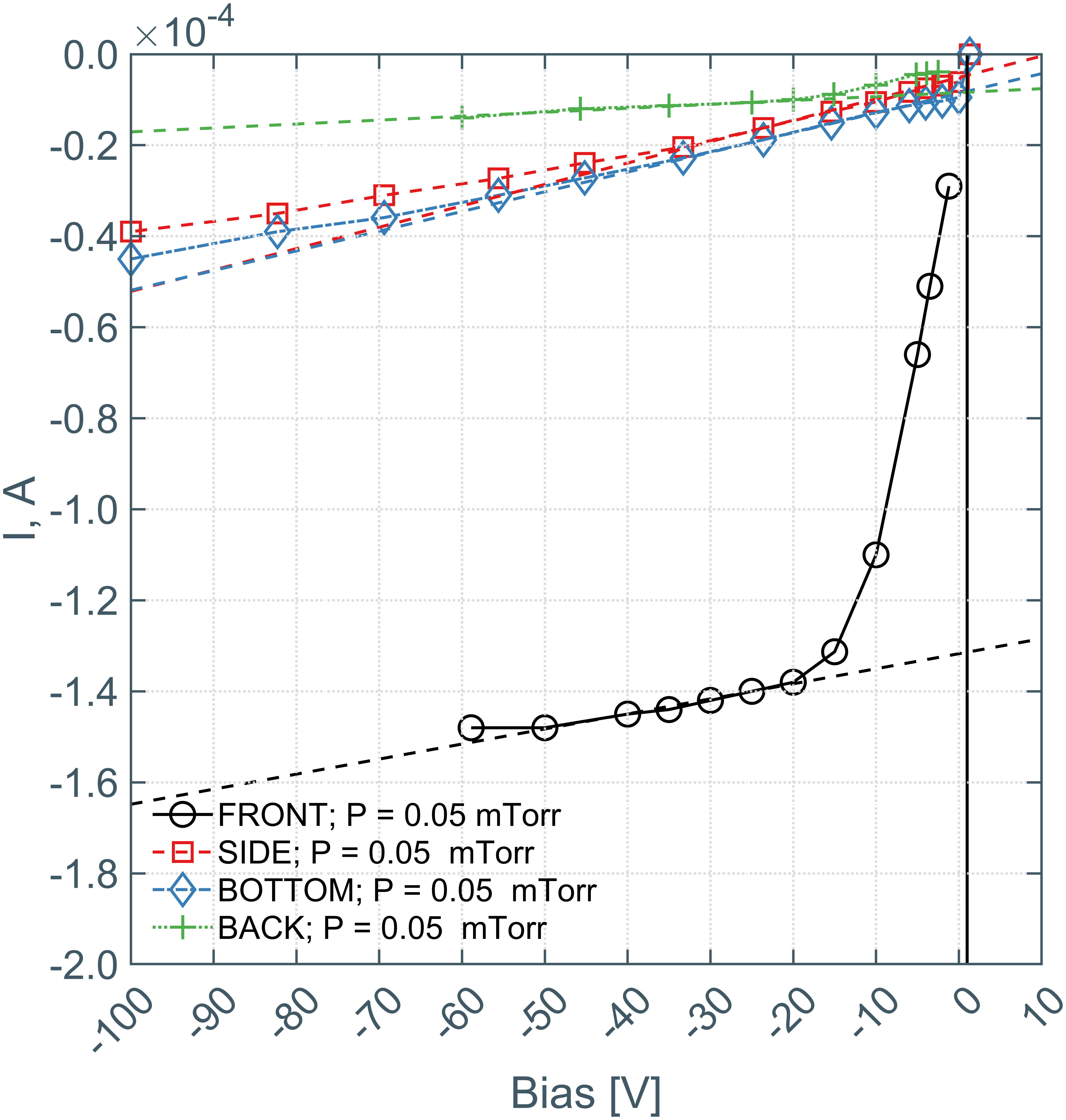}
\caption{\label{fig:4side_vi}Current--voltage characteristics of the four-sided probe collectors measured at $P=0.05$~mTorr. The ``front'' collector is oriented toward the ion source. The side, bottom, and back collectors quantify off-axis and background plasma ion currents at the same location.}
\end{figure}

The same V-I-correction approach that was applied for planar probe was used for four-sided probe: the ion-collection branch of $I(V)$ was fit and extrapolated to a bias-independent limit in order to define V-I-corrected currents for the background-facing collectors. Representative current--voltage characteristics of the four collectors were measured in order to verify ion-collection conditions and to select a fixed operating bias (Fig.~\ref{fig:4side_vi}). The front collector exhibited only a weak dependence of collected current on bias once the bias became sufficiently negative to repel plasma electrons. In contrast, the side, bottom, and back collectors showed no clear ion-saturation plateau over the explored bias range and displayed an approximately monotonic increase of ion current with increasingly negative bias. In such cases, current ratios such as $C_{\mathrm{back}}= I_{\mathrm{back}}/I_{\mathrm{front}}$ were evaluated using these V-I-corrected currents when needed.

\paragraph{Retarding potential analyzer measurements and processing.}
RPA measurements were used in order to resolve the IEDF\cite{dunaevsky2006plasma,granstedt2008cathode,simmonds2020application} in the ion-source plume and to quantify the fraction of low-energy ions associated with CEX and background plasma. The RPA was installed on a movable rod on the axis and translated to several axial locations downstream of the source.

The RPA consists of four grids with fixed biases except for the retarding grid. Grid~1 was held at ground. Grid~2 served as an electron-repelling grid and was biased to $-40$~V with respect to ground. Grid~3 was the retarding grid and was swept from 0 to 500~V. Grid~4 suppressed secondary electron emission and was biased to $-20$~V. The collector was connected to ground through a 10~k$\Omega$ shunt resistor. The collector current was recorded by a DAQ system from the voltage across the shunt. For each pressure condition and axial location, the collector current $I(V_r)$ was measured with 1~s averaging at each retarding-voltage point using 100-point forward and backward sweeps; the forward and backward traces were paired and averaged in order to reduce hysteresis. The overall transmission through the grid stack was $\approx~10\%$.

\begin{figure}[ht]
\centering
\begin{minipage}[t]{0.49\linewidth}
\centering
\includegraphics[height=4cm]{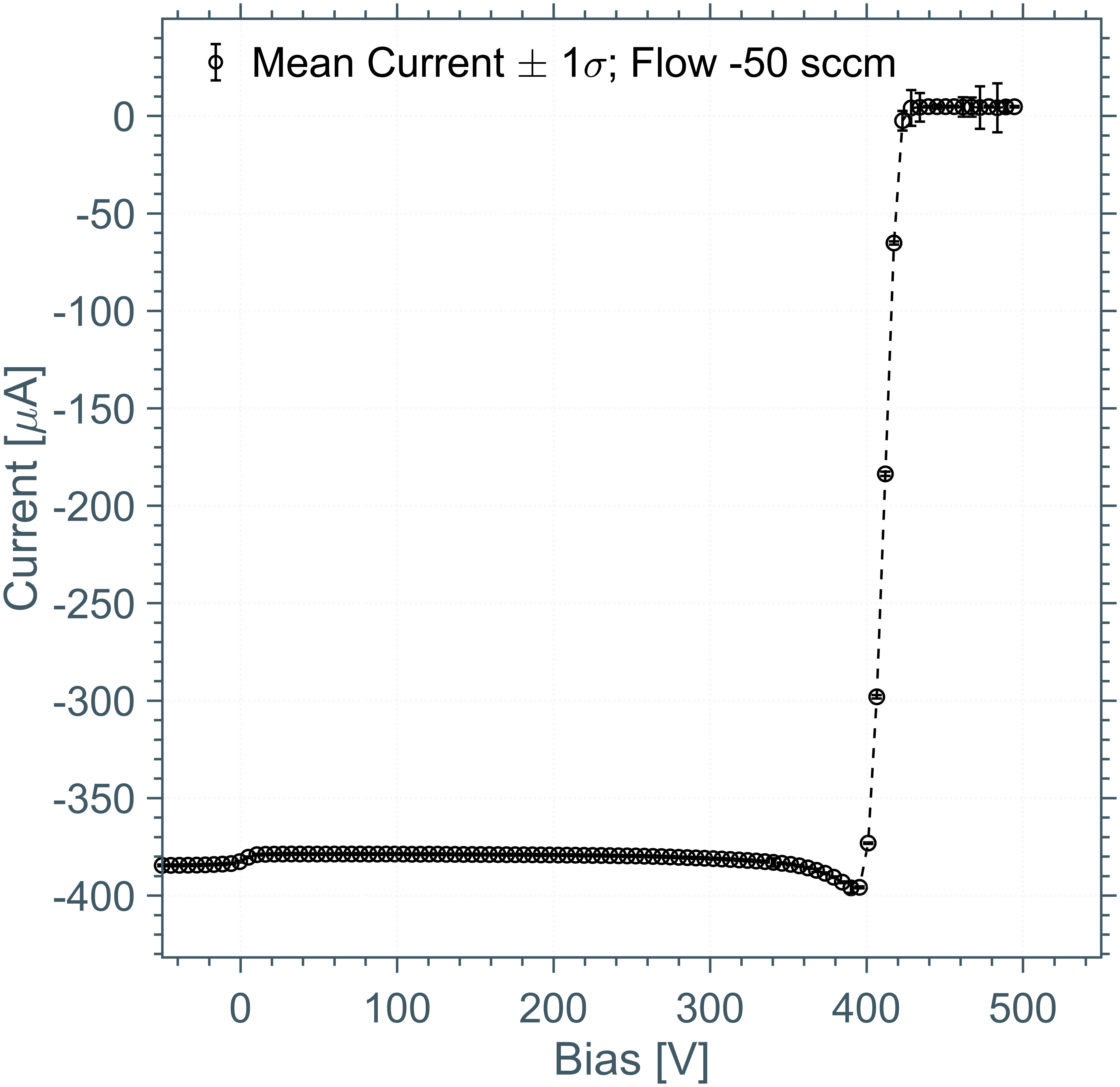}
\end{minipage}\hfill
\begin{minipage}[t]{0.49\linewidth}
\centering
\includegraphics[height=4cm]{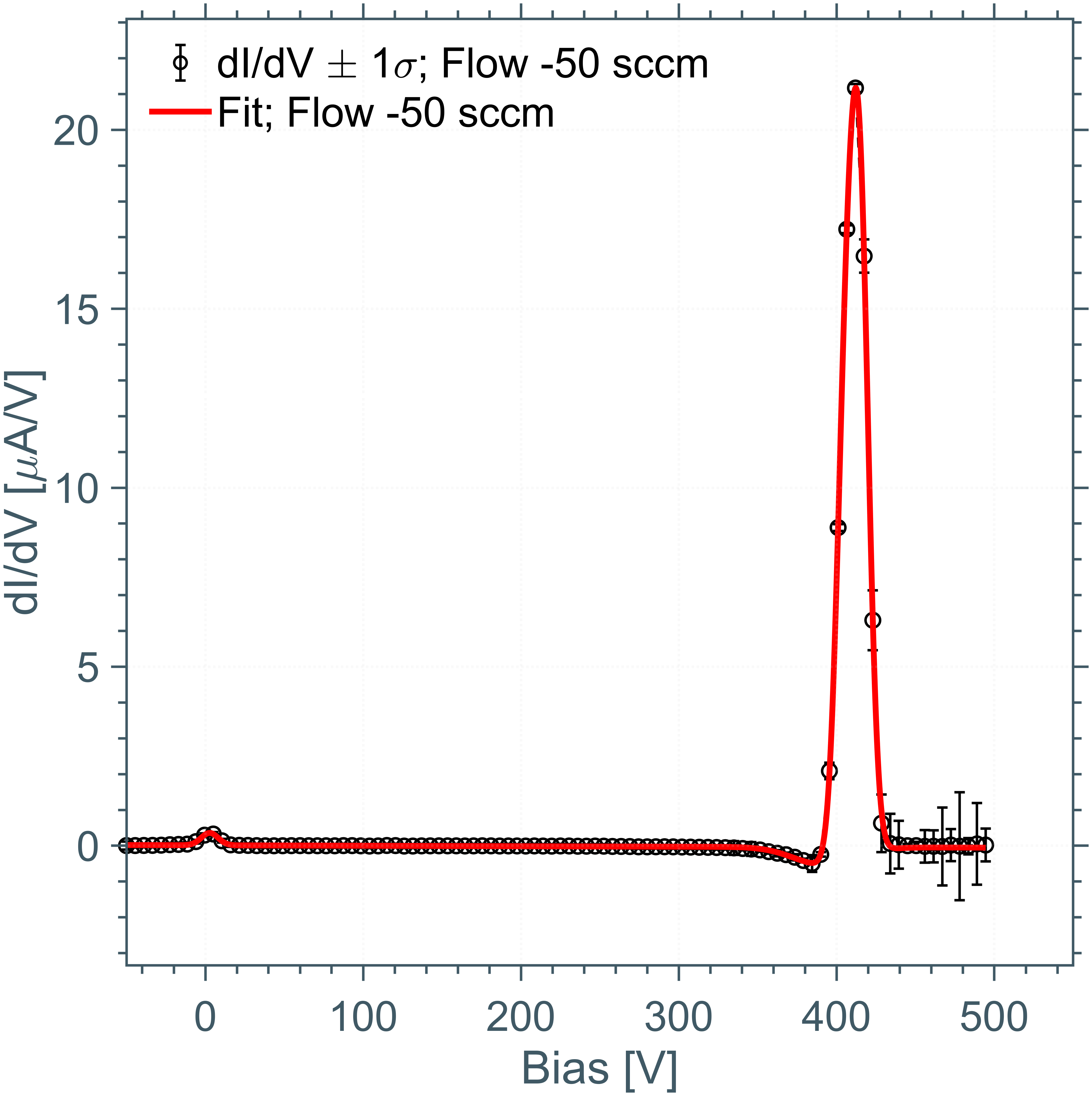}
\end{minipage}
\caption{\label{fig:rpa_processing}Representative RPA measurement and processing example: (a) collector current $I(V_r)$ from an averaged forward/backward sweep; (b) corresponding $dI/dV_r$ and the two-Gaussian peak decomposition used in order to define low-energy and fast-ion components. Integrating the fitted peaks yields $I_{\mathrm{low}}$, $I_{\mathrm{high}}$, and $I_{\mathrm{total}}$.}
\end{figure}

The IEDF was inferred from the first derivative $dI/dV_r$ computed from the averaged $I(V_r)$ using finite differences.\cite{benedikt2021foundations,Hutchinson_2002} The operational definition of the low-energy population used throughout this paper is the peak in $dI/dV_r$ near $V_r\approx 0$~V, which corresponds to ions with energies near $E\approx 0$ in the retarding analysis. This population is attributed primarily to low-energy ions produced by CEX and/or ions belonging to the low-density background plasma. In order to suppress numerical noise and to obtain a robust fractional partition between low-energy and fast components, $dI/dV_r$ was fit directly with a low-order background (linear trend) plus a sum of positive Gaussian peaks; only the two dominant peaks were retained, corresponding to (i) the low-energy peak near $V_r \approx 0$~V and (ii) the high-energy beam peak near $V_r \approx 400$~V (Fig.~\ref{fig:rpa_processing}). This two-peak representation was used in order to extract integrated component currents, and it is not intended to reproduce detailed IEDF shape changes beyond this partition. Component currents were obtained by numerical integration of the two retained Gaussian peaks in $dI/dV_r$, yielding $I_{\mathrm{low}}$, $I_{\mathrm{high}}$, and $I_{\mathrm{total}}=I_{\mathrm{high}}+I_{\mathrm{low}}$. With increasing background gas pressure, $I_{\mathrm{low}}/I_{\mathrm{total}}$ increased, while the high-energy peak broadened and shifted slightly. The low-energy fraction from the RPA was used in order to estimate the background plasma ion fraction in the total ion current, and it was defined as $C_{\mathrm{back}}=I_{\mathrm{low}}/I_{\mathrm{total}}$.

A systematic error associated with the RPA geometry was also assessed. The RPA forms a short dead-end tube with an entrance mesh, so neutral gas can accumulate inside because of neutralization of collected ions and finite mesh conductance. A conductance estimate was used to bound the resulting pressure rise inside the tube and the corresponding increase in the local CEX probability. Under representative conditions, the internal pressure rise was estimated to be about 0.04~mTorr. This effect can increase the apparent low-energy fraction measured by the RPA. Relative to the elevated-pressure operating conditions used in this work (>0.1 mTorr), this additional pressure is not sufficient to explain the observed pressure-dependent increase of the low-energy fraction, although its contribution can be less negligible near the lowest-pressure condition.

\paragraph{Thermal flux probe.}
The neutral flux was characterized with a thermal flux probe. The probe consisted of a $50~\mu$m tantalum foil, 20 mm in diameter, mounted on a ceramic holder for thermal and electrical insulation. A type K thermocouple was spot-welded to the back side of the foil in order to measure the transient temperature response when the beam acceleration was switched on and off. The foil, referred to here as the ``collector,'' was also biased with respect to ground and served as a planar probe for measurements of the ion flux. The thermal flux probe is capable of quantifying the total and individual heat fluxes transferred from a plasma to a surface. A detailed description of the probe operation and calibration is given in Refs.~\onlinecite{Thermal_Probe_AIAA,Stahl_calorimetric,gauter2018calorimetric}. In brief, the total thermal power deposited on the probe, $P_{\mathrm{in}}$, was inferred from an energy balance on the collector using an effective heat capacity $C_c$ obtained by calibration. Assuming that (i) the incident heat flux is approximately constant over a short heating interval and (ii) the heat-loss term depends only on the collector temperature, the incoming power at a given collector temperature $T_c$ is obtained by subtracting the temperature derivatives during heating and cooling,
\begin{equation}
P_{\mathrm{in}}(T_c)=
C_c\left[
\left(dT_c/dt\right)_{\mathrm{heating}}-
\left(dT_c/dt\right)_{\mathrm{cooling}}
\right]_{T_c}.
\end{equation}
Constant backgrounds such as heat radiation from hot filaments cancel in this differencing, while the negative probe bias suppresses electron collection. Therefore, the measured thermal power is attributed to energetic ions, fast neutrals, and ion recombination at the collector surface.

In the present analysis, the fast-neutral flux was not measured directly. Instead, it was inferred from the thermal flux probe measurements using a power balance. The incoming power to the collector, $P_{\mathrm{in}}$, was measured experimentally, while the fast-neutral contribution was obtained after subtracting the power delivered by fast ions, low-energy ions, and recombination from $P_{\mathrm{in}}$. For comparison of the thermal flux probe measurements with the CEX model described in Sec.~\ref{sec:model}, the fast-neutral contribution was expressed as an equivalent fast-neutral current.

With electrons being repelled, only ions bring charge and contribute to the current measured at the collector. Therefore, the total collector current was written as
\begin{equation}
I_{\mathrm{tot}} = I_{\mathrm{fast}} + I_{\mathrm{low-energy}},
\end{equation}
where $I_{\mathrm{fast}}$ and $I_{\mathrm{low-energy}}$ are the fast-ion and low-energy ion current contributions, respectively. The low-energy ion contribution was determined using an independently measured background fraction, $C_{\mathrm{back}}$. This fraction was obtained either from planar-probe current ratios or from the low-energy fraction measured by the RPA at the same operating condition. The characteristic energies of the fast-ion and low-energy ion populations, $E_{\mathrm{fast}}$ and $E_{\mathrm{low-energy}}$, were also taken from the RPA.

The collector was biased negatively to repel electrons. Therefore, the sheath acceleration at the collector was approximated as $\Delta V \simeq |V_{\mathrm{bias}}|$, assuming $\phi_{\mathrm{pl}} \approx 0$. The ion power contribution included both the particle energy and the sheath acceleration for the fast-ion and low-energy ion populations, whereas the fast neutrals contributed only their kinetic energy. The power due to recombination of ions at the surface was taken as a fixed energy per collected ion, $E_{\mathrm{rec}} = E_{\mathrm{ion}} - \phi$, where for argon $E_{\mathrm{ion}} = 15.8~\mathrm{eV}$ and for tantalum $\phi \approx 4.3~\mathrm{eV}$. This recombination term was applied to the total collected ion current, with $f_{\mathrm{rec}} = 1$. Therefore, the incoming power to the collector is given by
\begin{equation}
P_{\mathrm{in}} =
I_{\mathrm{fast}}\left(E_{\mathrm{fast}}/e + \Delta V\right) +
I_{\mathrm{low-energy}}\left(E_{\mathrm{low-energy}}/e + \Delta V\right) + 
I_{\mathrm{tot}}\left(f_{\mathrm{rec}} E_{\mathrm{rec}}/e\right) +
I_{\mathrm{eq,n}}\left(E_{\mathrm{fast}}/e\right).
\end{equation}
Here, the first term represents the power delivered by the directed fast-ion current, including both the fast-ion kinetic energy and the sheath acceleration at the collector. The second term represents the corresponding power delivered by the low-energy ion current, again including both the ion kinetic energy and the sheath acceleration. The third term represents the recombination power released at the collector surface by the total collected ion current. The fourth term represents the power delivered by the fast-neutral flux, expressed as an equivalent fast-neutral current $I_{\mathrm{eq,n}}$ with characteristic energy $E_{\mathrm{fast}}$.

Solving this expression for the equivalent fast-neutral current yields
\begin{equation}
I_{\mathrm{eq,n}} =
\frac{
P_{\mathrm{in}}-
I_{\mathrm{fast}}\left(E_{\mathrm{fast}}/e + \Delta V\right)-
I_{\mathrm{low-energy}}\left(E_{\mathrm{low-energy}}/e + \Delta V\right)-
I_{\mathrm{tot}}\left(f_{\mathrm{rec}} E_{\mathrm{rec}}/e\right)
}{
E_{\mathrm{fast}}/e
},
\label{eq:power_balance}
\end{equation}
where $I_{\mathrm{eq,n}}$ represents the fast-neutral contribution expressed as an equivalent fast-neutral current with characteristic energy $E_{\mathrm{fast}}$. The resulting $I_{\mathrm{eq,n}}(z,P)$ was then used to characterize the fast-neutral contribution as a function of distance and pressure and to compare the measurements with the model.

\newpage

\section{\label{sec:model}Semi-empirical quasi-2D fluid model for charge-exchange in a divergent plume}
A reduced, axisymmetric semi-empirical quasi-2D model was used to analyze background-gas-pressure effects on the resonant CEX process and on the plasma plume. The model evolves only the fast populations, namely the energetic ions extracted from the ion source and the energetic neutrals created by CEX. Plume divergence is prescribed through an expanding streamline geometry. Therefore, the densities decrease not only because of collisional attenuation, but also because of geometric reduction associated with plume spreading. The divergence was observed experimentally and was introduced into the model as a fitting parameter. In this way, the model is intended to capture the first-order effect of background gas pressure on fast-ion attenuation and fast-neutral formation, while retaining the dominant geometrical effect that is absent in a purely one-dimensional description.

Although the governing equations are written in time-dependent form, the analysis below is restricted to steady operation, $\partial/\partial t=0$. The following assumptions are used:
\begin{itemize}
\item The description is axisymmetric and quasi-2D in cylindrical coordinates $(r,z)$, where $z$ is the axial direction and $r$ is the radial direction.
\item The background gas is uniform (as shown with pressure measurements in Section~\ref{sec:exper}, with the important note that the measurements were performed without plasma), with spatially constant pressure $P$ and temperature $T_g$, so the neutral density $n_g$ is constant.
\item The fast ions are monoenergetic and have constant directed speed $v_b$. The ion--neutral relative speed is taken to be approximately equal to $v_b$.
\item Plume expansion is prescribed through a streamline mapping with a constant divergence half-angle $\theta$. The divergence is not solved self-consistently.
\item Each resonant CEX event produces one fast neutral, and this neutral is assumed to inherit the fast ion direction at the point of creation.
\item Subsequent collisions of fast neutrals are neglected. Low-energy ions, electrons, and secondary plasma are not evolved.
\end{itemize}

Let $n_i^f(r,z,t)$ be the number density of fast ions and $n_n^f(r,z,t)$ the number density of fast neutrals. Both fast species are advected by a prescribed velocity field $\mathbf{u}(r,z)$ that represents the plume expansion. Under these assumptions, the fast populations satisfy the continuity equations
\begin{equation}
\dfrac{\partial n_i^f}{\partial t}+\nabla\cdot\left(n_i^f \mathbf{u}\right)= -\nu_{\mathrm{cx}}n_i^f,
\label{eq:cont_ion}
\end{equation}
\begin{equation}
\dfrac{\partial n_n^f}{\partial t}+\nabla\cdot\left(n_n^f \mathbf{u}\right)= +\nu_{\mathrm{cx}}n_i^f.
\label{eq:cont_neu}
\end{equation}

The sink term in Eq.~\eqref{eq:cont_ion} represents removal of fast ions from the modeled population by resonant CEX collisions with background neutrals. The source term in Eq.~\eqref{eq:cont_neu} represents production of fast neutrals by the reaction
$
\mathrm{X}*{\mathrm{fast}}^+ + \mathrm{X}*{\mathrm{low-energy}}
\rightarrow
\mathrm{X}*{\mathrm{fast}} + \mathrm{X}*{\mathrm{low-energy}}^+,
$
under the assumption that one fast-neutral is created per CEX event and that it inherits the fast-ion direction at the point of creation. Only CEX is considered in this reduced model. Therefore, the total fast-ion loss frequency is $\nu_{\mathrm{tot}}\equiv \nu_{\mathrm{cx}}$. Since subsequent collisions of the fast neutrals are neglected, no sink term appears in Eq.~\eqref{eq:cont_neu}.

The background neutral density is obtained from the ideal gas law, $n_g=P/(k_B T_g)$, where $P$ is the chamber pressure and $T_g$ is the gas temperature. The fast-ion flux is treated as monoenergetic, with constant directed speed $v_b=\sqrt{2eE_b/m_i},$
where $m_i$ is the ion mass, $E_b$ is the ion energy, $e$ is the elementary charge, and only the singly charged case is considered here. The CEX collision frequency is then written as $\nu_{\mathrm{cx}}=n_g \sigma_{\mathrm{cx}}(E_b) v_b,$
which assumes that the ion speed is much larger than the thermal speed of the background gas, so the ion--neutral relative speed is approximated by $v_b$. The collisional cross section is evaluated at the mean fast ions energy and is treated as constant along the plume in this reduced model. In the present estimates, $\sigma_{\mathrm{cx}}(E_b)\approx 2.8\times 10^{-19}~\mathrm{m}^2$. The corresponding attenuation length of the fast-ion population is
\begin{equation*}
    \lambda_{\mathrm{CEX}}=v_b/\nu_{\mathrm{cx}}=1/\left(n_g \sigma_{\mathrm{cx}}\right).
\end{equation*}
The quasi-2D reduction prescribes plasma plume expansion by mapping streamlines from the ion-source exit plane at $z=0$. A streamline that originates at radius $r_0$ is written as
\begin{equation}
r(z)=f(z)r_0.
\label{eq:streamline}
\end{equation}
Therefore, the cross-sectional area of a streamtube scales as $f(z)^2$. In the absence of reactions, this implies geometric reduction of density as $1/f(z)^2$, while the radial density profile is stretched according to $r\rightarrow r/f(z)$. A velocity field consistent with this mapping is obtained by taking the axial velocity to be constant, $u_z=v_b$, and the radial velocity to be $u_r(r,z)=v_b f'(z) r/f(z)$, so that particles follow the prescribed streamlines in Eq.~\eqref{eq:streamline}. The expansion factor is parameterized through the characteristic plume width $w(z)$,
\begin{equation}
w(z)=w_0 f(z),
\qquad
f(z)=1+\dfrac{z}{z_v},
\qquad
z_v=\dfrac{w_0}{\tan\theta},
\label{eq:fz}
\end{equation}
which is equivalent to the linear growth law $w(z)=w_0+z\tan\theta$. In the limit $\theta\rightarrow 0$, one has $f(z)\rightarrow 1$, and the geometric reduction is removed.

\begin{figure}[h]
\centering
\includegraphics[height=4cm]{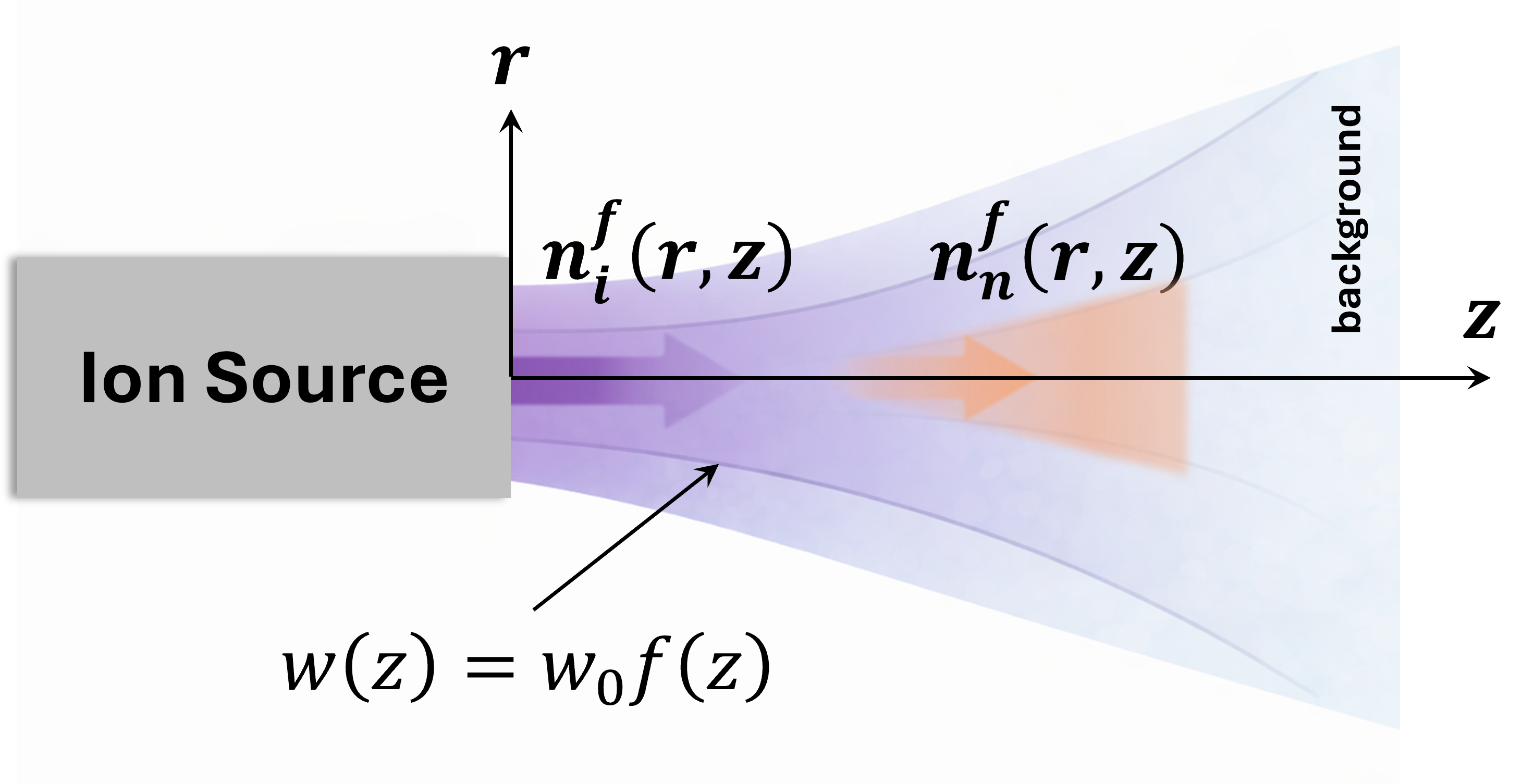}
\caption{\label{fig:model}
Schematic of the quasi-2D model geometry. The fast-ion flux propagates along the axial direction $z$, while $r$ is the radial direction. Fast ions, $n_i^f(r,z)$, leave the ion source and are attenuated by CEX collisions with the background gas. At the same time, fast neutrals, $n_n^f(r,z)$, are produced and continue to propagate downstream. The prescribed plume width increases as $w(z)=w_0 f(z)$, so the fast populations experience geometric reduction in addition to collisional conversion.}
\end{figure}

Under steady conditions and above assumptions, Eqs.~\eqref{eq:cont_ion} and \eqref{eq:cont_neu} reduce to coupled axial equations along streamlines. Using the streamtube area scaling, one obtains
\begin{equation}
\dfrac{d}{dz}\left[f(z)^2 n_i^f\right]=-\dfrac{1}{\lambda_{\mathrm{CEX}}}\left[f(z)^2 n_i^f\right],
\end{equation}
\begin{equation}
\dfrac{d}{dz}\left[f(z)^2 n_n^f\right]=+\dfrac{1}{\lambda_{\mathrm{CEX}}}\left[f(z)^2 n_i^f\right].
\end{equation}
For general initial profiles at $z=0$, namely $n_i^f(r,0)=n_{i0}^f(r)$ and $n_n^f(r,0)=n_{n0}^f(r)$, the quasi-2D solutions are
\begin{equation}
n_i^f(r,z)=\dfrac{1}{f(z)^2}
n_{i0}^f\left(\dfrac{r}{f(z)}\right)
\exp\left(-\dfrac{z}{\lambda_{\mathrm{CEX}}}\right),
\label{eq:ni_general}
\end{equation}
\begin{equation}
n_n^f(r,z)=\dfrac{1}{f(z)^2}
\left[
n_{n0}^f\left(\dfrac{r}{f(z)}\right)
+
n_{i0}^f\left(\dfrac{r}{f(z)}\right)
\left(1-\exp\left(-\dfrac{z}{\lambda_{\mathrm{CEX}}}\right)\right)
\right].
\label{eq:nn_general}
\end{equation}

These expressions make the physical behavior explicit. The fast-ion density decreases because two effects act simultaneously. First, ions are removed from the fast population by CEX collisions. Second, the same population is diluted geometrically as the plume expands. The fast-neutral density follows the same geometric reduction, because the neutrals are assumed to inherit the plume divergence at the point of creation. However, unlike the ions, the fast-neutral population is also supplied continuously by CEX conversion of the remaining fast ions. Therefore, the neutral density may either increase or decrease with axial distance, depending on which process is dominant locally. For the common case $n_{n0}^f=0$, the fast-neutral density approaches the geometrically diluted initial ion profile as $z/\lambda_{\mathrm{CEX}}$ becomes large.

For the Gaussian initial condition used in this work,
\begin{equation}
n_{i0}^f(r)=n_0\exp\left(-\dfrac{r^2}{w_0^2}\right),
\qquad
n_{n0}^f(r)=0,
\label{eq:ic_gauss}
\end{equation}
Eqs.~\eqref{eq:ni_general} and \eqref{eq:nn_general} become
\begin{equation}
n_i^f(r,z)=\dfrac{n_0}{f(z)^2}
\exp\left[-\dfrac{r^2}{\left(w_0 f(z)\right)^2}\right]
\exp\left(-\dfrac{z}{\lambda_{\mathrm{CEX}}}\right),
\label{eq:ni_gauss}
\end{equation}
\begin{equation}
n_n^f(r,z)=\dfrac{n_0}{f(z)^2}
\exp\left[-\dfrac{r^2}{\left(w_0 f(z)\right)^2}\right]
\left(1-\exp\left(-\dfrac{z}{\lambda_{\mathrm{CEX}}}\right)\right).
\label{eq:nn_gauss}
\end{equation}

To compare the model with probe measurements, the model fields can be integrated over the diagnostic collection area. For a circular aperture of radius $R_p$ normal to the $z$ axis, the particle fluxes are
\begin{equation}
\Phi_i(z)=v_b \int_{0}^{R_p} n_i^f(r,z)\,2\pi r\,dr,
\qquad
\Phi_n(z)=v_b \int_{0}^{R_p} n_n^f(r,z)\,2\pi r\,dr.
\label{eq:flux_integrals}
\end{equation}
The corresponding ion current is $I_i(z)=q\Phi_i(z)$ when ion saturation is achieved and sheath effects are treated separately. In practice, the RPA includes a transmission factor set mainly by the grid transparency, and the planar and thermal flux probes collect over finite surfaces rather than ideal apertures. However, when only normalized axial or pressure trends are compared, constant proportionality factors between model and measured fluxes cancel out.

\subsection{Effect of plume divergence on the density distribution}
\label{subsec:divergence_examples}

The closed-form solutions in Eqs.~\eqref{eq:ni_general} and \eqref{eq:nn_general} show directly how divergence enters the model. The factor $1/f(z)^2$ reduces the on-axis density through geometric reduction, while the transformed argument $r/f(z)$ broadens the radial profile. At the same time, CEX causes exponential attenuation of the fast ions and continuous production of fast neutrals over the length scale $\lambda_{\mathrm{CEX}}$.

\begin{figure}[ht]
\centering
\begin{minipage}[t]{0.49\linewidth}
\centering
\includegraphics[height=4cm]{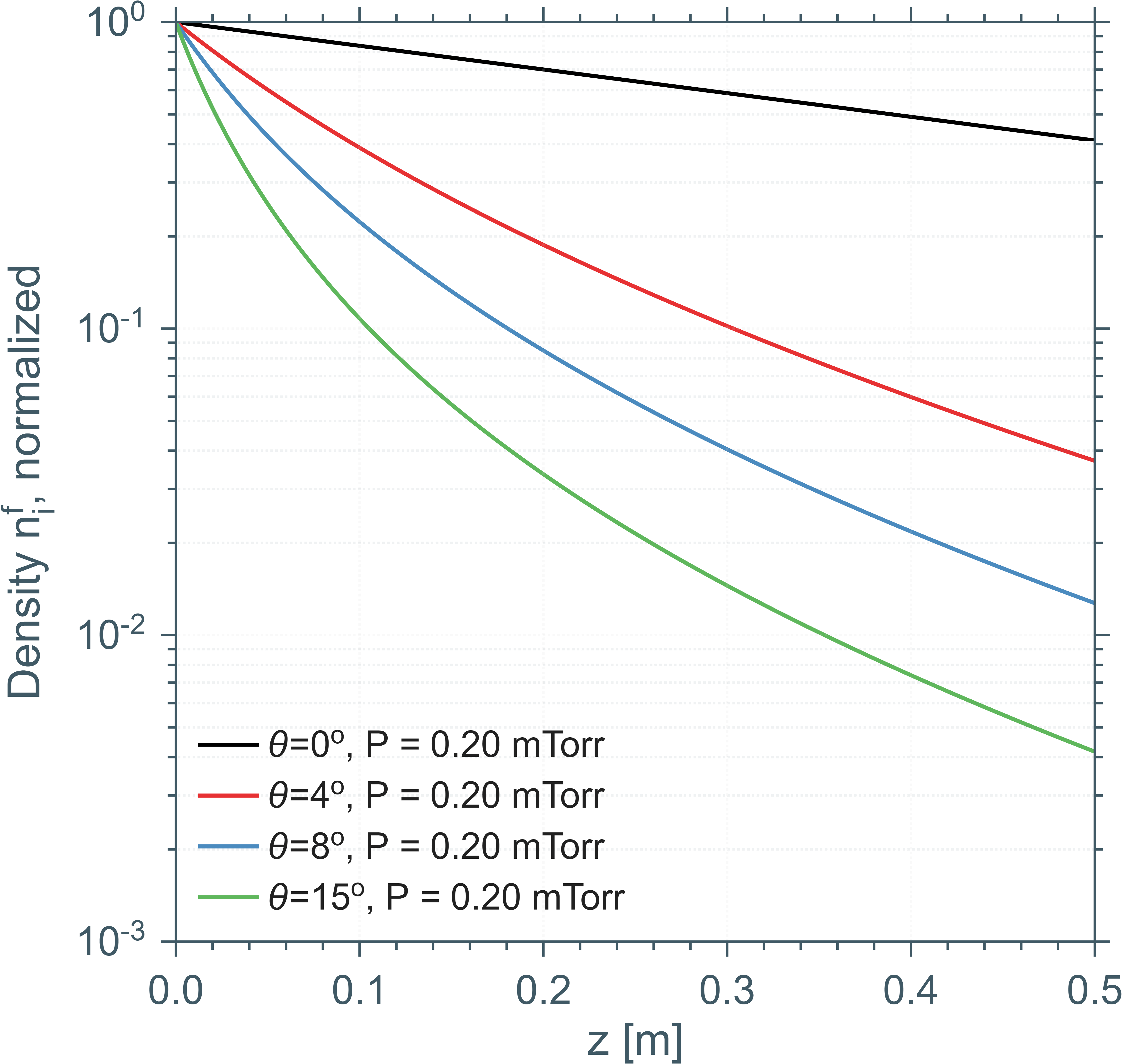}
\end{minipage}\hfill
\begin{minipage}[t]{0.49\linewidth}
\centering
\includegraphics[height=4cm]{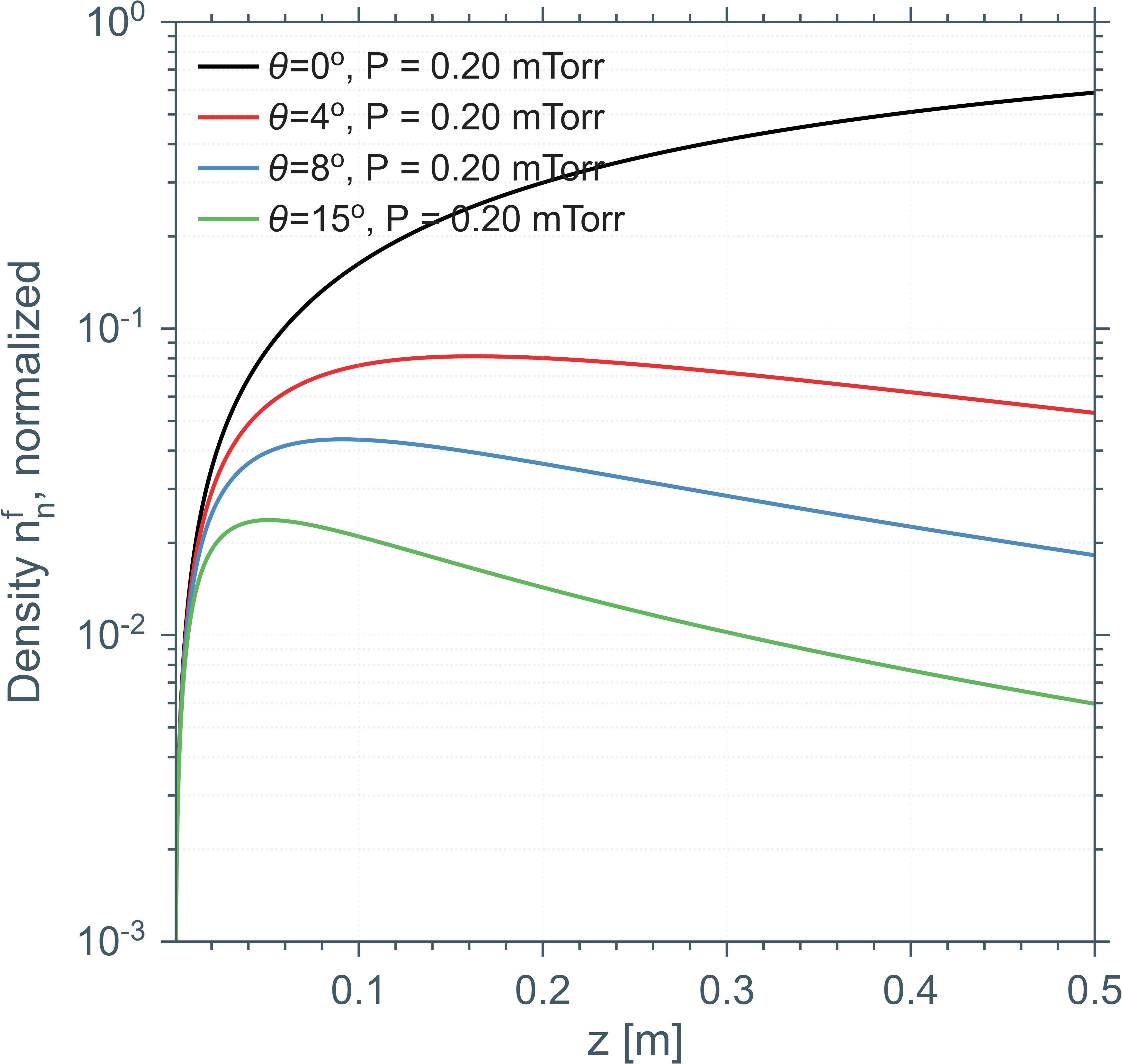}
\end{minipage}
\caption{\label{fig:onaxis_divergence}
Normalized on-axis fast-ion density, $n_i^f(0,z)$, and fast-neutral density, $n_n^f(0,z)$, predicted by the quasi-2D model at $P=0.2$ mTorr for several prescribed divergence half-angles $\theta$. The fast-ion density decreases more rapidly as $\theta$ increases because collisional attenuation is supplemented by geometric reduction. The fast-neutral density is monotonic only for $\theta=0^\circ$. For all divergent cases shown here, it first rises due to CEX production and then falls because plume divergence reduces the on-axis density.}
\end{figure}

Figure~\ref{fig:onaxis_divergence} shows the normalized on-axis densities, $n_i^f(0,z)$ and $n_n^f(0,z)$, for several prescribed divergence half-angles at fixed pressure, $P=0.20$ mTorr. The left panel shows that when $\theta=0^\circ$, the on-axis fast-ion density decreases only because of CEX attenuation, and therefore the curve is a simple exponential. When $\theta>0^\circ$, the decay becomes substantially faster because geometric reduction acts together with collisional loss. This is why the on-axis ion density decreases more rapidly as $\theta$ increases from $4^\circ$ to $15^\circ$.

The right panel shows the corresponding on-axis fast-neutral density. In the non-divergent case, $\theta=0^\circ$, the neutral density increases monotonically because neutrals are produced continuously, while no geometric reduction is present. However, once divergence is included, the behavior changes qualitatively. In that case, neutral production is strongest near the source, where the fast-ion density is still high, but farther downstream the on-axis density is reduced by plume divergence. Therefore, for all divergent cases shown in the figure, the neutral density first increases, reaches a maximum, and then decreases with $z$. This non-monotonic behavior is a direct consequence of the competition between CEX production and geometric reduction.

\begin{figure}[ht]
\centering
\begin{minipage}[t]{0.49\linewidth}
\centering
\includegraphics[height=4cm]{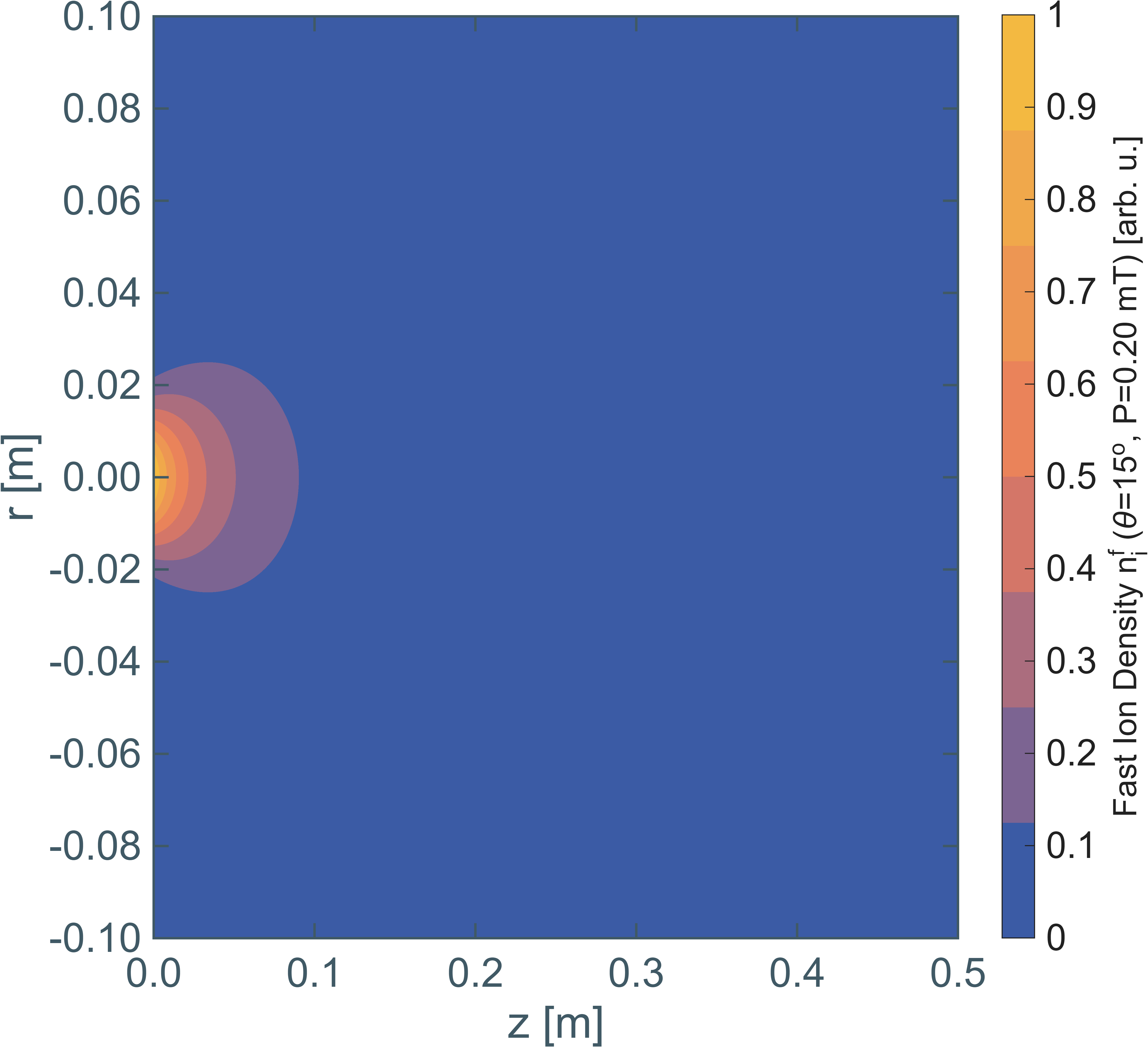}
\end{minipage}\hfill
\begin{minipage}[t]{0.49\linewidth}
\centering
\includegraphics[height=4cm]{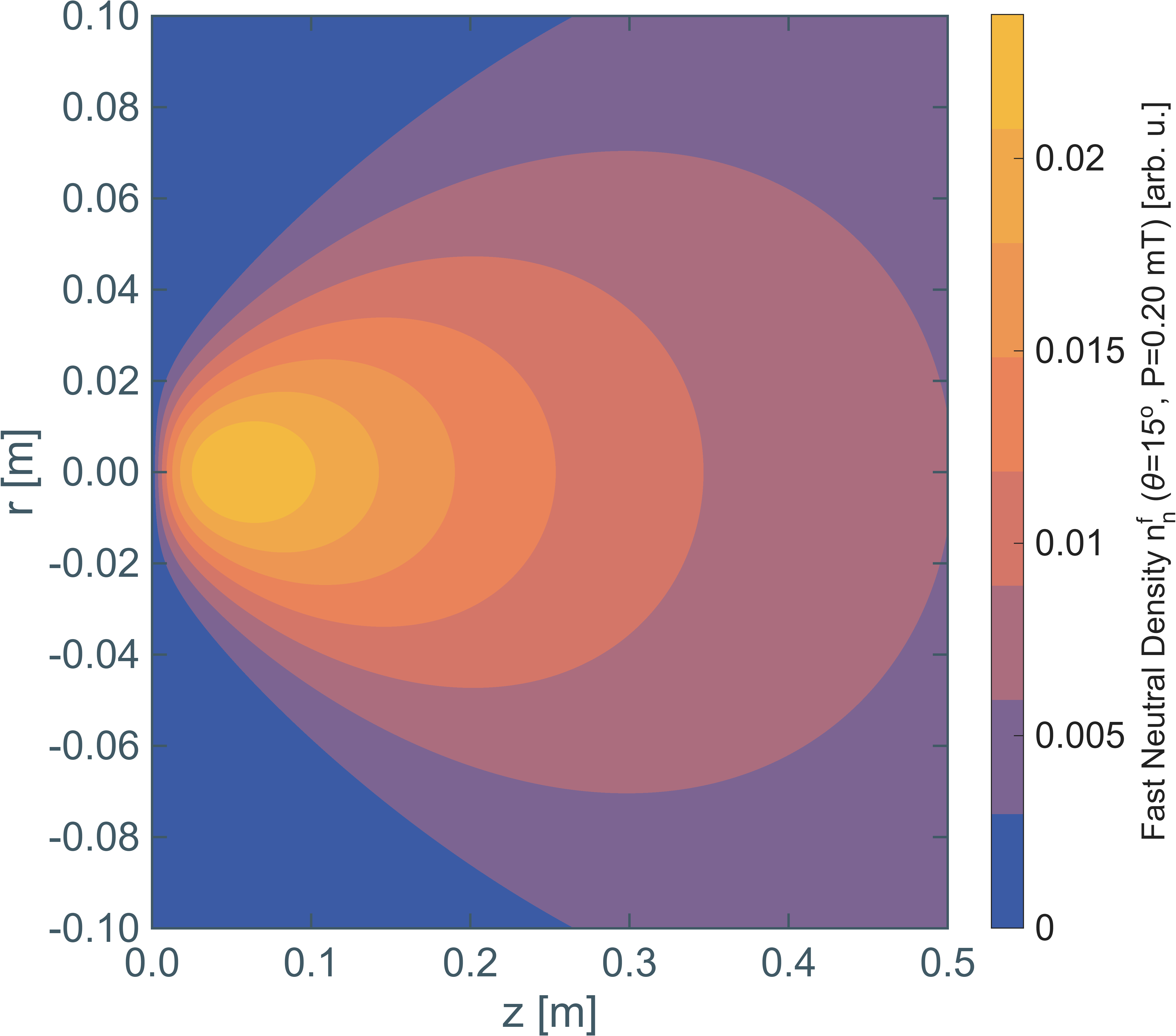}
\end{minipage}
\caption{\label{fig:maps_divergence}
Example quasi-2D density fields of fast ions (left) and fast neutrals (right) in the $(r,z)$ plane for $\theta=15^\circ$ and $P=0.2$ mTorr. The fast-ion density is highest near the source and decreases downstream, while the plume width increases because of the prescribed divergence. The fast-neutral density is zero at $z=0$ in the model, then builds up downstream through CEX conversion and spreads radially with the same imposed divergence. The color scales differ between the two panels.}
\end{figure}

Figure~\ref{fig:maps_divergence} shows the corresponding quasi-2D density fields in the $(r,z)$ plane for one representative case, namely $\theta=15^\circ$ and $P=0.2$ mTorr corresponding to ion source plume divergence and one of the cases of elevated pressure. The fast-ion map in the left panel is largest near the source and then decreases downstream. At the same time, the radial extent of the ion plume increases because of the prescribed streamline expansion. The fast-neutral map in the right panel starts from zero at the source plane in this model, then builds up downstream as CEX proceeds, and finally broadens radially in the same prescribed geometry. Therefore, the peak neutral density appears downstream of the source rather than at $z=0$.

These maps clarify why a one-dimensional interpretation is not sufficient when divergence is present. Even if the total fast-particle content along a streamtube follows a simple axial law, the local density measured on axis or over a finite collection area depends strongly on radial spreading. Thus, in this case, geometric effects are not a secondary correction. They are an essential part of the interpretation of probe data.

\newpage

\section{\label{sec:results}Experimental Results and Comparison with the Model}

\subsection{Low-energy ion contribution estimates from probe current ratios and RPA}
\label{sec:results_background_fraction}
The contribution of low-energy CEX ions and background plasma ions was estimated using the following current ratios: the RPA low-energy to total-energy current ratio, the double-sided planar-probe back/front and the four-sided planar-probe back/front current ratios. For the RPA, the low-energy ratio was defined as
\begin{equation}
C_{\mathrm{low}}^{\mathrm{RPA}}=I_{\mathrm{low}}/I_{\mathrm{total}},
\end{equation}
where $I_{\mathrm{low}}$ is the low-energy component and $I_{\mathrm{total}}$ is the total current. Both were obtained from the two-peak decomposition of the differentiated RPA characteristic. For the planar probes, the back/front current ratios were defined as
\begin{equation}
C_{\mathrm{back,2side}}=I_{\mathrm{back}}/I_{\mathrm{front}}, \qquad
C_{\mathrm{back,4side}}=I_{\mathrm{back}}/I_{\mathrm{front}},
\end{equation}
where the ion currents were obtained from the measured $VI$ characteristics when required. 

\begin{figure}[ht]
\centering
\begin{minipage}[t]{0.32\linewidth}
\centering
\includegraphics[height=4cm]{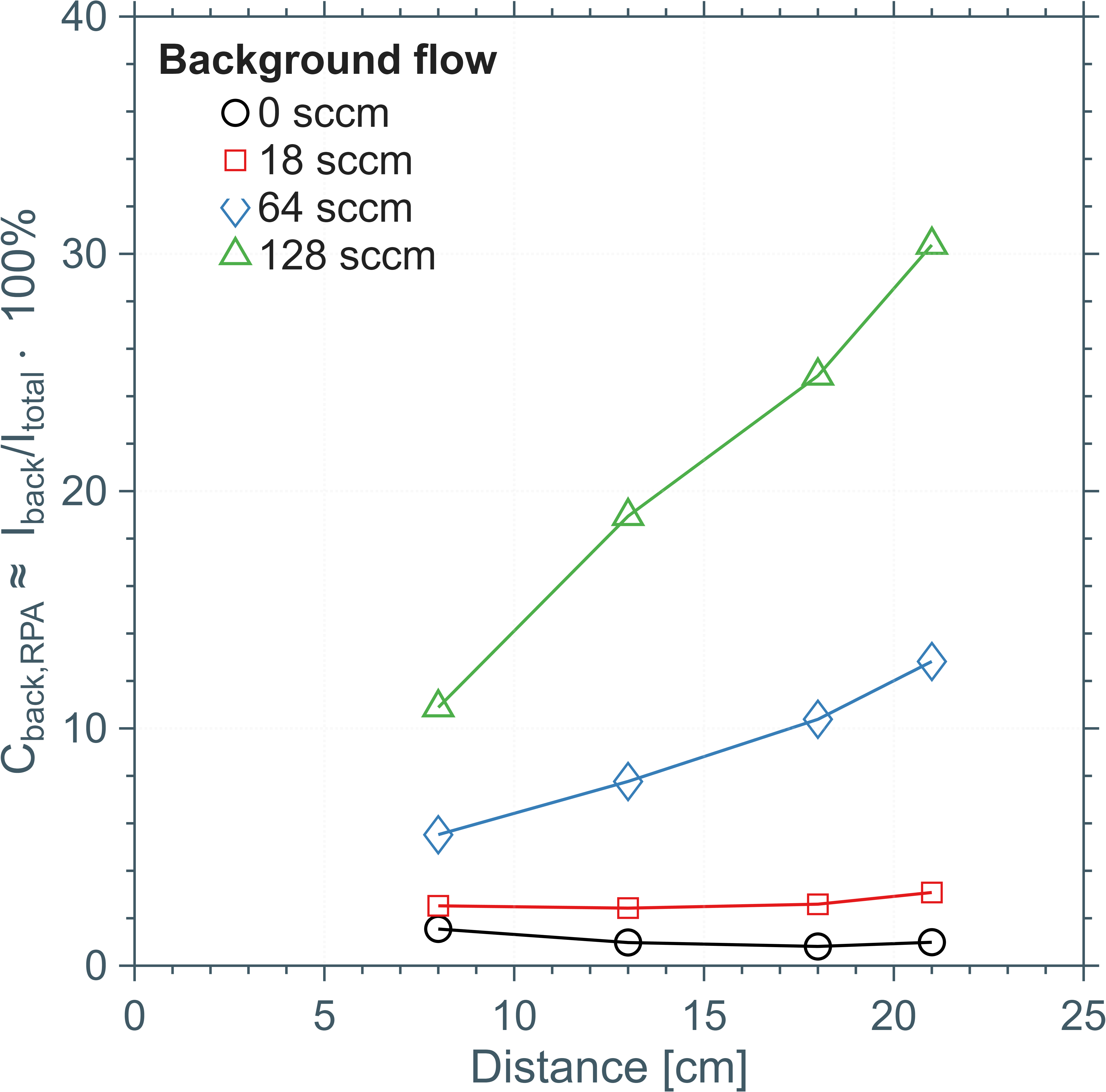}
\end{minipage}
\hfill
\begin{minipage}[t]{0.32\linewidth} \centering
\includegraphics[height=4cm]{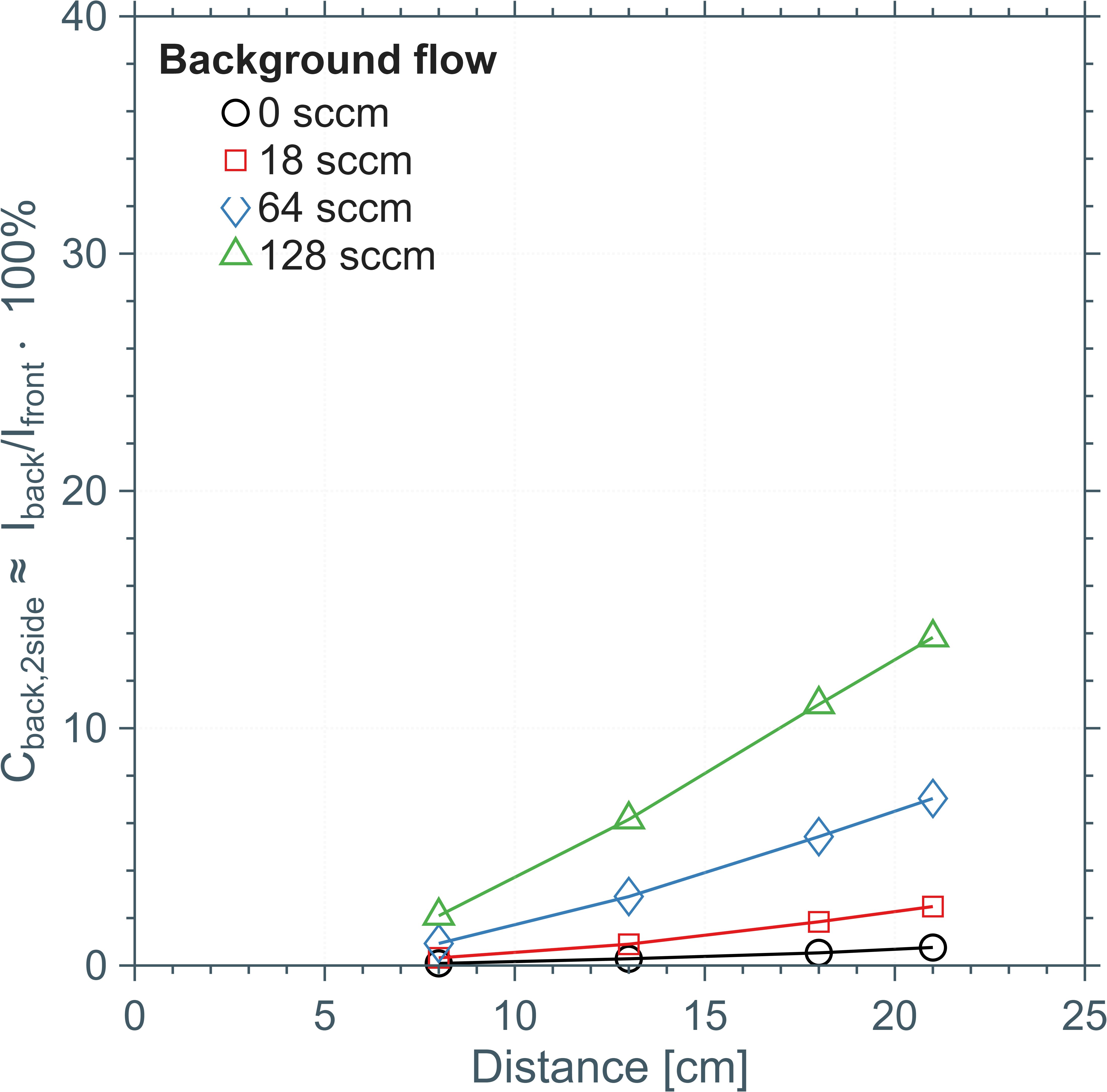}
\end{minipage}\hfill \begin{minipage}[t]{0.32\linewidth}
\centering
\includegraphics[height=4cm]{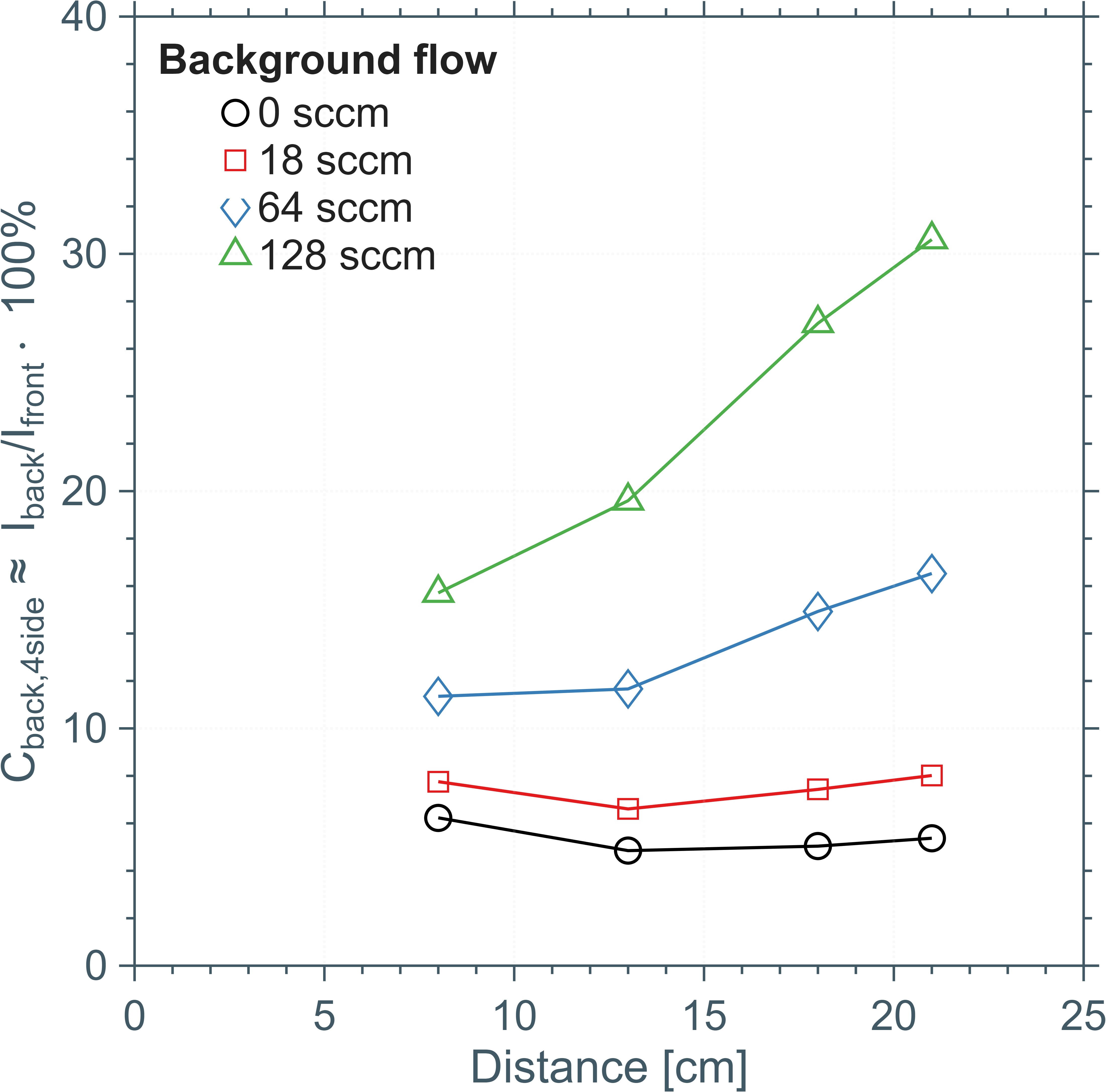}
\end{minipage}
\caption{\label{fig:bg_fraction_compare}Background fraction versus distance for four added chamber argon flows (0, 18, 64, and 128~sccm), estimated using three diagnostics: (left) RPA low-energy ratio $I_{\mathrm{low}}/I_{\mathrm{total}}$; (middle) double-sided planar-probe current ratio $I_{\mathrm{back}}/I_{\mathrm{front}}$; (right) four-sided planar-probe current ratio $I_{\mathrm{back}}/I_{\mathrm{front}}$.} \end{figure}

Figure~\ref{fig:bg_fraction_compare} shows that all three observables increase with added chamber flow. In all three panels, the ordering of the curves is preserved as $128>64>18>0$~sccm over the full axial range. The separation between the curves also grows with distance, especially for $Q_{\mathrm{add}}=64$ and 128~sccm. This trend is consistent with the increase of the charge-exchange effect with background gas density and distance. As the density of the background gas and the propagation length increase, a larger fraction of the collected current appears in the low-energy ion component, while the contribution of the fast ions from the ion source decreases. At the two lowest additional gas flows, $Q_{\mathrm{add}}=0$ and $18~\mathrm{sccm}$, the distance dependence is weak. This suggests that, under these conditions, when the background contribution is small, probe-related collection effects can have a larger relative effect on the measurements.

Although the three diagnostics show the same overall trend, they do not yield the same numerical values. This can be explained as a geometry effect when each diagnostic samples a different combination of ion energy and arrival angle. In the planar probes, the measured ratio is additionally affected by finite collector geometry, sheath expansion, and wake formation, as suggested by the V-I characteristics in Section~\ref{sec:diagn}, Fig.~\ref{fig:vi_planar}.

To assess whether sheath expansion is negligible, order-of-magnitude estimates were made using the local directed ion density, $J_i = I_i/A_c = e n_i v_i$, with $v_i=\sqrt{2eE_i/m_i}$. Using representative ion energies of $E_{\mathrm{front}} \sim 400~\mathrm{eV}$ for the front-facing collector and $E_{\mathrm{back}} \sim 1~\mathrm{eV}$ for the back-facing collector gives ion densities of order $10^{14}$--$10^{15}~\mathrm{m}^{-3}$ on both sides. For $T_e \sim 0.2~\mathrm{eV}$, taken as the temperature of electrons supplied by the neutralizer filament for plume neutralization, the corresponding Debye length,
$\lambda_D = \sqrt{\varepsilon_0 k_B T_e / n_e e^2},$
is of order $0.1$--$0.4~\mathrm{mm}$. The sheath thickness was then estimated independently from a planar Child--Langmuir-type, scaling\cite{lieberman1994principles} $s_{\mathrm{sh}} \sim \left[(4 \varepsilon_0)/(9 J_i)\sqrt{2e/m_i}(\Delta \phi)^{3/2}\right]^{1/2},$
with $\Delta \phi \sim 20~\mathrm{V}$. This gives sheath thicknesses of order $0.5$--$1.0~\mathrm{mm}$ on the front side and $2.5$--$3.0~\mathrm{mm}$ on the back side for the double-sided probe, and $0.6$--$1.5~\mathrm{mm}$ on the front side and $1.8$--$3.5~\mathrm{mm}$ on the back side for the four-sided probe. The corresponding area expansion, defined here as the ratio of the sheath-expanded collection area to the geometric collector area, $A_{\mathrm{eff}}/A_c$, is of order $1.1$--$1.2$ on the front side and $1.5$--$1.6$ on the back side for the double-sided probe, and $1.3$--$1.6$ on the front side and $1.9$--$2.7$ on the back side for the four-sided probe. These estimates show that sheath expansion is not negligible compared to the probe dimensions. Therefore, the effective collection area and the corresponding angular acceptance can differ substantially between the collector plates and between the two probe geometries.

The back-facing collector does not measure the full low-energy ion population, but only the subset able to reach the back surface under the local sheath, probe geometry, wake, and ion arrival-angle distribution; it is therefore both energy- and angle-selective. By contrast, the RPA low-energy ratio is defined by retarding energy and is independent of geometric effects such as shadowing or wake (but dependent on grid parameters like transparency). The front-facing collector can still collect low-energy ions with a velocity component close to the surface normal. Therefore, the low-energy ion population sampled by the back-facing planar collector is not identical to the low-energy population identified by the RPA, and the planar-probe ratios should be interpreted as operational correction parameters rather than direct measurements of the total low-energy ion population.
\subsection{Fast-ion attenuation from planar probes and RPA}
\label{sec:results_fast_ion_atten}

Using the low-energy ion fractions defined in the previous subsection, the axial attenuation of the fast-ion flux was compared across the planar probes and the RPA. At each added flow, the measured current was normalized to its value at the nearest common axial location to the ion-source exit plane, $z=z_{\mathrm{ref}}$,
\begin{equation}
A(z;Q_{\mathrm{add}})=\frac{I(z;Q_{\mathrm{add}})}{I(z_{\mathrm{ref}};Q_{\mathrm{add}})}.
\end{equation}
This normalization removes differences in absolute current level between diagnostics. For the planar probes, the fast-ion estimate was taken as
\begin{equation}
I_{\mathrm{front,fast}}=I_{\mathrm{front}}\left(1-C_{\mathrm{back}}\right),
\end{equation}
using $C_{\mathrm{back,2side}}$ or $C_{\mathrm{back,4side}}$ from Fig.~\ref{fig:bg_fraction_compare}. For the RPA, the fast-ion flux was taken as the high-energy current $I_{\mathrm{high}}$.

\begin{figure}[ht]
\centering
\begin{minipage}[t]{0.49\linewidth}
\centering
\includegraphics[height=4cm]{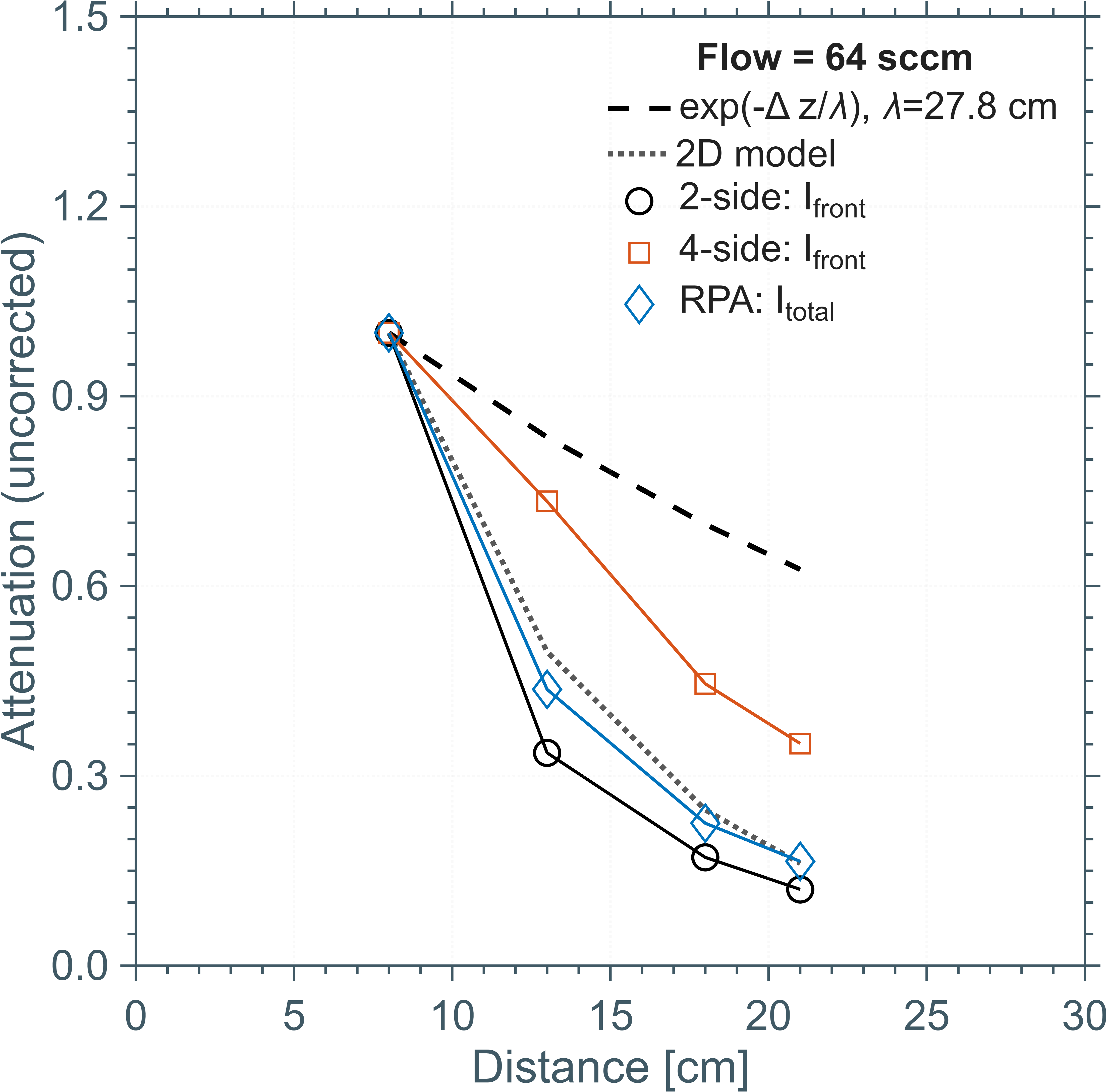}
\end{minipage}\hfill
\begin{minipage}[t]{0.49\linewidth}
\centering
\includegraphics[height=4cm]{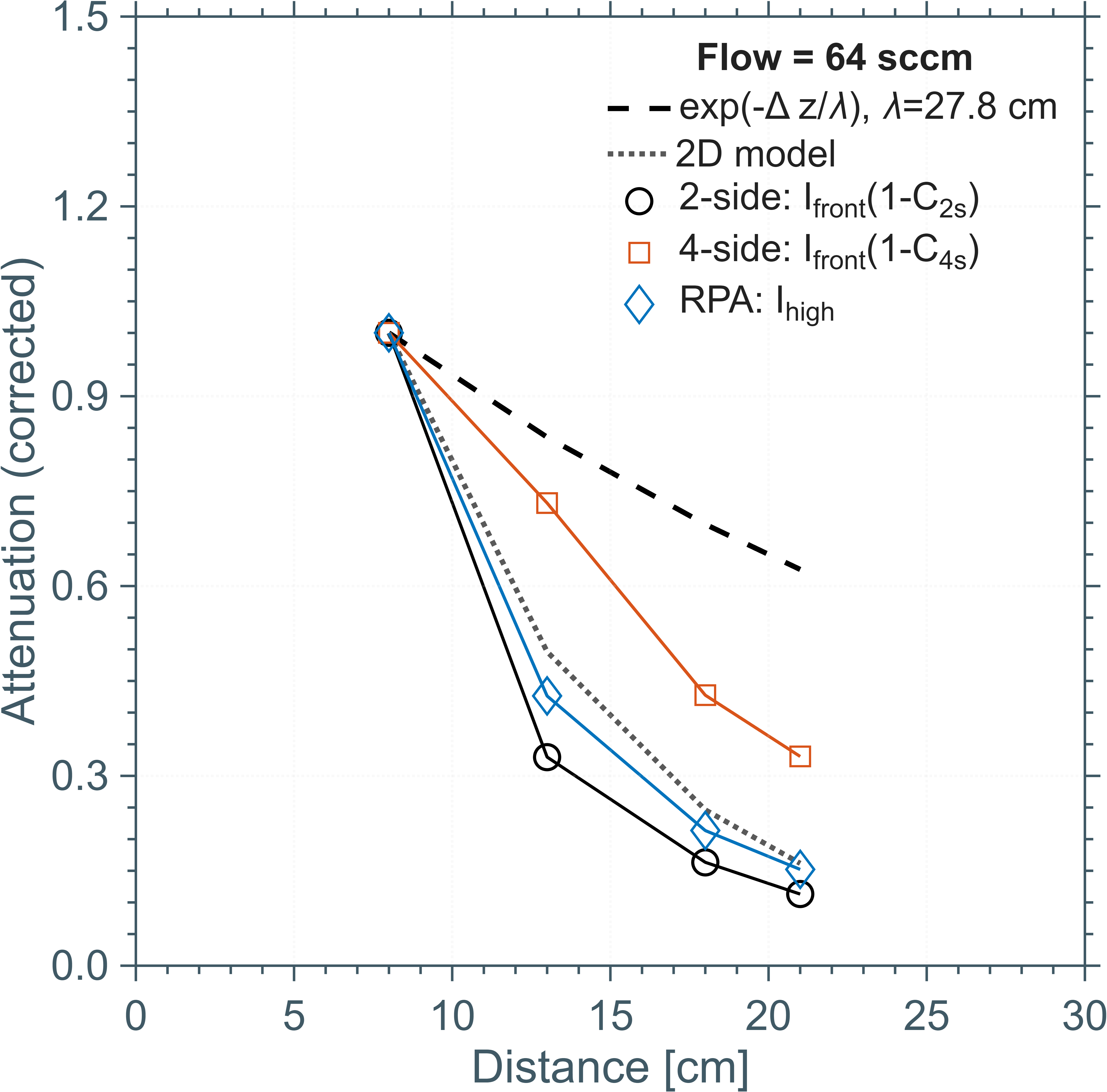}
\end{minipage}
\caption{\label{fig:atten_uncorr_corr_flow64}Representative comparison of the attenuation at $Q_{\mathrm{add}}=64$~sccm. Left: normalized attenuation measured with the uncorrected front current of the double-sided and four-sided planar probes and with the total RPA current $I_{\mathrm{total}}$, together with the exponential attenuation and the quasi-2D model. Right: the same comparison after correction of the planar-probe front current using the background fractions, $I_{\mathrm{front}}(1-C_{\mathrm{back}})$, together with the RPA high-energy current $I_{\mathrm{high}}$. Here $C_{2s}\equiv C_{\mathrm{back,2side}}$ and $C_{4s}\equiv C_{\mathrm{back,4side}}$. All points were normalized to $z=z_{\mathrm{ref}}$.}
\end{figure}

Figure~\ref{fig:atten_uncorr_corr_flow64} presents a representative case for $Q_{\mathrm{add}}=64$~sccm. Similar qualitative behavior was observed at all other added-gas flow rates (Appendix~\ref{app:fast_ions_atten}). The dashed black curve denotes the exponential attenuation predicted by the Beer--Lambert model [Eq.~\ref{eq:BLL}], while the dashed gray curve denotes the quasi-2D model. The two- and four-sided planar-probe measurements differ substantially from one another, consistent with discrepancies reported previously for similar probes (Ref. \onlinecite{RSI_probe}), and they also differ from the RPA data and the theoretical predictions.

The background correction does not significantly improve the agreement. The double-sided probe yields the closest planar-probe approximation to the quasi-2D model, whereas the four-sided probe attenuates more slowly and remains elevated at larger distances. This difference may arise from the smaller geometric size of the four-sided probe, which is expected to produce stronger edge effects (Ref. \onlinecite{RSI_probe}). The attenuation behavior is also consistent with the sheath estimates in Sec.~\ref{sec:results_background_fraction}. The estimated values of $A_{\mathrm{eff}}/A_c$ are $1.1$--$1.2$ on the front side and $1.5$--$1.6$ on the back side for the double-sided probe, and $1.3$--$1.6$ and $1.9$--$2.7$, respectively, for the four-sided probe. Although approximate, these estimates support a larger effective angular acceptance for the four-sided front collector. Moreover, sheath expansion may become stronger at larger distances and for lower-energy ions.

For the present comparison, the high-energy RPA current provides the most direct reference for the directed fast-ion component among the diagnostics used here. Because the RPA is energy selective and accepts a relatively collimated ion flux, only a small fraction of ions entering at an angle are interpreted as having lower energy than their true energy. It is therefore expected to track the directed on-axis fast-ion component and to remain close to the quasi-2D model.

\subsection{Normalized fast-ion attenuation and fast-neutral equivalent current from the thermal flux probe}
\label{sec:results_norm_flux_model}

The heat flux measured by the thermal flux probe can be decomposed into a fast-ion contribution and into an equivalent fast-neutral current using Eq.~(\ref{eq:power_balance}). Since the thermal flux probe does not directly measure this background contribution, the correction must be obtained from complementary diagnostics. It is therefore necessary to verify that the fast-ion flux inferred from the thermal flux probe remains consistent with the ion-flux measurements from the other probes. For this purpose, three correction choices were examined: the RPA low-energy fraction $C_{\mathrm{low}}^{\mathrm{RPA}}$, the double-sided planar-probe ratio $C_{\mathrm{back,2side}}$, and the four-sided planar-probe ratio $C_{\mathrm{back,4side}}$, as defined in Section~\ref{sec:results_background_fraction}. The resulting thermal-flux-probe fast-ion current can then be evaluated for its sensitivity to the selected correction by comparison with the quasi-2D model in Section~\ref{sec:model}.

\begin{figure}[ht]
\centering
\begin{minipage}[t]{0.25\linewidth}
\centering
\includegraphics[height=4cm]{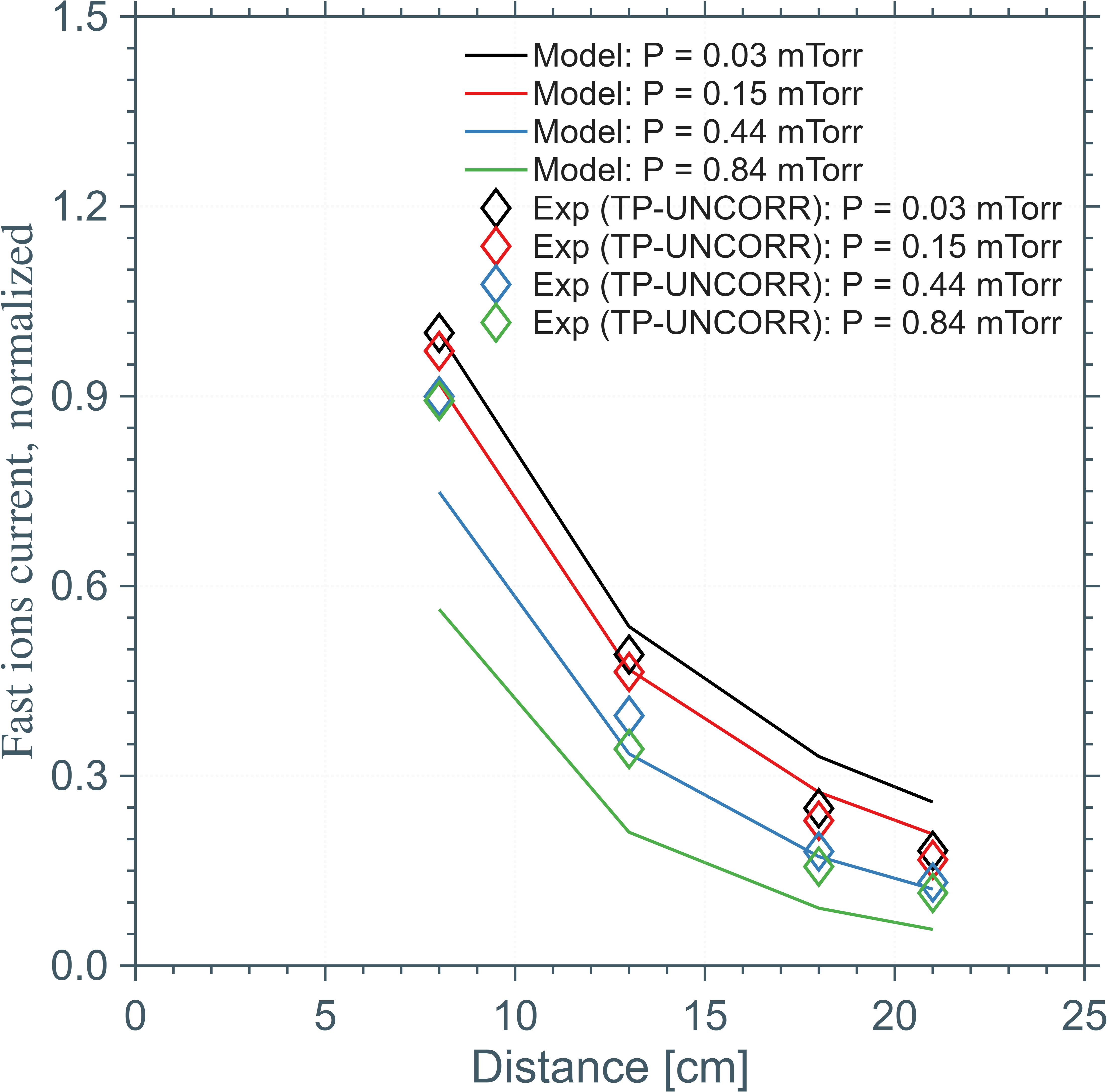}
\end{minipage}\hfill
\begin{minipage}[t]{0.25\linewidth}
\centering
\includegraphics[height=4cm]{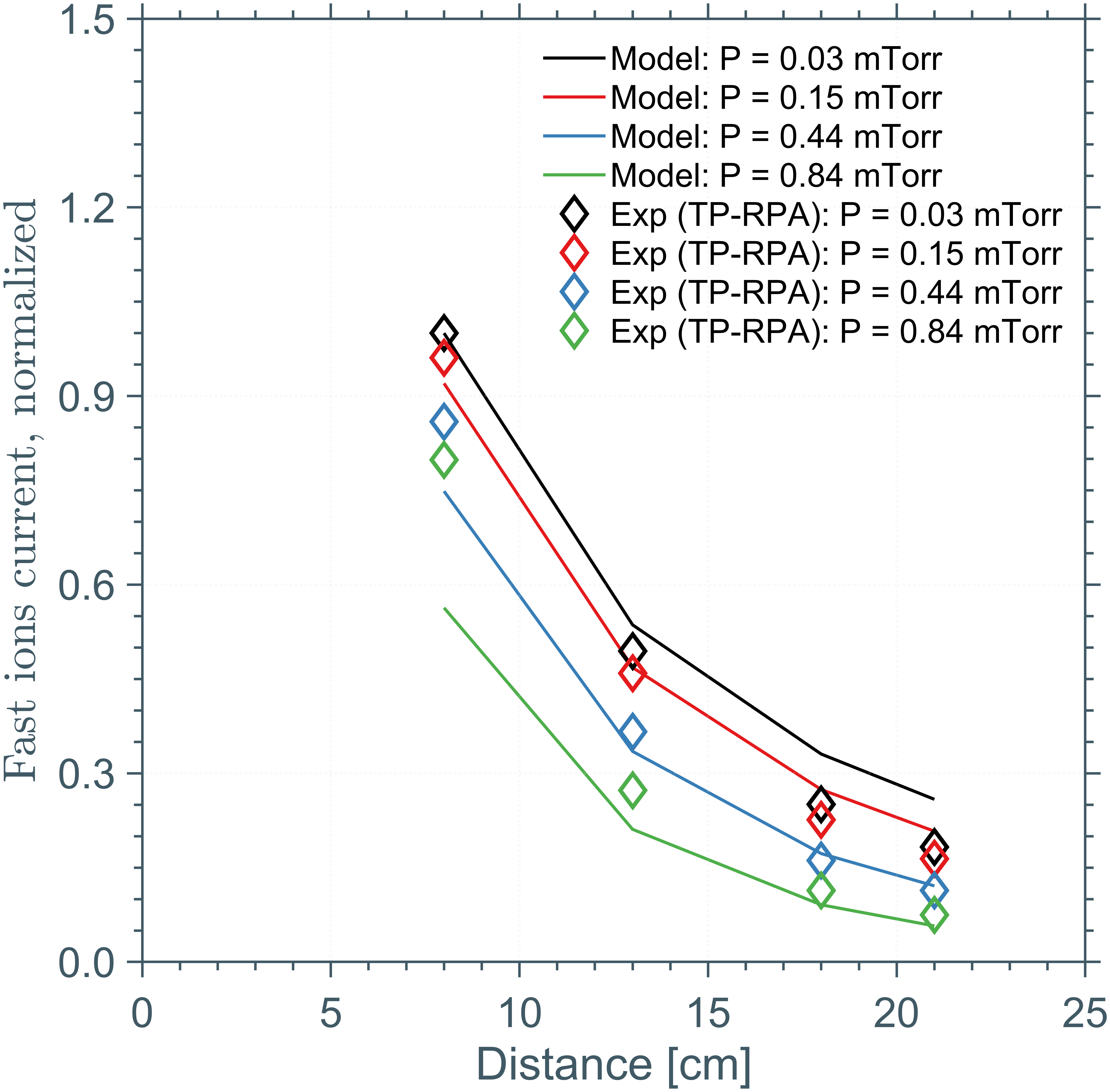}
\end{minipage}\hfill
\begin{minipage}[t]{0.25\linewidth}
\centering
\includegraphics[height=4cm]{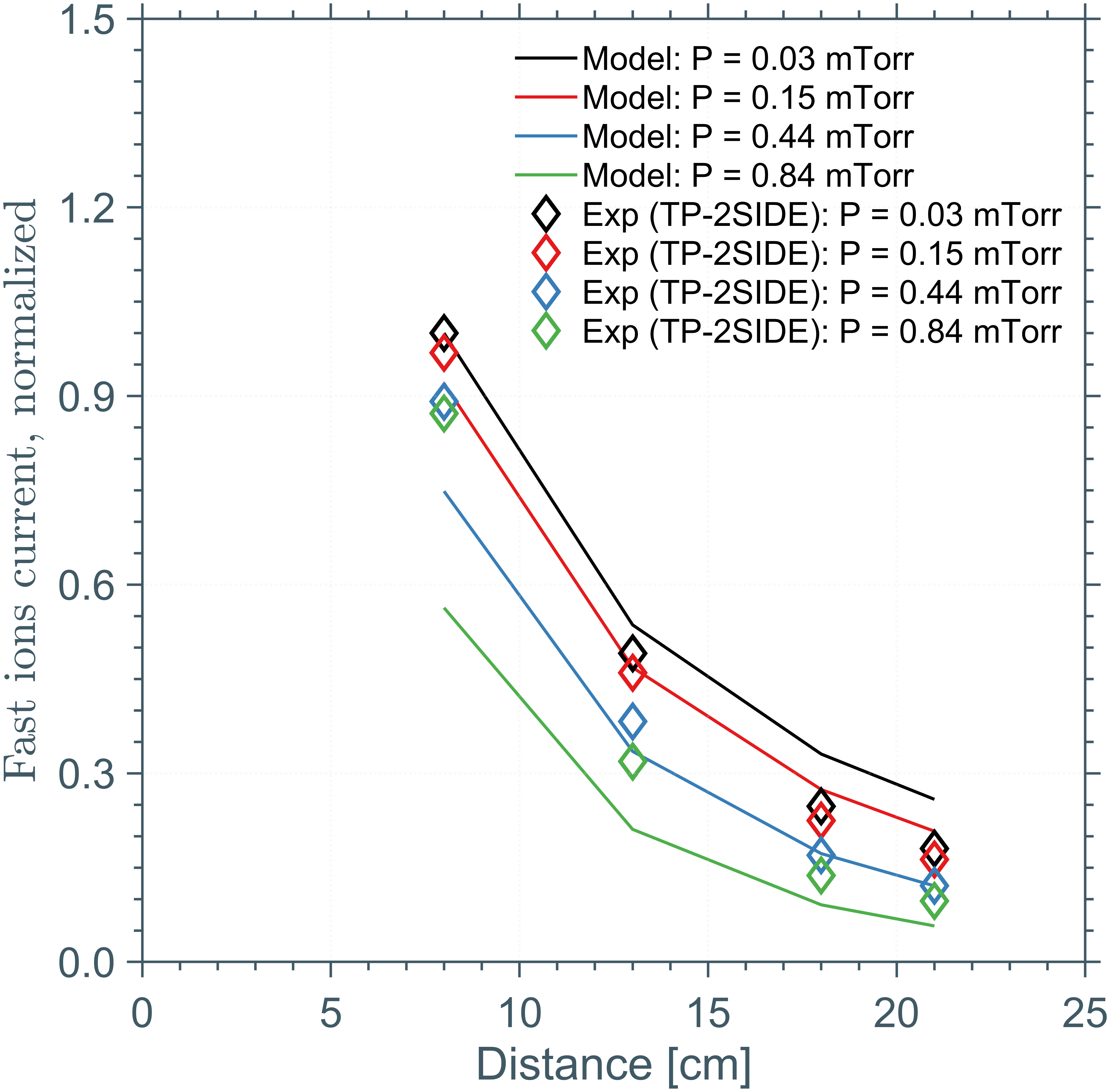}
\end{minipage}\hfill
\begin{minipage}[t]{0.25\linewidth}
\centering
\includegraphics[height=4cm]{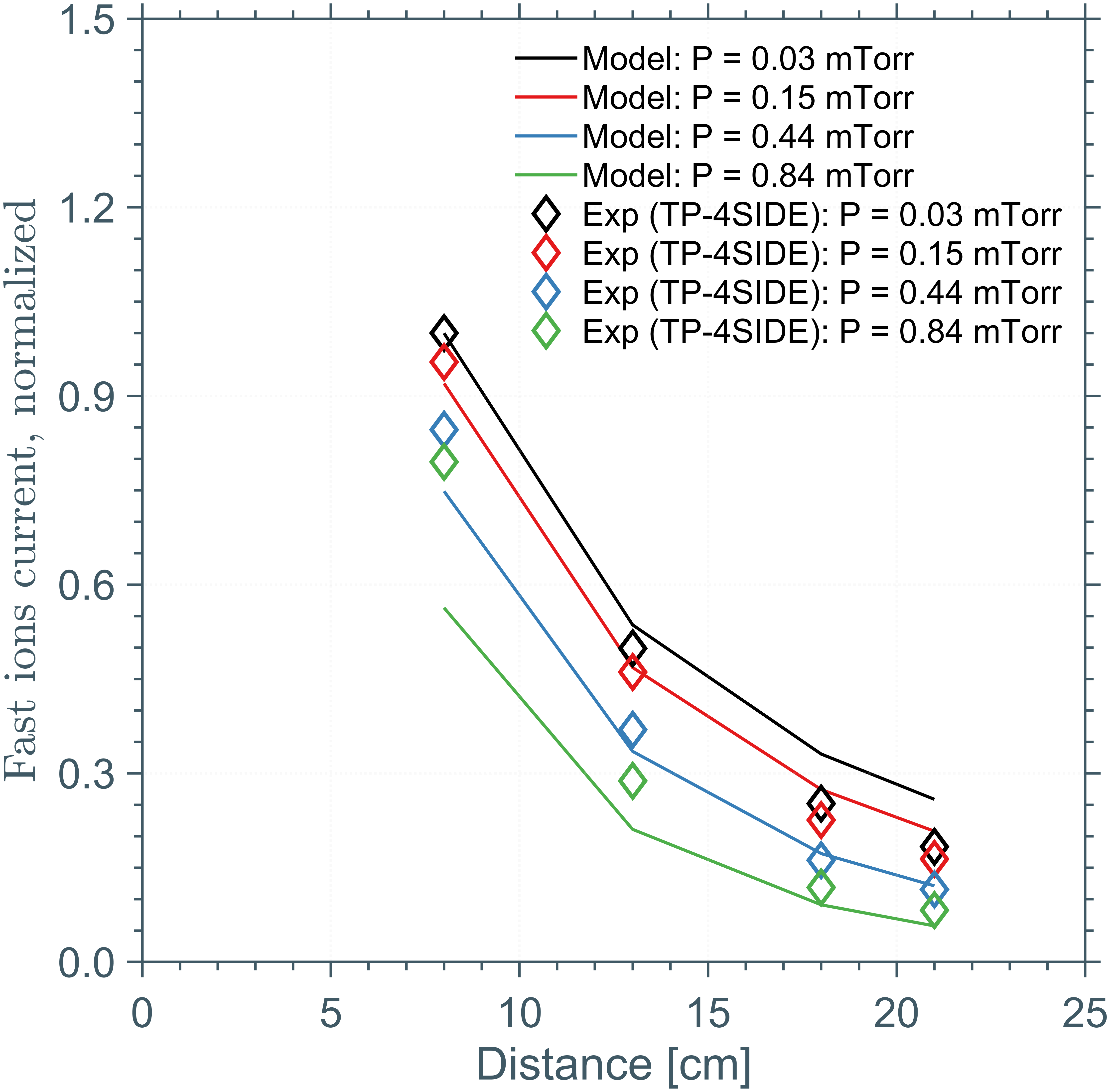}
\end{minipage}
\caption{\label{fig:tp_ion_norm_model}Normalized fast-ion current from the thermal flux probe versus distance at four background gas pressures ($P=0.03$, 0.15, 0.44, and 0.84~mTorr). Symbols show the fast-ion current obtained from the thermal flux probe for four cases: (left) without low-energy correction, (center-left) using $C_{\mathrm{low}}^{\mathrm{RPA}}$, (center-right) using $C_{\mathrm{back,2side}}$, and (right) using $C_{\mathrm{back,4side}}$. Solid lines show the quasi-2D model for the same pressures. All curves are normalized to $I_i(z_0,P_0)$ with $z_0=8$~cm and $P_0=0.03$~mTorr.}
\end{figure}

Figure~\ref{fig:tp_ion_norm_model} shows that the normalized fast-ion current decreases with both distance and pressure for all correction choices.  Using the root-mean-square error between theoretical curves and measured points over all conditions gives RMSE$=0.11$ for the uncorrected case, RMSE$=0.08$ for the $C_{\mathrm{low}}^{\mathrm{RPA}}$ case, RMSE$=0.10$ for the $C_{\mathrm{back,2side}}$ case, and RMSE$=0.08$ for the $C_{\mathrm{back,4side}}$ case. Here the RMSE is used only as a relative indicator of change between cases, not as an absolute measure of improvement level. In that relative sense, the RPA-based and four-sided corrections give the closest agreement with the model, while the double-sided correction remains slightly higher, mainly at the larger pressures. This is consistent with the results in Fig.~\ref{fig:bg_fraction_compare} that the RPA and four-sided probe gave more similar estimates of the low-energy background contribution. Overall, the ion flux measurements with the thermal flux probe are consistent with the other probes.

After the ion term is estimated, a fast-neutral equivalent current is obtained from Eq.~\ref{eq:power_balance}. This then allows the neutral component to be examined under the same operating conditions and with the same three low-energy correction choices.

\begin{figure}[ht]
\centering
\begin{minipage}[t]{0.25\linewidth}
\centering
\includegraphics[height=4cm]{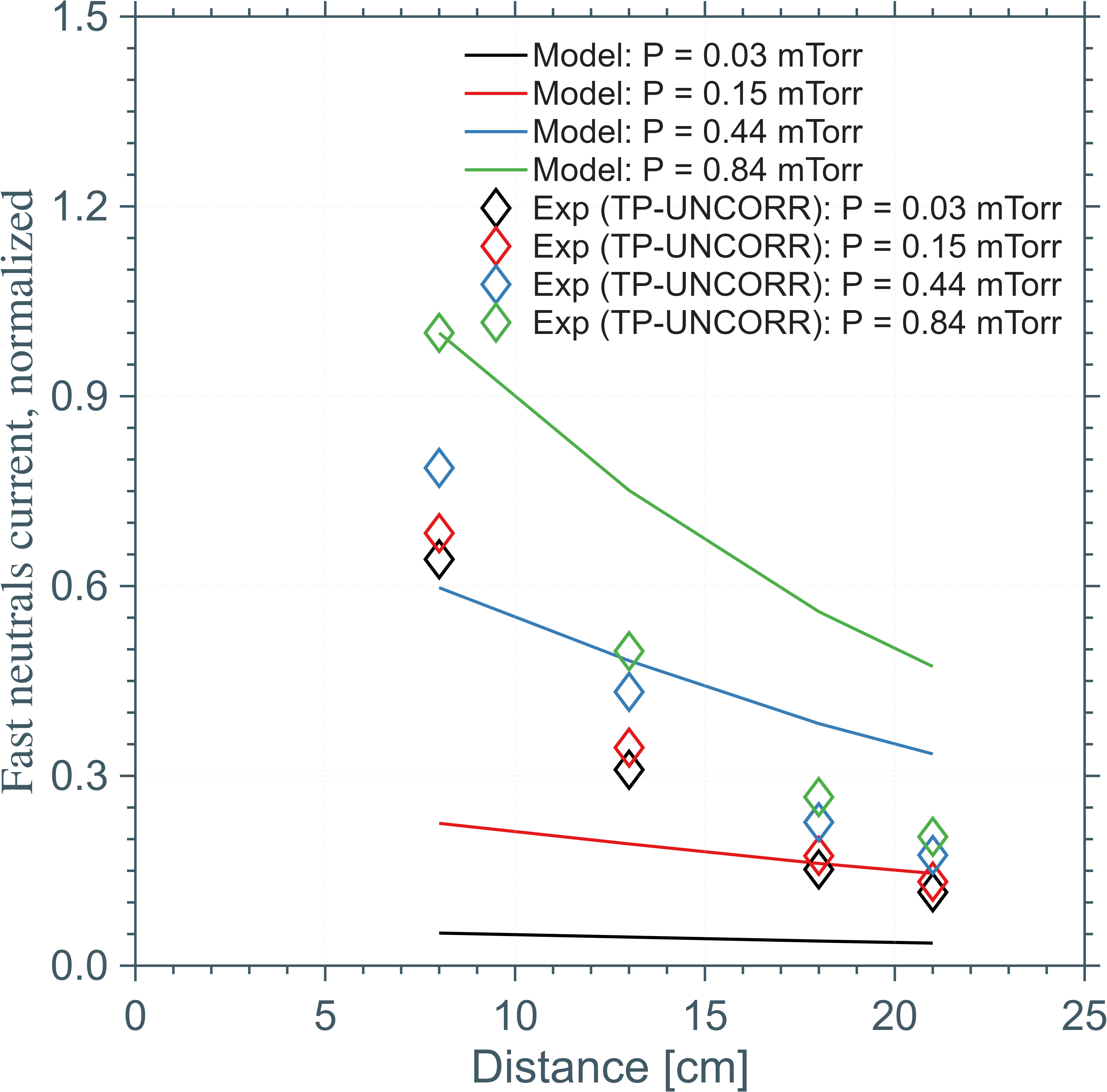}
\end{minipage}\hfill
\begin{minipage}[t]{0.25\linewidth}
\centering
\includegraphics[height=4cm]{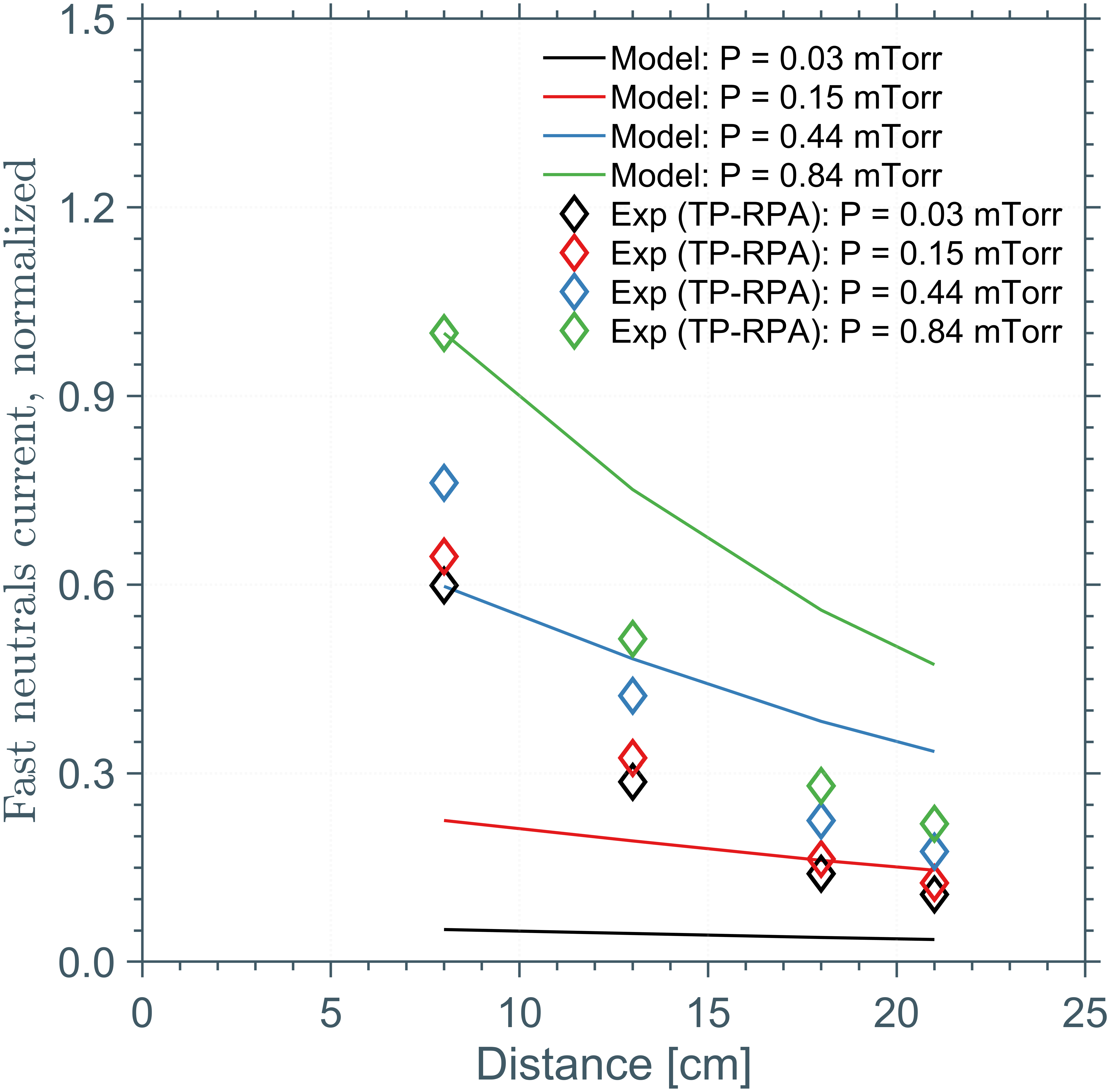}
\end{minipage}\hfill
\begin{minipage}[t]{0.25\linewidth}
\centering
\includegraphics[height=4cm]{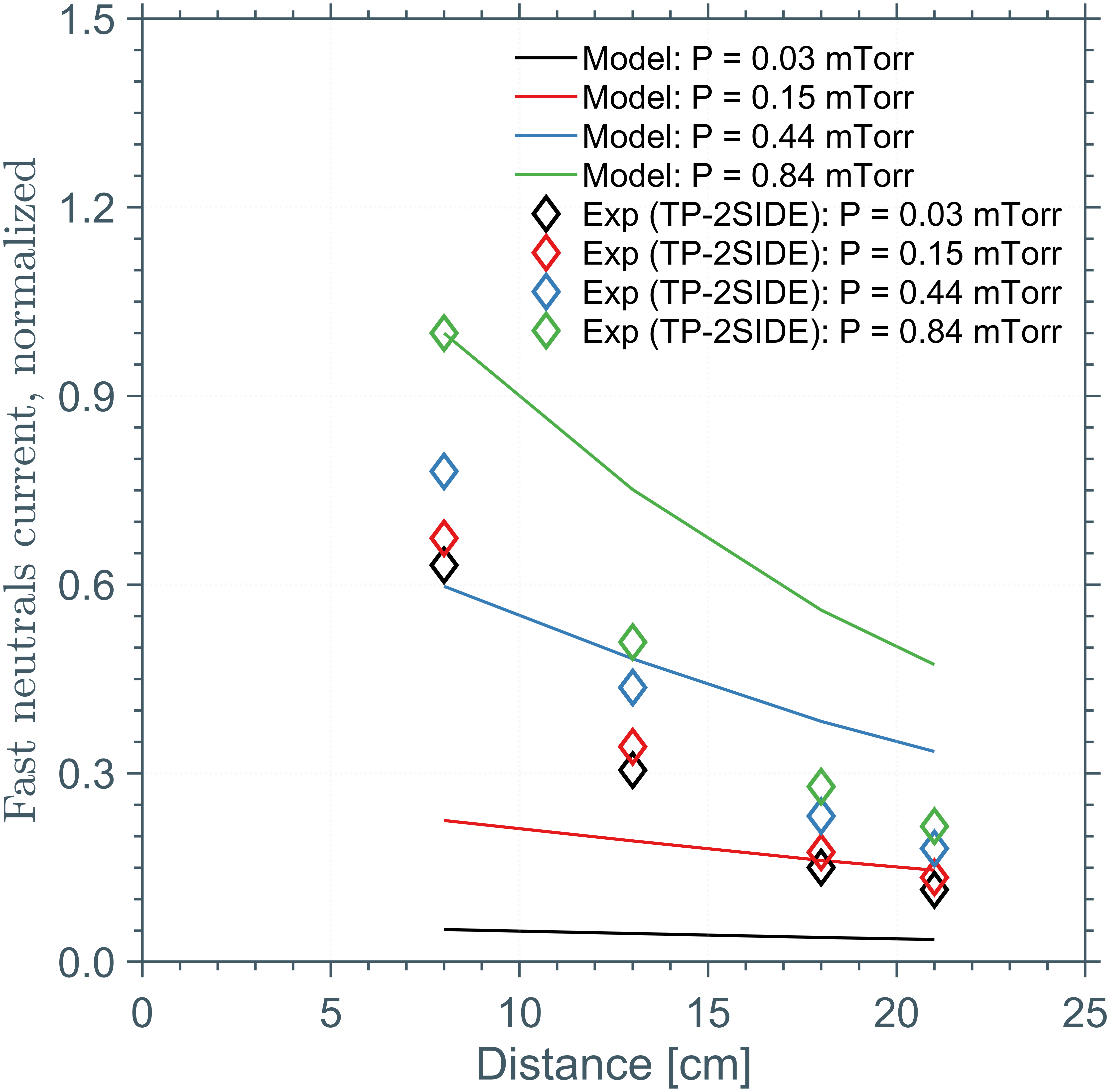}
\end{minipage}\hfill
\begin{minipage}[t]{0.25\linewidth}
\centering
\includegraphics[height=4cm]{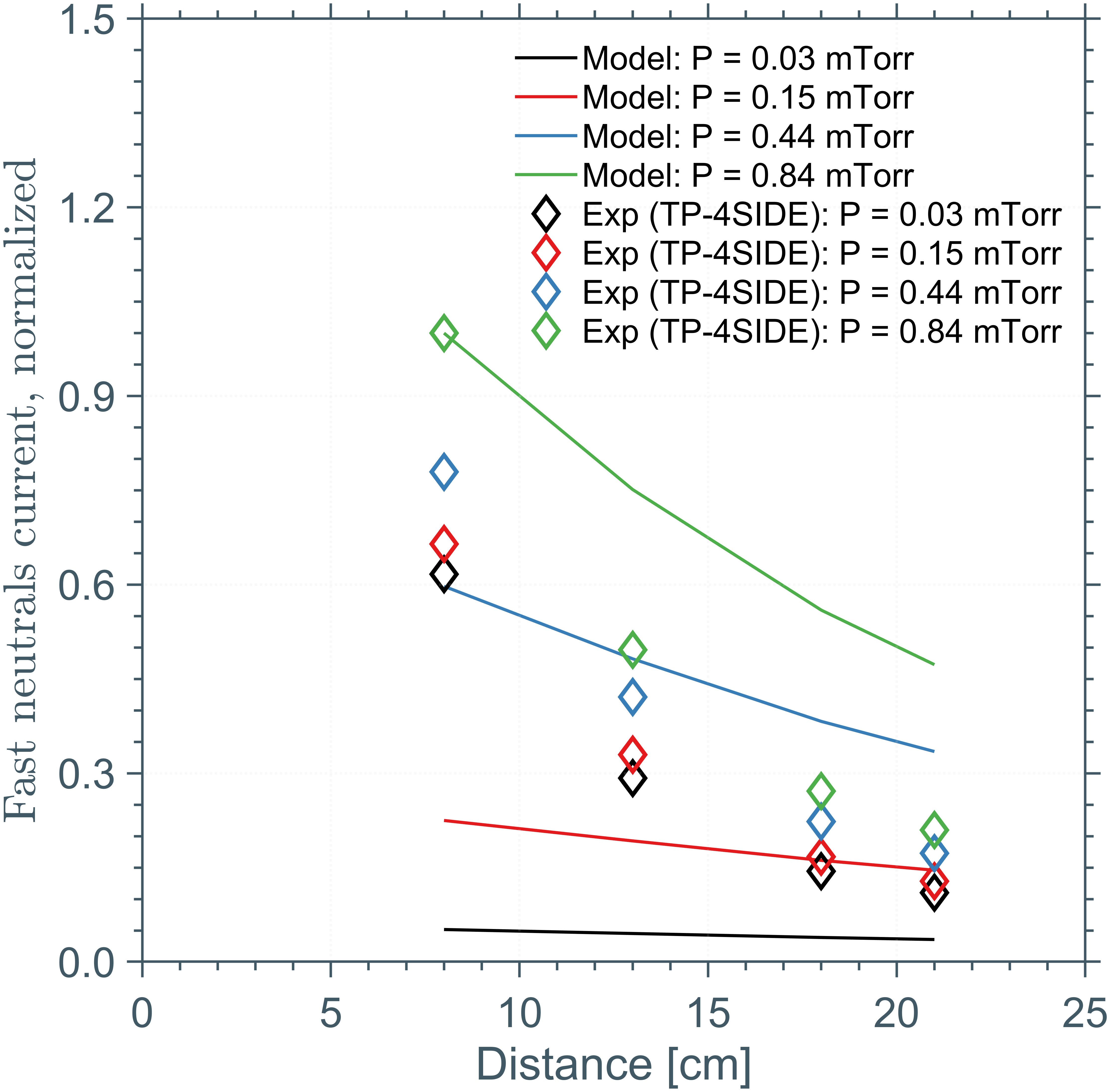}
\end{minipage}
\caption{\label{fig:tp_neutral_norm_model}Normalized fast-neutral equivalent current from the thermal flux probe versus distance at four background gas pressures ($P=0.03$, 0.15, 0.44, and 0.84~mTorr). Symbols show the fast-neutral equivalent current obtained from the thermal flux probe for four cases: (left) without low-energy correction, (center-left) using $C_{\mathrm{low}}^{\mathrm{RPA}}$, (center-right) using $C_{\mathrm{back,2side}}$, and (right) using $C_{\mathrm{back,4side}}$. Solid lines show the quasi-2D model for the same pressures. All curves are normalized to $I_n(z_0,P_{\max})$ with $z_0=8$~cm and $P_{\max}=0.84$~mTorr.}
\end{figure}

Figure~\ref{fig:tp_neutral_norm_model} shows the fast-neutral equivalent current Eq.~(\ref{eq:power_balance}). The quantitative mismatch between the experimental data and the quasi-2D model was evaluated using the RMSE, as for the fast-ion currents: RMSE$=0.25$ for the uncorrected case, RMSE$=0.23$ for $C_{\mathrm{low}}^{\mathrm{RPA}}$, RMSE$=0.24$ for $C_{\mathrm{back,2side}}$, and RMSE$=0.24$ for $C_{\mathrm{back,4side}}$. Thus, the discrepancy with the model remains systematic in both the corrected and uncorrected cases, and the choice of correction has only a weak effect on the overall agreement. Qualitatively, however, the model captures the main trends, indicating that the neutral distribution is shaped by both CEX and plume divergence. Specifically, the normalized current increases strongly with pressure, consistent with enhanced CEX production at higher background density, and decreases with distance at fixed pressure, indicating the additional influence of plume divergence after formation.

The discrepancy between the model and the measured data may arise from several factors. First, as discussed above, the inferred fast-ion flux does not fully follow the model prediction, which indicates uncertainty in the estimated ion-power term. Second, the treatment of the neutral flux is likely oversimplified, since the model assumes that the fast neutrals have the same energy and angular distribution as the fast ions and that they are produced only in the plume downstream of the ion source. However, the fast-neutral flux inferred from the measurements (Fig.~\ref{fig:tp_neutral_norm_model}) is already much larger than predicted at the $z_0$ position. This suggests that a substantial fraction of the fast neutrals may originate inside the ion source or in the gap between the acceleration grids, where the pressure is expected to be higher than in the plume because of the finite transparency of the grids and charge-exchange processes are more likely because the neutral mean free path is shorter. Such neutrals would be expected to have a broad spread in energy and larger divergence, since they are formed from ions accelerated through different potentials and with different initial angular distributions. This contribution is not included in the present model and requires further confirmation.

Overall, the normalized thermal-flux-probe results separate into two cases. The fast-ion attenuation is physically consistent, only weakly sensitive to the selected correction, and reasonably close to the quasi-2D model. The fast-neutral equivalent current is similarly insensitive to the correction choice in normalized form, but it shows a systematic disagreement with the model. Under the present assumptions, this indicates that the main limitation is not the choice of correction itself, but the present treatment of the fast-neutral contribution in the thermal-flux-probe analysis and model comparison.

\newpage

\section{\label{sec:conclusion}Conclusions}
In this work, we characterize effects of background gas pressure on CEX within the plume of a 400 eV argon ion beam. Measurements were conducted using a suite of diagnostics, including a custom-developed thermal flux probe, planar probes, and RPA. The results show that increasing background gas pressure changes not only the plume through CEX, but also the measured probe response through low-energy ions and effective collection geometry changes due to sheath effects.

For fast ions, the main conclusion is that the measured on-axis attenuation is controlled by both CEX and plume divergence. An analytical solution of the reduced semi-empirical quasi-2D model, which includes these effects, reproduces the attenuation more accurately than a simple 1D Beer--Lambert model. Among the diagnostics, the RPA provides the most direct reference for the fast-ion component, the corrected double-sided probe gives the closest planar-probe approximation, and the four-sided probe is less selective because of its broader effective angular acceptance.

For fast neutrals, the equivalent current derived from thermal flux probe measurements shows the expected increase with pressure, but the same quasi-2D model is not sufficient to reproduce the measured behavior. The mismatch is systematic and remains similar for all correction methods, so it is unlikely to be controlled mainly by the choice of the low-energy ion correction. This indicates that the main limitation lies in the present treatment of the fast-neutral term, including angular redistribution after CEX, multiple collisions, uncertainty in the neutral impact energy used in the power balance, and possible fast neutrals produced inside the ion source not included in the model initial condition.

Overall, these results show that effects of background gas pressure in plasma plumes are not adequately described in all cases by a simple Beer--Lambert attenuation law. That approximation is most appropriate when attenuation is dominated by line-of-sight loss along a prescribed path. However, when the plume is not well collimated, when collisions or other processes modify the angular distributions of ions and neutrals along the plume, and when plume divergence and probe geometry affect the measured response, a more explicit treatment of plume geometry is required. The thermal flux probe developed here can be used to directly track fast-neutral trends as the background gas pressure varies. Further work is needed to improve quantitative agreement with theory through self-consistent coupled ion and neutral transport modeling and additional validation of the diagnostic technique.

\begin{acknowledgments}
The authors acknowledge Drs. Hokuto Sekine,  Alexander Khrabrov, Igor Kaganovich for fruitful discussions.  This research was supported in part by Samsung Electronics Co., Ltd. (IO230718‑06770‑01) and the U.S. Department of Energy through the Princeton Collaborative Research Facility (PCRF) under Contract No. DEAC02-09CH11466.
\end{acknowledgments}

\section*{Conflict of Interest Statement}
The authors have no conflicts to disclose.

\section*{Data Availability Statement}
The data that support the findings of this study are available from the corresponding author upon reasonable request.

\newpage

\appendix
\section{Pressure distribution along the plasma plume and probe path}\label{app:pressure}

For the present analysis, the pressure distribution along the plasma plume and probe path is important because the local background neutral density sets the strength of facility-induced interactions. To determine whether the pressure can be approximated as uniform over the measurement region, the axial pressure profile was measured using a nude ion gauge mounted on a movable rod and translated along the chamber centerline between the ion source and the downstream diagnostics. For this scan, position was referenced to the ion source exit plane $z_s$.

\begin{figure}[ht]
\centering
\begin{minipage}[t]{0.32\linewidth}
\centering
\includegraphics[height=4cm]{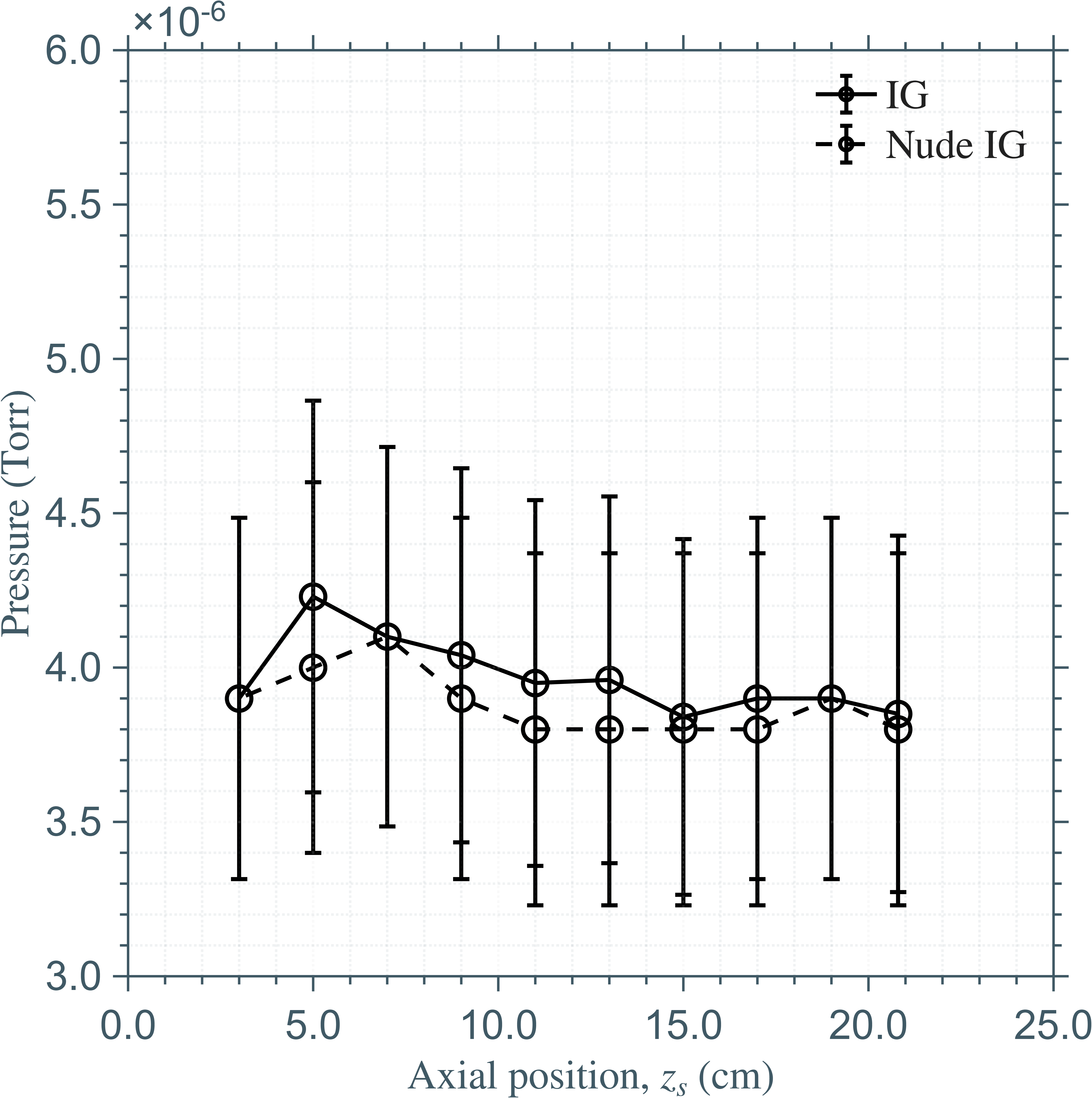}
\end{minipage}\hfill
\begin{minipage}[t]{0.32\linewidth}
\centering
\includegraphics[height=4cm]{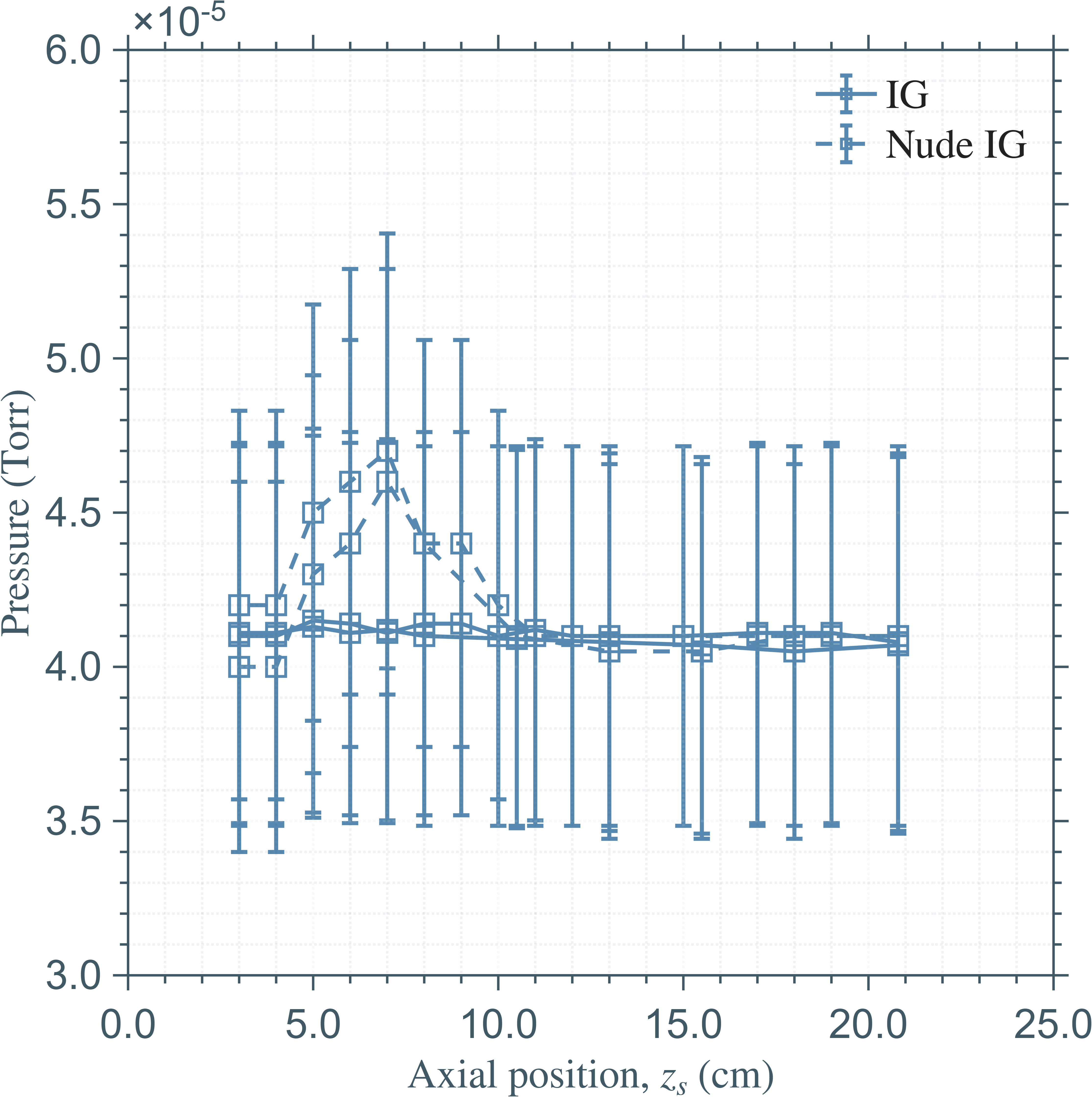}
\end{minipage}\hfill
\begin{minipage}[t]{0.32\linewidth}
\centering
\includegraphics[height=4cm]{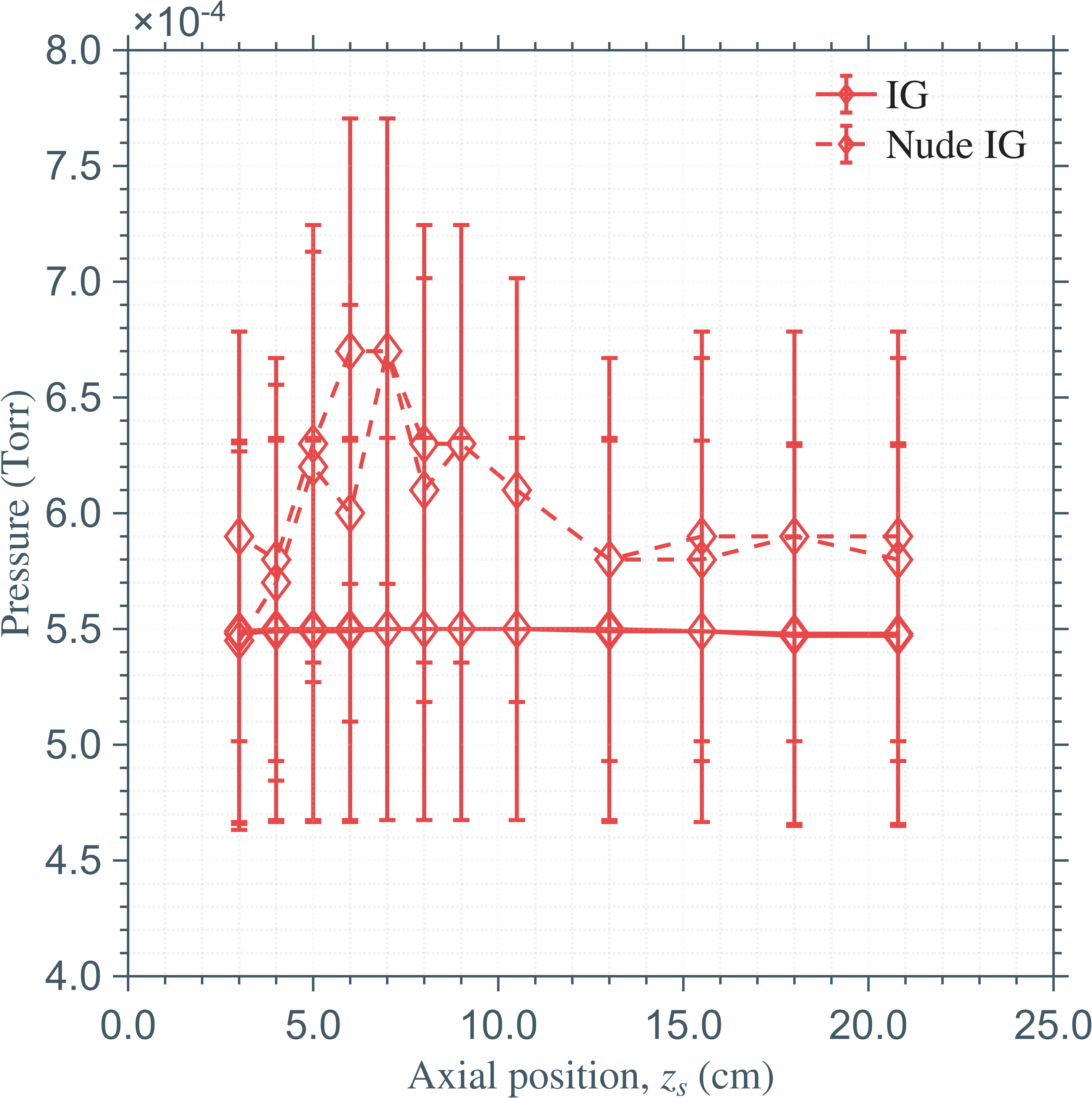}
\end{minipage}
\caption{\label{fig:pressure_uniformity}Axial background gas pressure profiles measured with a stationary ion gauge (IG) and a movable nude ion gauge (Nude IG). The horizontal axis shows the translation-stage coordinate $z_s$ along the chamber centerline; reported axial coordinates satisfy $z=z_s+3$~cm. Operating conditions are (left) $Q_{\mathrm{IS}}/Q_{\mathrm{add}}=0/0$~sccm, (middle) $Q_{\mathrm{IS}}/Q_{\mathrm{add}}=4.6/0$~sccm, and (right) $Q_{\mathrm{IS}}/Q_{\mathrm{add}}=4.6/64$~sccm. Error bars indicate an assumed ion-gauge uncertainty of $\pm 15\%$. The ion source was not operated during measurements.}
\end{figure}

Figure~\ref{fig:pressure_uniformity} compares the pressure measured by the stationary chamber ion gauge (IG) and the movable nude gauge (Nude IG) for three representative flow conditions: (a) $Q_{\mathrm{IS}}/Q_{\mathrm{add}}=0/0$~sccm, (b) $Q_{\mathrm{IS}}/Q_{\mathrm{add}}=4.6/0$~sccm, and (c) $Q_{\mathrm{IS}}/Q_{\mathrm{add}}=4.6/64$~sccm. The axial variation over the scanned range is weak, and the two gauges agree within the assumed uncertainty of $\pm 15\%$. Thus, any axial dependence is too small to affect the CEX characterization, and the background pressure is treated as spatially uniform and represented by the stationary gauge reading $P$.

Because the ion source was inactive during the pressure measurements, any ion beam induced pressure rise was not measured directly. Although such effects have been reported in ion-beam facilities, an upper-bound estimate indicates that they are negligible here. Ref.~\onlinecite{Thermal_Probe_AIAA} shows that the expected contribution is well below both the operating pressure range and the ion-gauge uncertainty. This justifies the assumption of a spatially uniform background neutral density $n_g$ in the reduced model (Sec.~\ref{sec:model}).

\section{Probe and RPA diagnostics with background correction}\label{app:probe_rpa_correction}

\begin{figure}[ht]
\centering
\begin{minipage}[t]{0.32\linewidth}
\centering
\includegraphics[height=4cm]{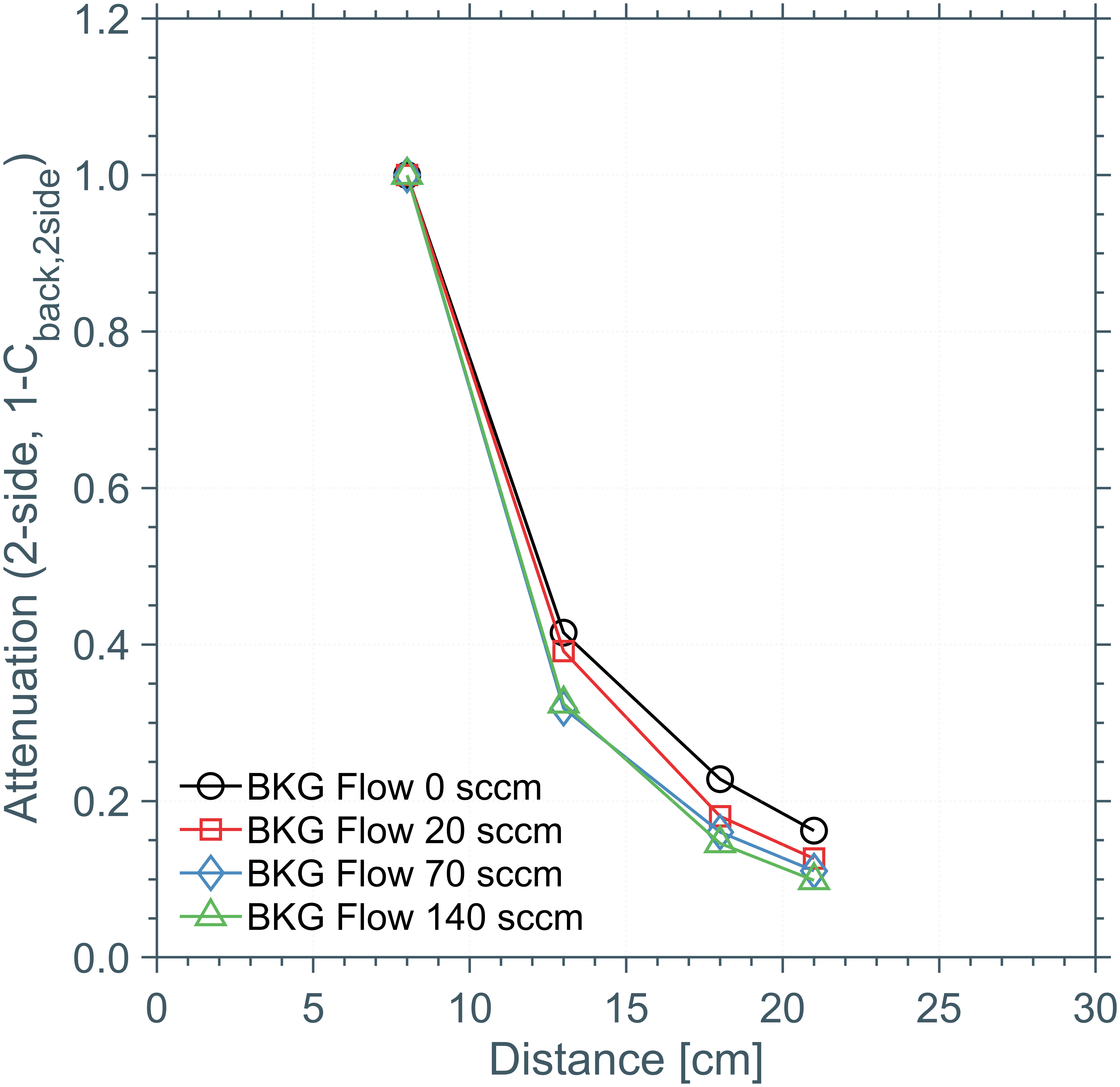}
\end{minipage}\hfill
\begin{minipage}[t]{0.32\linewidth}
\centering
\includegraphics[height=4cm]{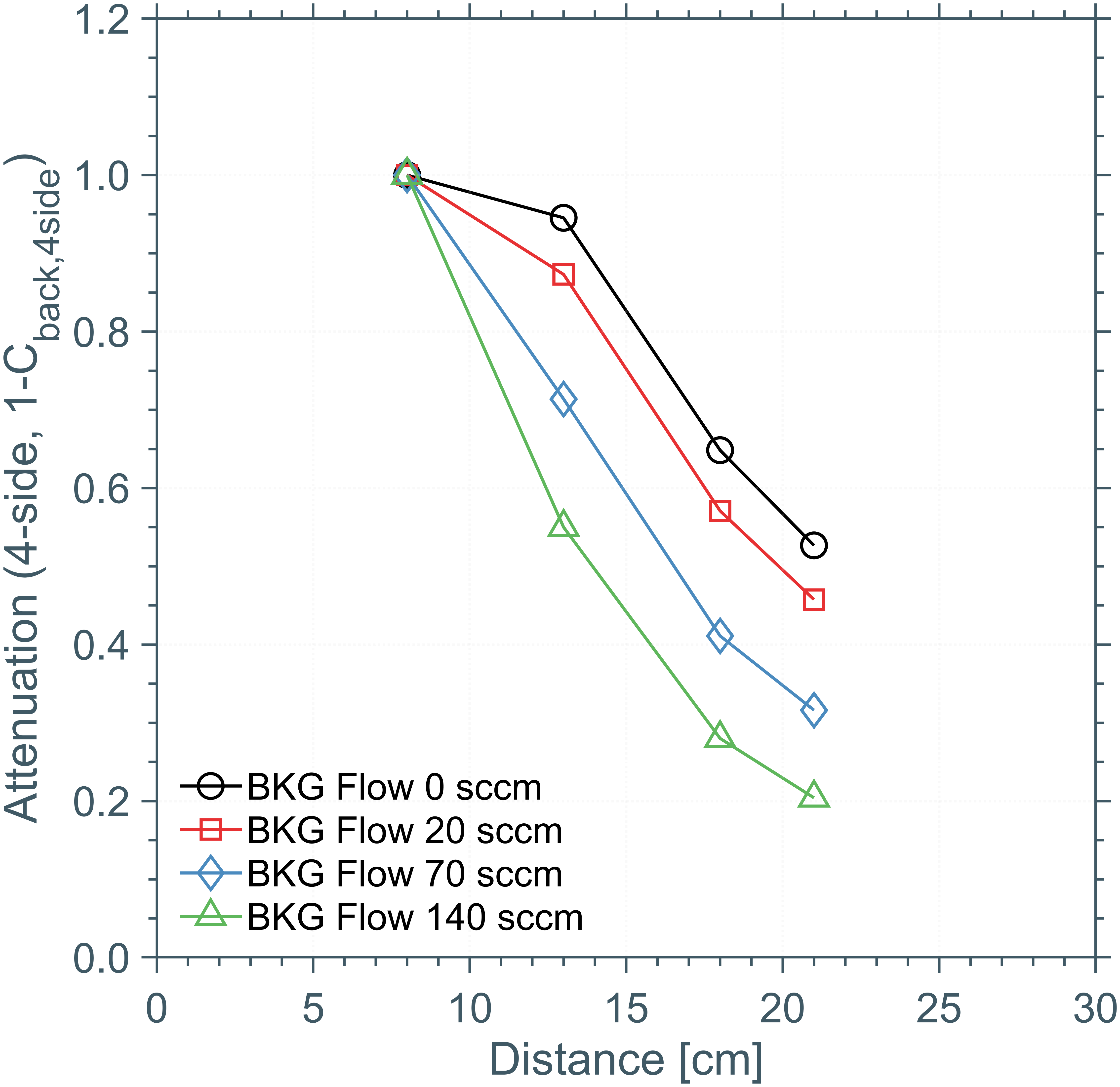}
\end{minipage}\hfill
\begin{minipage}[t]{0.32\linewidth}
\centering
\includegraphics[height=4cm]{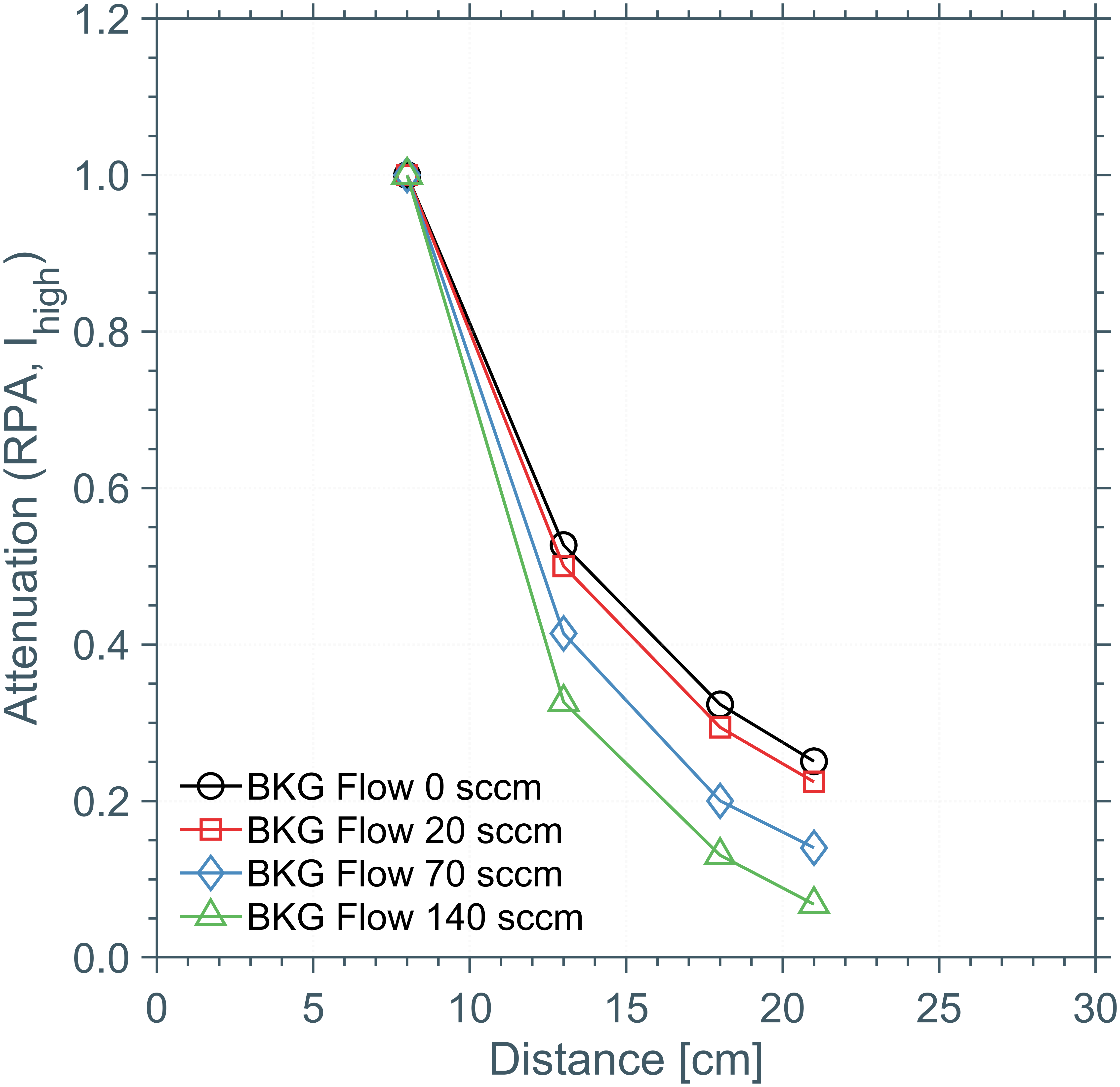}
\end{minipage}
\caption{\label{fig:atten_fast_front_allflows}Background-corrected fast-ion attenuation versus distance for four added chamber argon flows (0, 18, 64, and 128~sccm). (Left) double-sided probe $I_{\mathrm{front}}\left(1-C_{\mathrm{back,2side}}\right)$. (Middle) Four-sided probe $I_{\mathrm{front}}\left(1-C_{\mathrm{back,4side}}\right)$. (Right) RPA high-energy current $I_{\mathrm{high}}$. Each curve is normalized to $A(z_{\mathrm{ref}};Q_{\mathrm{add}})=1$.}
\end{figure}

The background-corrected fast-ion attenuation for all added flows is summarized in Fig.~\ref{fig:atten_fast_front_allflows}. For all three fast-ion estimates, attenuation increases with distance and becomes stronger as $Q_{\mathrm{add}}$ increases. The RPA-based $I_{\mathrm{high}}$ provides an energy-resolved reference trend for the directed fast-ion component. Across the full flow matrix, the double-sided corrected attenuation remains closest to $I_{\mathrm{high}}$, while the four-sided corrected attenuation remains systematically higher at larger distances.

The persistent separation between diagnostics indicates that correcting each probe current by its own background fraction, $I_{\mathrm{front,fast}}=I_{\mathrm{front}}(1-C_{\mathrm{back}})$, is not sufficient to remove all facility- and diagnostic-related effects. This correction targets the additive contribution associated with $C_{\mathrm{back}}$, but it does not correct for distance- and pressure-dependent collection efficiency and angular acceptance. As the plume propagates, divergence and collisions broaden the ion angular distribution. The RPA has a restricted angular acceptance set by its entrance aperture and grid stack, so ions that are deflected from the axis contribute less to the transmitted current, which steepens the measured attenuation of $I_{\mathrm{high}}$. The double-sided probe can also preferentially sample the near-axis directed component because its back-to-back geometry and larger obstruction can distort the local sheath and produce a wake as the density decreases downstream, reducing the collected front current beyond the background contribution. In contrast, the four-sided probe has a smaller body, and its sheath can expand to intercept ions arriving at larger angles. This increases the effective angular acceptance of the front collector and partially offsets the divergence-driven reduction of the on-axis directed component, yielding a weaker apparent attenuation and therefore systematically higher values at larger distance.

\section{Thermal-flux-probe ion attenuation with background correction}\label{app:thermal_flux_probe_correction}

\begin{figure}[ht]
\centering
\begin{minipage}[t]{0.32\linewidth}
\centering
\includegraphics[height=4cm]{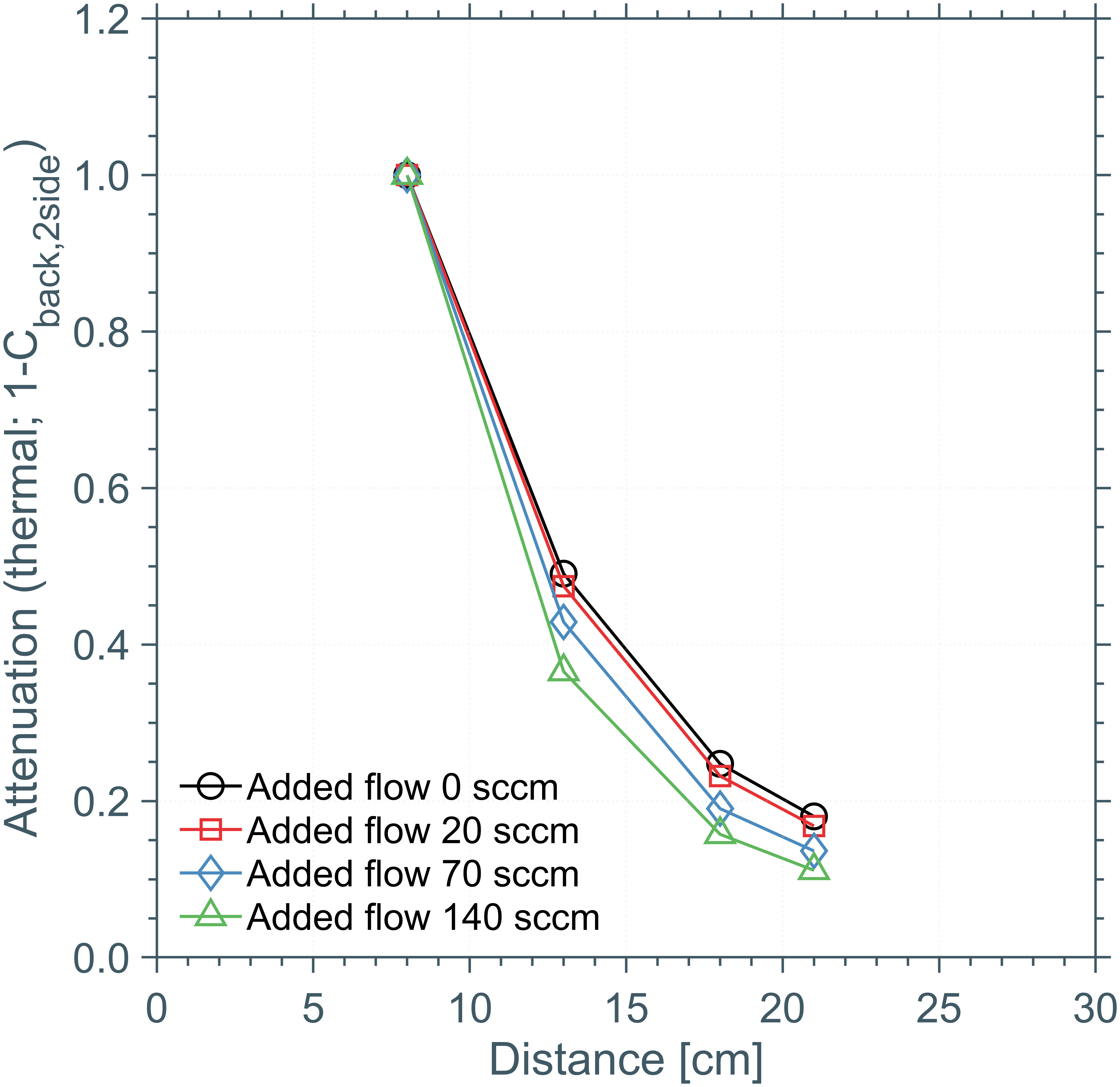}
\end{minipage}\hfill
\begin{minipage}[t]{0.32\linewidth}
\centering
\includegraphics[height=4cm]{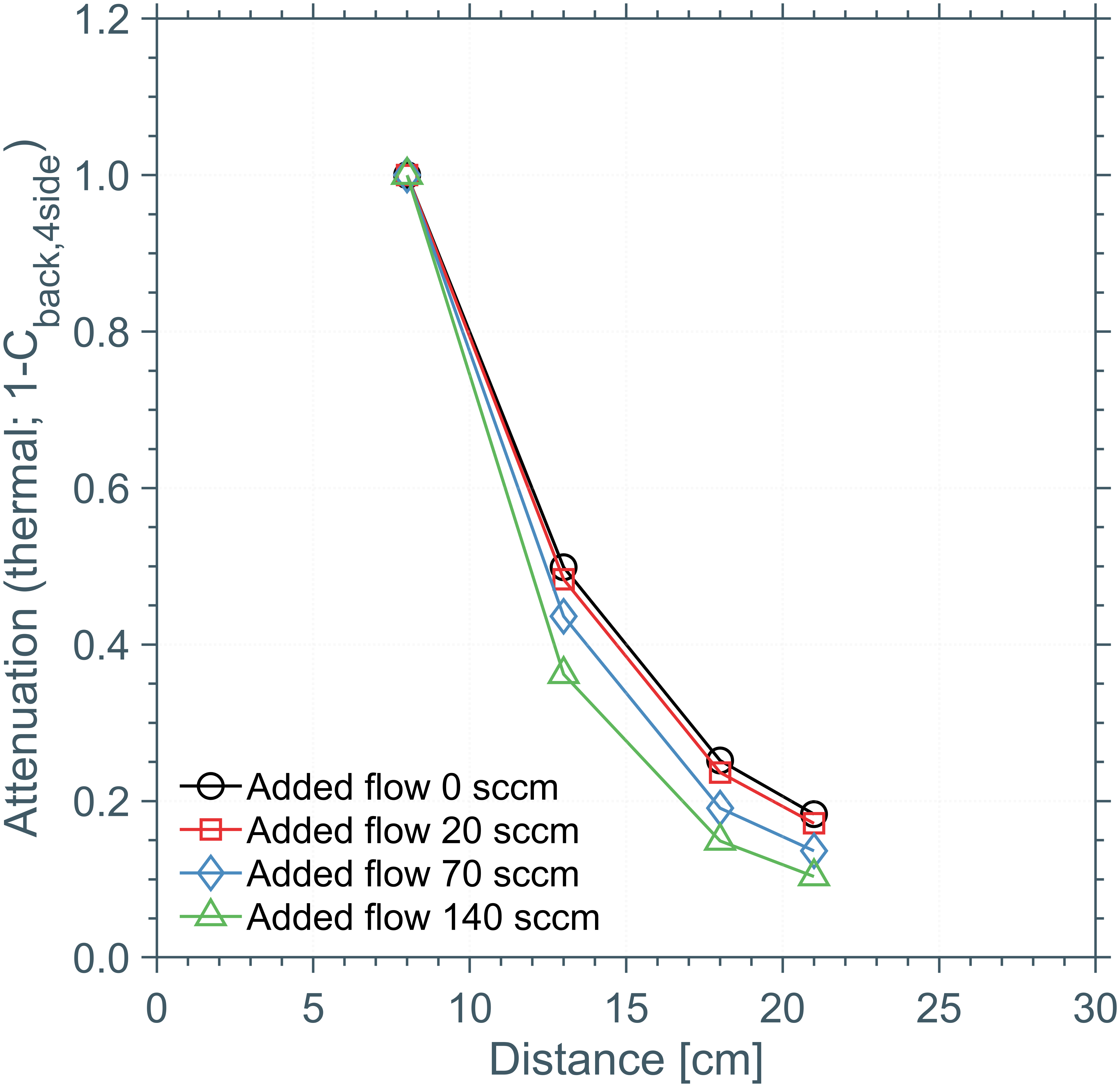}
\end{minipage}\hfill
\begin{minipage}[t]{0.32\linewidth}
\centering
\includegraphics[height=4cm]{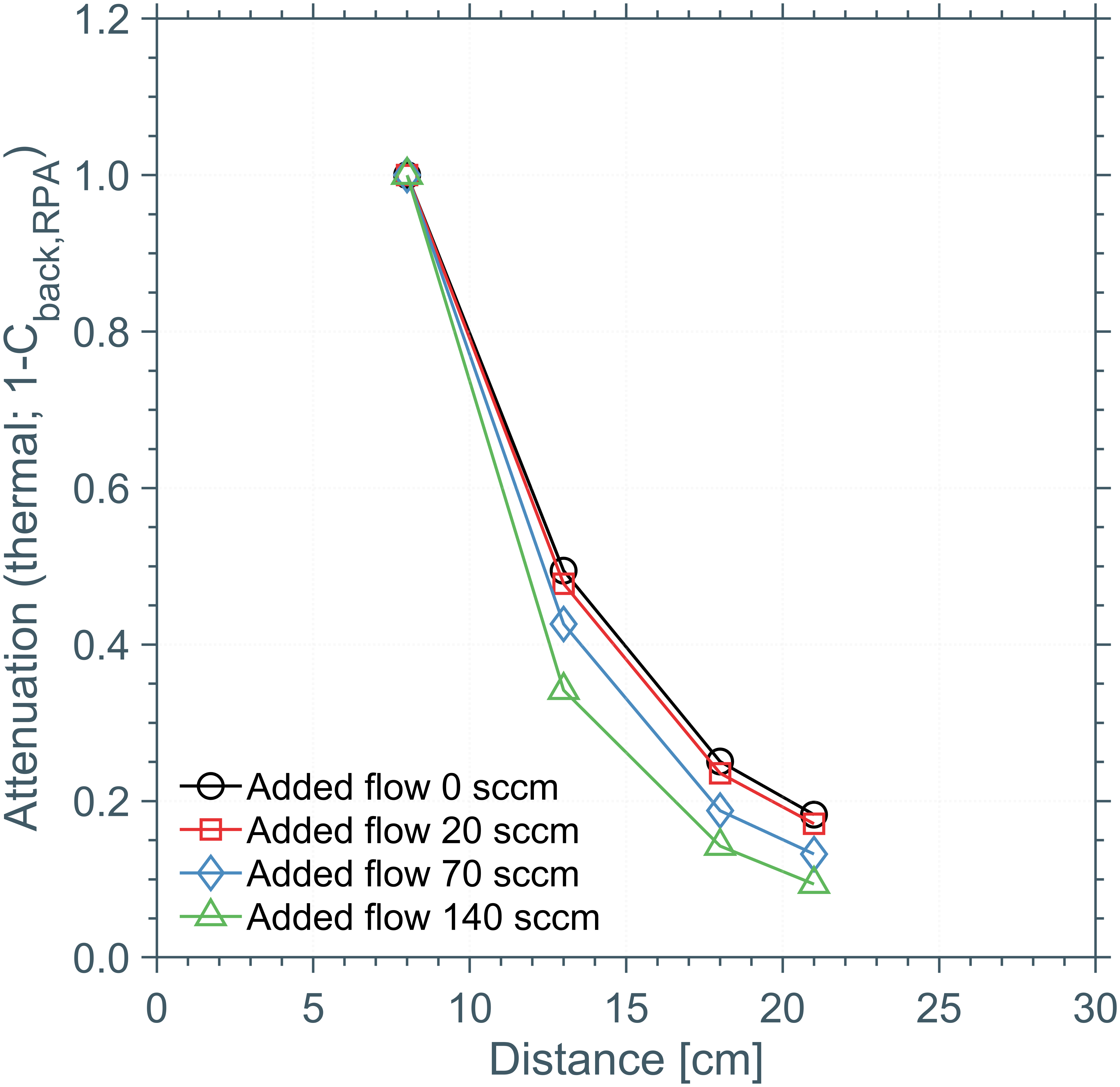}
\end{minipage}
\caption{\label{fig:atten_thermal_allflows}Thermal-flux-probe background-corrected ion-current attenuation versus distance for four added chamber argon flows (0, 18, 64, and 128~sccm). The correction uses $I_{\mathrm{ion,fast}}=I_{\mathrm{ion}}\left(1-C_{\mathrm{back}}\right)$ with $C_{\mathrm{back}}$ taken from (left) double-sided probe ratios, (middle) four-sided probe ratios, and (right) the RPA low-energy fraction. Each curve is normalized to $A(z_{\mathrm{ref}};Q_{\mathrm{add}})=1$.}
\end{figure}

The thermal flux probe also collects an ion current, $I_{\mathrm{ion}}$, which can include both fast-ion flux and facility-generated low-energy ions. A corrected thermal flux probe ion current was defined as
\begin{equation}
I_{\mathrm{ion,fast}}=I_{\mathrm{ion}}\left(1-C_{\mathrm{back}}\right),
\end{equation}
where $C_{\mathrm{back}}$ was taken independently from $C_{\mathrm{back,2side}}$, $C_{\mathrm{back,4side}}$, and $C_{\mathrm{low}}^{\mathrm{RPA}}$.
Here, $I_{\mathrm{ion,fast}}$ is a corrected on-axis estimate used in the thermal flux probe power balance. It depends on the selected $C_{\mathrm{back}}$, the thermal flux probe bias, and collection effects, and therefore it is not an independent energy-resolved measurement equivalent to $I_{\mathrm{high}}$ from the RPA. Figure~\ref{fig:atten_thermal_allflows} shows the corresponding corrected thermal flux probe attenuation for all added flows. Using different estimates of $C_{\mathrm{back}}$ has a weak effect on the magnitude of the corrected attenuation, while the same flow ordering and monotonic axial decay are preserved.

\newpage

\section{Fast-ion flux attenuation for different flows}\label{app:fast_ions_atten}

\begin{figure}[ht]
\centering
\begin{minipage}[t]{0.49\linewidth}
\centering
\includegraphics[height=4cm]{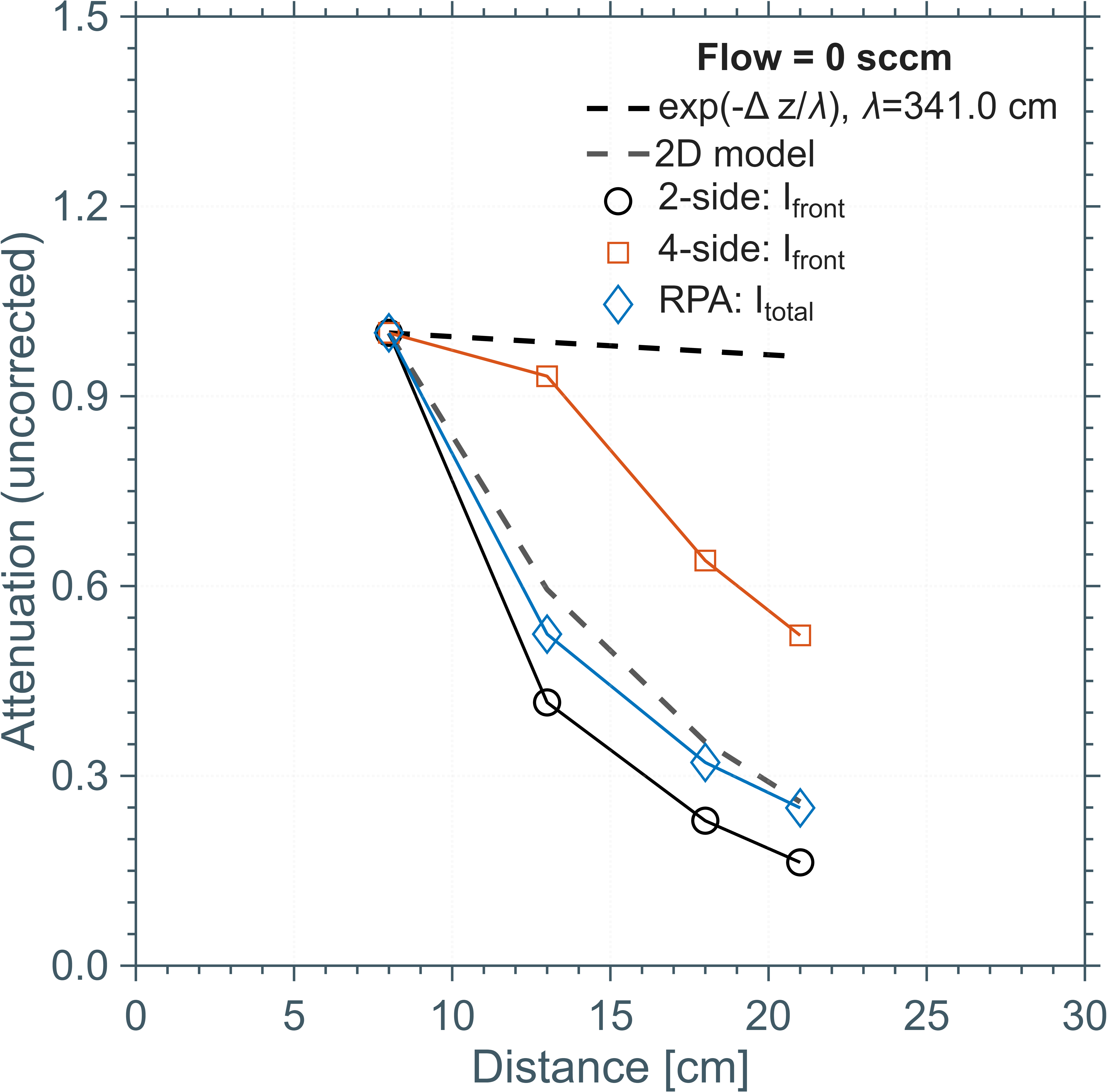}
\end{minipage}\hfill
\begin{minipage}[t]{0.49\linewidth}
\centering
\includegraphics[height=4cm]{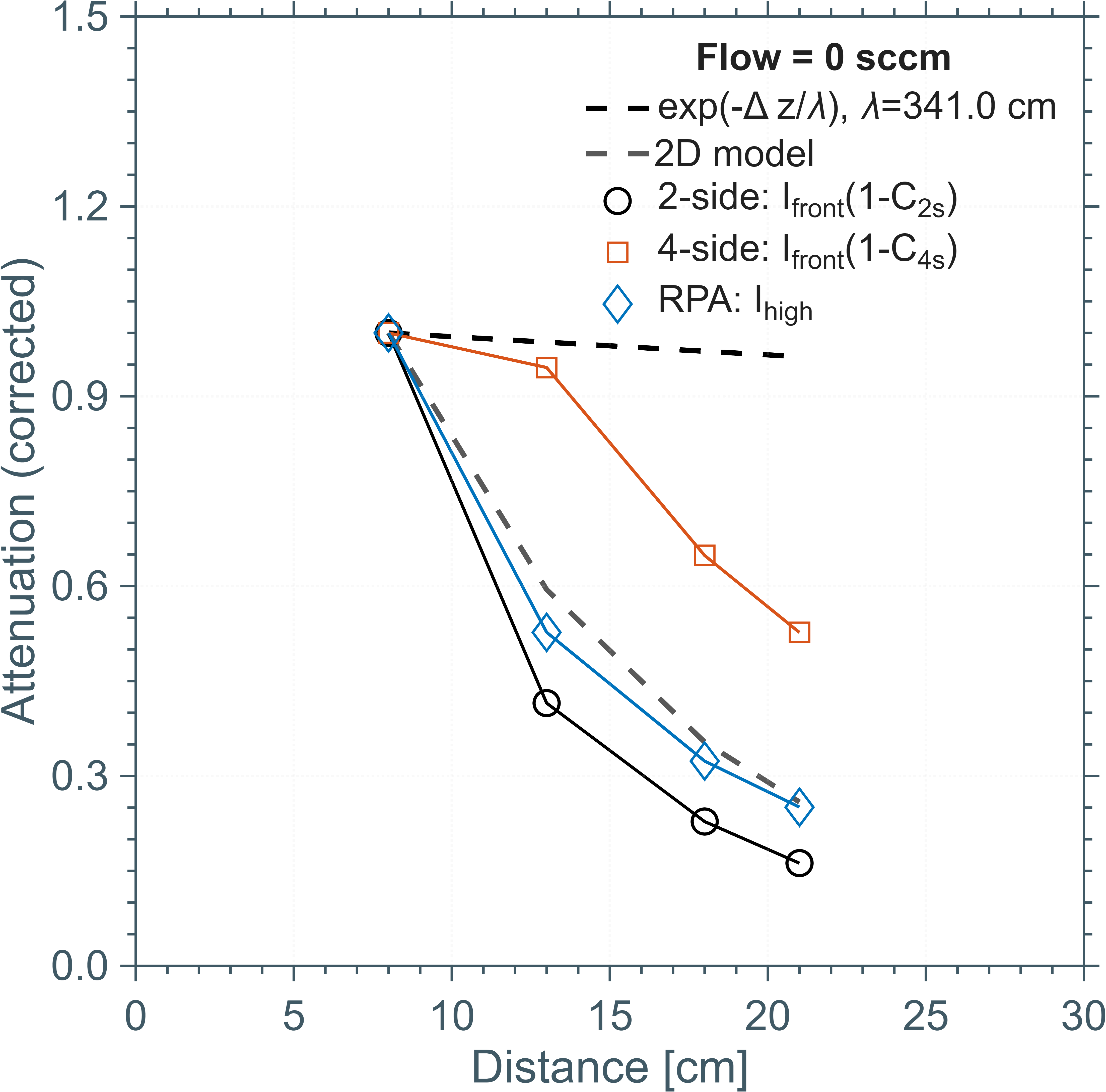}
\end{minipage}
\caption{\label{fig:atten_uncorr_corr_flow0}Representative attenuation comparison at $Q_{\mathrm{add}}=0$~sccm. Left: uncorrected attenuation of the double-sided and four-sided planar-probe front currents and the RPA total current, with the exponential reference curve and the quasi-2D model curve. Right: background-corrected attenuation for the planar probes, $I_{\mathrm{front}}\left(1-C_{\mathrm{back}}\right)$, compared with the RPA high-energy current $I_{\mathrm{high}}$ (with $C_{2s}\equiv C_{\mathrm{back,2side}}$ and $C_{4s}\equiv C_{\mathrm{back,4side}}$ as labeled in the figure), and with the same exponential and quasi-2D reference curves. All curves are normalized to $A(z_{\mathrm{ref}};Q_{\mathrm{add}})$.}
\end{figure}

\begin{figure}[ht]
\centering
\begin{minipage}[t]{0.49\linewidth}
\centering
\includegraphics[height=4cm]{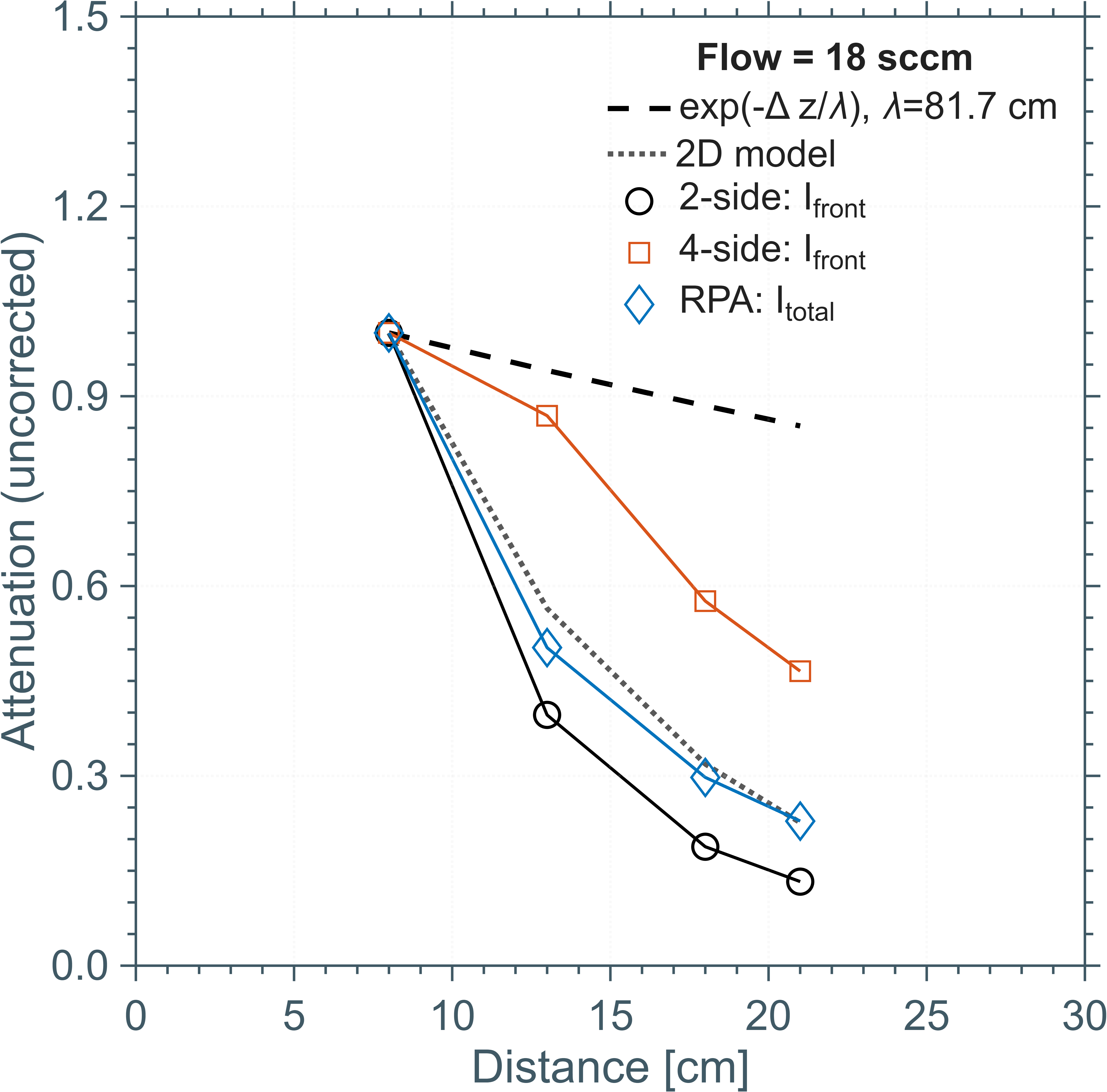}
\end{minipage}\hfill
\begin{minipage}[t]{0.49\linewidth}
\centering
\includegraphics[height=4cm]{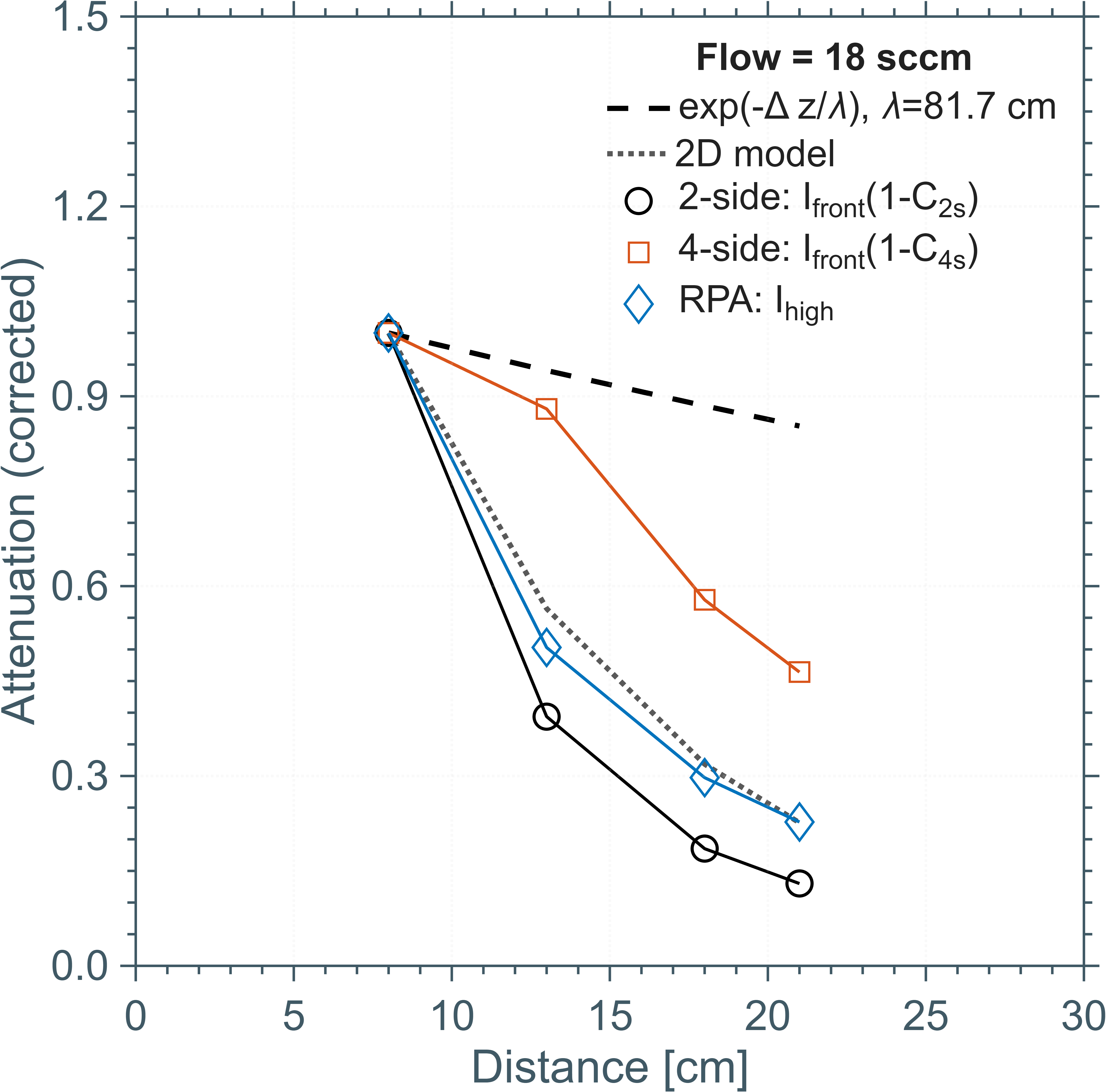}
\end{minipage}
\caption{\label{fig:atten_uncorr_corr_flow18}Representative attenuation comparison at $Q_{\mathrm{add}}=18$~sccm. Left: uncorrected attenuation of the double-sided and four-sided planar-probe front currents and the RPA total current, with the exponential reference curve and the quasi-2D model curve. Right: background-corrected attenuation for the planar probes, $I_{\mathrm{front}}\left(1-C_{\mathrm{back}}\right)$, compared with the RPA high-energy current $I_{\mathrm{high}}$ (with $C_{2s}\equiv C_{\mathrm{back,2side}}$ and $C_{4s}\equiv C_{\mathrm{back,4side}}$ as labeled in the figure), and with the same exponential and quasi-2D reference curves. All curves are normalized to $A(z_{\mathrm{ref}};Q_{\mathrm{add}})$.}
\end{figure}

\begin{figure}[ht]
\centering
\begin{minipage}[t]{0.49\linewidth}
\centering
\includegraphics[height=4cm]{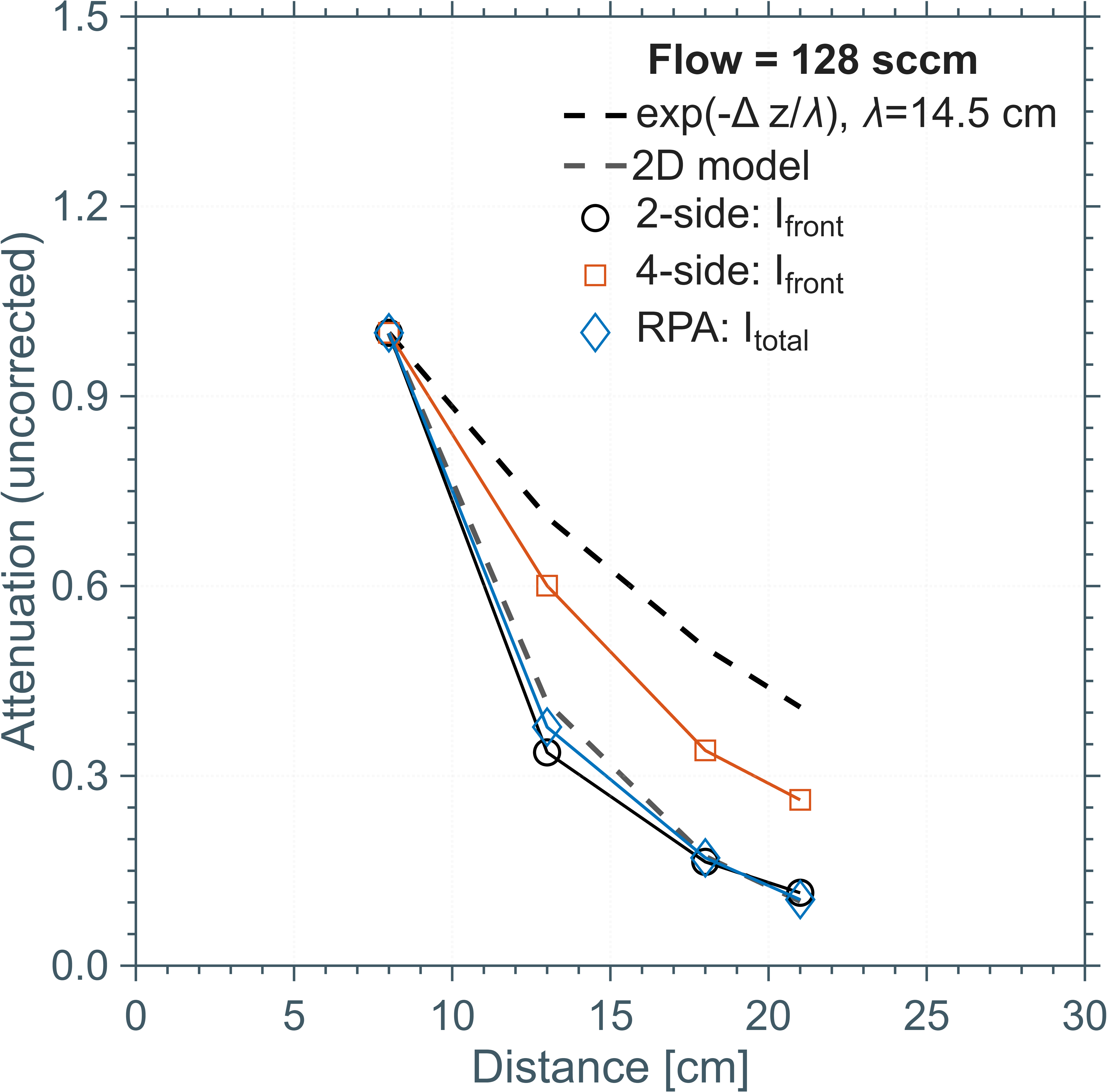}
\end{minipage}\hfill
\begin{minipage}[t]{0.49\linewidth}
\centering
\includegraphics[height=4cm]{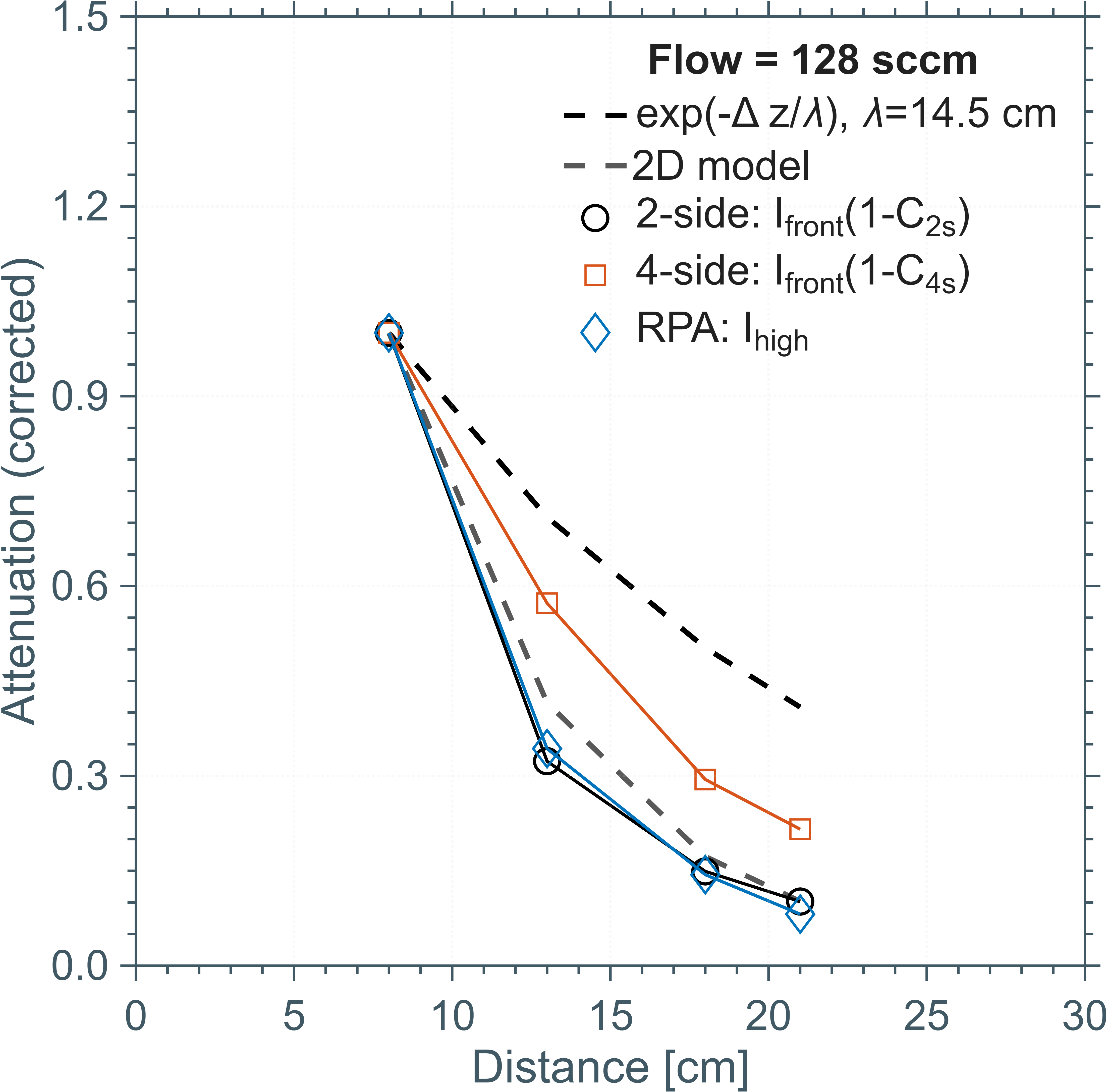}
\end{minipage}
\caption{\label{fig:atten_uncorr_corr_flow128}Representative attenuation comparison at $Q_{\mathrm{add}}=128$~sccm. Left: uncorrected attenuation of the double-sided and four-sided planar-probe front currents and the RPA total current, with the exponential reference curve and the quasi-2D model curve. Right: background-corrected attenuation for the planar probes, $I_{\mathrm{front}}\left(1-C_{\mathrm{back}}\right)$, compared with the RPA high-energy current $I_{\mathrm{high}}$ (with $C_{2s}\equiv C_{\mathrm{back,2side}}$ and $C_{4s}\equiv C_{\mathrm{back,4side}}$ as labeled in the figure), and with the same exponential and quasi-2D reference curves. All curves are normalized to $A(z_{\mathrm{ref}};Q_{\mathrm{add}})$.}
\end{figure}

\newpage
\nocite{*}
\bibliography{aipsamp}

@inbook{Thermal_Probe_AIAA,
author = {Stanislav Musikhin and Ivan Romadanov and Yevgeny Raitses},
title = {Background Pressure Effects on the Ion Source/Thruster Plume},
booktitle = {AIAA SCITECH 2026 Forum},
chapter = {},
pages = {},
doi = {10.2514/6.2026-2134}
}

@article{Brown_Facility,
author = {Brown, Daniel L. and Gallimore, Alec D.},
title = {Evaluation of Facility Effects on Ion Migration in a Hall Thruster Plume},
journal = {Journal of Propulsion and Power},
volume = {27},
number = {3},
pages = {573-585},
year = {2011},
doi = {10.2514/1.B34068},
}

@inproceedings{kahn2005,
  title = {Low-{{Energy End-Hall Ion Source Characterization}} at {{Millitorr Pressures}}},
  booktitle = {48th {{Annual Technical Conference Proceedings}}},
  author = {Kahn, J R},
  year = 2005,
  langid = {english}
}

@article{Walker_Backpressure,
author = {Walker, Mitchell L. R. and Victor, Allen L. and Hofer, Richard R. and Gallimore, Alec D.},
title = {Effect of Backpressure on Ion Current Density Measurements in Hall Thruster Plumes},
journal = {Journal of Propulsion and Power},
volume = {21},
number = {3},
pages = {408-415},
year = {2005},
doi = {10.2514/1.7713},
}

@article{Stahl_calorimetric,
    author = {Stahl, Marc and Trottenberg, Thomas and Kersten, Holger},
    title = {A calorimetric probe for plasma diagnostics},
    journal = {Review of Scientific Instruments},
    volume = {81},
    number = {2},
    pages = {023504},
    year = {2010},
    month = {02},
    issn = {0034-6748},
    doi = {10.1063/1.3276707},
}

@article{Foster_ground_test_impact,
    author = {Foster, John E. and Topham, Tyler J.},
    title = {A review of the impact of ground test-related facility effects on gridded ion thruster operation and performance},
    journal = {Physics of Plasmas},
    volume = {31},
    number = {3},
    pages = {030501},
    year = {2024},
    month = {03},
    issn = {1070-664X},
    doi = {10.1063/5.0173655},
}

@article{Nakles_Background_IVDF,
author = {Nakles, Michael R. and Hargus, William A.},
title = {Background Pressure Effects on Ion Velocity Distribution Within a Medium-Power Hall Thruster},
journal = {Journal of Propulsion and Power},
volume = {27},
number = {4},
pages = {737-743},
year = {2011},
doi = {10.2514/1.48027},
}

@article{Langmuir_Johnson,
    author = {Johnson, J. D. and Holmes, A. J. T.},
    title = {Edge effect correction for small planar Langmuir probes},
    journal = {Review of Scientific Instruments},
    volume = {61},
    number = {10},
    pages = {2628-2631},
    year = {1990},
    month = {10},
    issn = {0034-6748},
    doi = {10.1063/1.1141849},
}

@article{Langmuir_Sheridan,
    author = {Sheridan, T. E.},
    title = {How big is a small Langmuir probe?},
    journal = {Physics of Plasmas},
    volume = {7},
    number = {7},
    pages = {3084-3088},
    year = {2000},
    month = {07},
    issn = {1070-664X},
    doi = {10.1063/1.874162},
}

@article{Langmuir_Hershkowitz,
    author = {Lee, Dongsoo and Hershkowitz, Noah},
    title = {Ion collection by planar Langmuir probes: Sheridan’s model and its verification},
    journal = {Physics of Plasmas},
    volume = {14},
    number = {3},
    pages = {033507},
    year = {2007},
    month = {03},
    issn = {1070-664X},
    doi = {10.1063/1.2715557},
}

@article{Smirnov_CHT_Plasma,
    author = {Smirnov, A. and Raitses, Y. and Fisch, N. J.},
    title = {Plasma measurements in a 100 W cylindrical Hall thruster},
    journal = {Journal of Applied Physics},
    volume = {95},
    number = {5},
    pages = {2283-2292},
    year = {2004},
    month = {03},
    issn = {0021-8979},
    doi = {10.1063/1.1642734}
}

@inproceedings{raitses2007aedc,
  title={AEDC plume measurements using bi-directional ion flux probes},
  author={Raitses, Yevgeny and Moeller, Trevor and Szabo, James},
  booktitle={the proceedings of the 30th International Electric Propulsion Conference, Florence, Italy},
  pages={2007--334},
  year={2007}
}

@article{lieberman1994principles,
  title={Principles of plasma discharges and materials processing},
  author={Lieberman, Michael A and Lichtenberg, Allan J},
  journal={MRS Bulletin},
  volume={30},
  number={12},
  pages={899--901},
  year={1994}
}

@article{Boyd_Plume_Model,
author = {Boyd, Iain D.},
title = {Review of Hall Thruster Plume Modeling},
journal = {Journal of Spacecraft and Rockets},
volume = {38},
number = {3},
pages = {381-387},
year = {2001},
doi = {10.2514/2.3695},
}

@article{Economou_2008,
doi = {10.1088/0022-3727/41/2/024001},
url = {https://doi.org/10.1088/0022-3727/41/2/024001},
year = {2008},
month = {jan},
publisher = {},
volume = {41},
number = {2},
pages = {024001},
author = {Economou, Demetre J},
title = {Fast (tens to hundreds of eV) neutral beams for materials processing},
journal = {Journal of Physics D: Applied Physics}
}

@article{Hopf_2021,
doi = {10.1088/1741-4326/ac227a},
url = {https://doi.org/10.1088/1741-4326/ac227a},
year = {2021},
month = {sep},
publisher = {IOP Publishing},
volume = {61},
number = {10},
pages = {106032},
author = {Hopf, C. and Starnella, G. and den Harder, N. and Fantz, U.},
title = {Neutral beam injection for fusion reactors: technological constraints versus functional requirements},
journal = {Nuclear Fusion}
}

@inbook{Tajmar_SMART,
author = {Martin Tajmar and Bernard Foing and Jose Gonzalez and Giovanni Noci and Walter Schmidt and Franck Darnon},
title = {Charge-Exchange Plasma Contamination on SMART-1: First Measurements and Model Verification},
booktitle = {40th AIAA/ASME/SAE/ASEE Joint Propulsion Conference and Exhibit},
chapter = {},
pages = {},
doi = {10.2514/6.2004-3437},
URL = {https://arc.aiaa.org/doi/abs/10.2514/6.2004-3437},
eprint = {https://arc.aiaa.org/doi/pdf/10.2514/6.2004-3437}
}

@article{Stotler_NSTX,
    author = {Stotler, D. P. and Scotti, F. and Bell, R. E. and Diallo, A. and LeBlanc, B. P. and Podestà, M. and Roquemore, A. L. and Ross, P. W.},
    title = {Midplane neutral density profiles in the National Spherical Torus Experiment},
    journal = {Physics of Plasmas},
    volume = {22},
    number = {8},
    pages = {082506},
    year = {2015},
    month = {08},
    issn = {1070-664X},
    doi = {10.1063/1.4928372},
}

@inproceedings{huang2016facility,
  title={Facility effect characterization test of NASA’s HERMeS Hall thruster},
  author={Huang, Wensheng and Kamhawi, Hani and Haag, Thomas},
  booktitle={52nd AIAA/SAE/ASEE Joint Propulsion Conference},
  pages={4828},
  year={2016}
}

@article{snyder2020effects,
  title={Effects of background pressure on SPT-140 Hall thruster performance},
  author={Snyder, John Steven and Lenguito, Giovanni and Frieman, Jason D and Haag, Thomas W and Mackey, Jonathan A},
  journal={Journal of Propulsion and Power},
  volume={36},
  number={5},
  pages={668--676},
  year={2020},
  publisher={American Institute of Aeronautics and Astronautics}
}

@inproceedings{passaro2004plasma,
  title={Plasma thruster plume simulation: Effect of vacuum chamber environment},
  author={Passaro, Andrea and Vicini, Alessandro and Biagioni, Leonardo},
  booktitle={35th AIAA Plasmadynamics and Lasers Conference},
  pages={2357},
  year={2004}
}

@article{jin2022vacuum,
  title={Vacuum control system for the space plasma environment research facility},
  author={Jin, Chenggang and Zhang, Yongqi and Ling, Wenbin and Liu, Manxing and Chen, Chunxi and Li, Yunxuan and Peng, Zhiyong and Lu, Yaowen and Li, Liyi and others},
  journal={Journal of Vacuum Science \& Technology B},
  volume={40},
  number={3},
  year={2022},
  publisher={AIP Publishing}
}

@article{fisch2008plasma,
  title={Plasma plume of annular and cylindrical Hall thrusters},
  author={Fisch, NJ and Raitses, Y},
  journal={IEEE transactions on plasma science},
  volume={36},
  number={4},
  pages={1204--1205},
  year={2008},
  publisher={IEEE}
}

@phdthesis{azziz2007experimental,
  title={Experimental and theoretical characterization of a Hall thruster plume},
  author={Azziz, Yassir},
  year={2007},
  school={Massachusetts Institute of Technology}
}

@article{raitses2007enhanced,
  title={Enhanced performance of cylindrical Hall thrusters},
  author={Raitses, Y and Smirnov, A and Fisch, Nathaniel J},
  journal={Applied physics letters},
  volume={90},
  number={22},
  year={2007},
  publisher={AIP Publishing}
}

@inproceedings{frieman2018long,
  title={Long duration wear test of the NASA HERMeS Hall thruster},
  author={Frieman, Jason D and Kamhawi, Hani and Williams, George and Huang, Wensheng and Herman, Daniel A and Peterson, Peter Y and Gilland, James H and Hofer, Richard R},
  booktitle={2018 Joint Propulsion Conference},
  pages={4645},
  year={2018}
}

@article{dorf2004electrostatic,
  title={Electrostatic probe apparatus for measurements in the near-anode region of Hall thrusters},
  author={Dorf, L and Raitses, Y and Fisch, NJ},
  journal={Review of Scientific Instruments},
  volume={75},
  number={5},
  pages={1255--1260},
  year={2004},
  publisher={American Institute of Physics}
}

@article{staack2004shielded,
  title={Shielded electrostatic probe for nonperturbing plasma measurements in Hall thrusters},
  author={Staack, D and Raitses, Y and Fisch, NJ},
  journal={Review of Scientific Instruments},
  volume={75},
  number={2},
  pages={393--399},
  year={2004},
  publisher={American Institute of Physics}
}

@article{farnell2017recommended,
  title={Recommended practice for use of electrostatic analyzers in electric propulsion testing},
  author={Farnell, Casey C and Farnell, Cody C and Farnell, Shawn C and Williams, John D},
  journal={Journal of Propulsion and Power},
  volume={33},
  number={3},
  pages={638--658},
  year={2017},
  publisher={American Institute of Aeronautics and Astronautics}
}

@article{rosenfeldt2021use,
  title={The Use of Passive Thermal Probes for the Determination of Energy Fluxes in Atmospheric Pressure Plasmas},
  author={Rosenfeldt, Lukas and Hansen, Luka and Kersten, Holger},
  journal={IEEE Transactions on Plasma Science},
  volume={49},
  number={11},
  pages={3325--3335},
  year={2021},
  publisher={IEEE}
}

@article{thornton1978substrate,
  title={Substrate heating in cylindrical magnetron sputtering sources},
  author={Thornton, John A},
  journal={Thin solid films},
  volume={54},
  number={1},
  pages={23--31},
  year={1978},
  publisher={Elsevier}
}

@article{wiese2015energy,
  title={Energy influx measurements with an active thermal probe in plasma-technological processes},
  author={Wiese, Ruben and Kersten, Holger and Wiese, Georg and Bartsch, Ren{\'e}},
  journal={EPJ Techniques and Instrumentation},
  volume={2},
  number={1},
  pages={2},
  year={2015},
  publisher={Springer}
}

@article{bornholdt2013transient,
  title={Transient calorimetric diagnostics for plasma processing},
  author={Bornholdt, Sven and Kersten, Holger},
  journal={The European Physical Journal D},
  volume={67},
  number={8},
  pages={176},
  year={2013},
  publisher={Springer}
}

@article{schlichting2023retarding,
  title={A retarding field thermal probe for combined plasma diagnostics},
  author={Schlichting, Felix and Kersten, Holger},
  journal={EPJ Techniques and Instrumentation},
  volume={10},
  number={1},
  pages={1--17},
  year={2023},
  publisher={Springer}
}

@article{schlichting2022energy,
  title={Energy-dependent film growth of Cu and NiTi from a tilted DC magnetron sputtering source determined by calorimetric probe analysis},
  author={Schlichting, Felix and Thormaehlen, Lars and Cipo, Julia and Meyners, Dirk and Kersten, Holger},
  journal={Surface and Coatings Technology},
  volume={450},
  pages={129000},
  year={2022},
  publisher={Elsevier}
}

@article{el2022correlation,
  title={On the correlation between the TiN thin film properties and the energy flux of neutral sputtered atoms in direct current magnetron discharge},
  author={El Farsy, Abderzak and Pierson, Jean-Fran{\c{c}}ois and Gries, Thomas and de Poucques, Ludovic and Bougdira, Jamal},
  journal={Journal of Physics D: Applied Physics},
  volume={55},
  number={50},
  pages={505203},
  year={2022},
  publisher={IOP Publishing}
}

@article{van1986characterization,
  title={Characterization of a 3 cm Kaufman ion source with nitrogen feed gas},
  author={Van Vechten, D and Hubler, GK and Donovan, EP},
  journal={Vacuum},
  volume={36},
  number={11-12},
  pages={841--845},
  year={1986},
  publisher={Elsevier}
}

@article{lichten1963resonant,
  title={Resonant charge exchange in atomic collisions},
  author={Lichten, William},
  journal={Physical Review},
  volume={131},
  number={1},
  pages={229},
  year={1963},
  publisher={APS}
}

@article{swinehart1962beer,
  title={The beer-lambert law},
  author={Swinehart, Donald F},
  journal={Journal of chemical education},
  volume={39},
  number={7},
  pages={333},
  year={1962},
  publisher={ACS Publications}
}

@article{godyak2015comparative,
  title={Comparative analyses of plasma probe diagnostics techniques},
  author={Godyak, VA and Alexandrovich, BM},
  journal={Journal of Applied Physics},
  volume={118},
  number={23},
  year={2015},
  publisher={AIP Publishing}
}

@article{benedikt2021foundations,
  title={Foundations of measurement of electrons, ions and species fluxes toward surfaces in low-temperature plasmas},
  author={Benedikt, Jan and Kersten, Holger and Piel, Alexander},
  journal={Plasma sources science and technology},
  volume={30},
  number={3},
  pages={033001},
  year={2021},
  publisher={IOP Publishing}
}

@article{bromley2019symmetric,
  title={Symmetric charge exchange for intermediate velocity noble gas projectiles},
  author={Bromley, Steven and Sosolik, CE and Marler, JP},
  journal={Journal of Physics B: Atomic, Molecular and Optical Physics},
  volume={52},
  number={21},
  pages={215203},
  year={2019},
  publisher={IOP Publishing}
}

@inproceedings{chen2003langmuir,
  title={Langmuir probe diagnostics},
  author={Chen, Francis F},
  booktitle={IEEE-ICOPS Meeting, Jeju, Korea},
  volume={2},
  number={6},
  year={2003},
  organization={Citeseer}
}

@phdthesis{gauter2018calorimetric,
  title={Calorimetric investigation on plasma and ion beam sources used for thin film deposition},
  author={Gauter, Sven},
  year={2018},
  school={Christian-Albrechts Universit{\"a}t Kiel}
}

@article{dunaevsky2006plasma,
  title={Plasma acceleration from radio-frequency discharge in dielectric capillary},
  author={Dunaevsky, A and Raitses, Yevgeny and Fisch, NJ},
  journal={Applied physics letters},
  volume={88},
  number={25},
  year={2006},
  publisher={AIP Publishing}
}

@article{granstedt2008cathode,
  title={Cathode effects in cylindrical Hall thrusters},
  author={Granstedt, EM and Raitses, Y and Fisch, NJ},
  journal={Journal of Applied Physics},
  volume={104},
  number={10},
  year={2008},
  publisher={AIP Publishing}
}

@inproceedings{simmonds2020application,
  title={Application of Hall thrusters with modulated oscillations},
  author={Simmonds, Jacob B and Raitses, Yevgeny and Smolyakov, Andrei and Chapurin, Oleksandr and Chaplin, Vernon H},
  booktitle={AIAA Propulsion and Energy 2020 Forum},
  pages={3618},
  year={2020}
}

@inproceedings{Vawter2000IonBE,
  title={Ion Beam Etching of Compound Semiconductors},
  author={Gregory A. Vawter},
  year={2000},
  url={https://api.semanticscholar.org/CorpusID:136961949}
}

@inbook{Zhurin_Industrial_Ion_Sources,
publisher = {John Wiley \& Sons, Ltd},
isbn = {9783527635726},
title = {Ion and Plasma Sources for Science and Technology},
booktitle = {Industrial Ion Sources},
chapter = {11},
pages = {255-267},
doi = {https://doi.org/10.1002/9783527635726.ch11},
url = {https://onlinelibrary.wiley.com/doi/abs/10.1002/9783527635726.ch11},
eprint = {https://onlinelibrary.wiley.com/doi/pdf/10.1002/9783527635726.ch11},
year = {2011}
}

@article{RSI_probe,
    author = {Rovey, Joshua L. and Walker, Mitchell L. R. and Gallimore, Alec D. and Peterson, Peter Y.},
    title = {Magnetically filtered Faraday probe for measuring the ion current density profile of a Hall thruster},
    journal = {Review of Scientific Instruments},
    volume = {77},
    number = {1},
    pages = {013503},
    year = {2006},
    month = {01},
    issn = {0034-6748},
    doi = {10.1063/1.2149006},
}

@article{SPT_probe,
author = {MacDonald-Tenenbaum, Natalia and Pratt, Quinn and Nakles, Michael and Pilgram, Nickolas and Holmes, Michael and Hargus, William},
title = {Background Pressure Effects on Ion Velocity Distributions in an SPT-100 Hall Thruster},
journal = {Journal of Propulsion and Power},
volume = {35},
number = {2},
pages = {403-412},
year = {2019},
}

@article{Gahan2008,
    author = {Gahan, D. and Dolinaj, B. and Hopkins, M. B.},
    title = {Retarding field analyzer for ion energy distribution measurements at a radio-frequency biased electrode},
    journal = {Review of Scientific Instruments},
    volume = {79},
    number = {3},
    pages = {033502},
    year = {2008},
    month = {03},
    issn = {0034-6748},
    doi = {10.1063/1.2890100},
    url = {https://doi.org/10.1063/1.2890100},
}

@article{Ellmer2003,
  author  = {K. Ellmer and R. Wendt and K. Wiesemann},
  title   = {Interpretation of ion distribution functions measured by a combined energy and mass analyzer},
  journal = {International Journal of Mass Spectrometry},
  year    = {2003},
  volume  = {223--224},
  pages   = {679--693},
  doi     = {10.1016/S1387-3806(02)00940-5}
}

@book{Hutchinson_2002, place={Cambridge}, edition={2}, title={Principles of Plasma Diagnostics}, publisher={Cambridge University Press}, author={Hutchinson, I. H.}, year={2002}}

\end{document}